\documentclass[fleqn,usenatbib]{mnras}

\usepackage{newtxtext,newtxmath}

\usepackage[T1]{fontenc}

\DeclareRobustCommand{\VAN}[3]{#2}
\let\VANthebibliography\thebibliography
\def\thebibliography{\DeclareRobustCommand{\VAN}[3]{##3}\VANthebibliography}

\usepackage{graphicx}	
\usepackage{amsmath}	
\usepackage{xcolor}
\usepackage{longtable}
\usepackage{multirow}
\usepackage{pdflscape}

\title[Deep radius valley with \textit{Kepler} short cadence]{A deep radius valley revealed by \textit{Kepler} short cadence observations}

\author[Ho \& Van Eylen]{
Cynthia S. K. Ho\thanks{E-mail: sze.ho.20@ucl.ac.uk}
and Vincent Van Eylen
\\
Mullard Space Science Laboratory, University College London, Dorking RH5 6NT, UK \\
}

\date{Accepted 2022 December 19. Received 2022 December 2; in original form 2022 October 17}

\pubyear{2022}

\begin{document}
\label{firstpage}
\pagerange{\pageref{firstpage}--\pageref{lastpage}}
\maketitle

\begin{abstract}
    The characteristics of the radius valley, i.e., an observed lack of planets between 1.5-2 Earth radii at periods shorter than about 100 days, provide insights into the formation and evolution of close-in planets. We present a novel view of the radius valley by refitting the transits of 431 planets using Kepler 1-minute short cadence observations, the vast majority of which have not been previously analysed in this way. In some cases, the updated planetary parameters differ significantly from previous studies, resulting in a deeper radius valley than previously observed. This suggests that planets are likely to have a more homogeneous core composition at formation. Furthermore, using support-vector machines, we find that the radius valley location strongly depends on orbital period and stellar mass and weakly depends on stellar age, with $\partial \log {\left(R_{p, \text{valley}} \right)}/ \partial \log{P} = -0.096_{-0.027}^{+0.023}$, $\partial \log {\left(R_{p, \text{valley}} \right)}/ \partial \log{M_{\star}} = 0.231_{-0.064}^{+0.053}$, and $\partial \log {\left(R_{p, \text{valley}} \right)}/ \partial \log{\left( \text{age} \right)} = 0.033_{-0.025}^{+0.017}$. These findings favour thermally-driven mass loss models such as photoevaporation and core-powered mass loss, with a slight preference for the latter scenario. Finally, this work highlights the value of transit observations with short photometric cadence to precisely determine planet radii, and we provide an updated list of precisely and homogeneously determined parameters for the planets in our sample.
\end{abstract}



\defcitealias{fulton2018california}{F18}
\defcitealias{vaneylen2018asteroseismic}{V18}

\section{Introduction} \label{sect:intro}

The `radius valley', also known as the `radius gap', is the relative paucity of planets with sizes between about 1.5 and 2 Earth radii at orbital periods less than about 100 days. This phenomenon has been predicted theoretically due to the heavy radiation these close-in planets receive from their host star \citep[e.g.][]{owen2013kepler, lopez2013role} and was subsequently seen observationally \citep[e.g.][]{fulton2017california,  vaneylen2018asteroseismic, fulton2018california}. Several theories have been suggested to explain the physical origin of the radius valley. On one hand, thermally-driven mass loss scenarios have been proposed, which include photoevaporation \citep[e.g.][]{owen2013kepler,lopez2013role, owen2017evaporation} and core-powered mass loss \citep[e.g.][]{ginzburg2018core,gupta2019sculpting, gupta2020signatures} models. In these scenarios, the valley separates planets that have lost their atmosphere from those that have retained it. Alternatively, late gas-poor formation, where planets below the valley have formed atmosphere-free, may also be able to explain the origin of the valley \citep[e.g.][]{lee2014make, lee2016breeding, lopez2018how, cloutier2020evolution}. 

Observed characteristics of the radius valley can therefore reveal the properties of these close-in planets and their formation history. For example, in photoevaporation models, the location of the radius valley and its slope as a function of orbital period depend on the planetary composition and photoevaporation physics \citep{owen2017evaporation,mordasini2020planetary}. The valley's location and relative emptiness can therefore be used to infer the composition of planets surrounding it and their relative homogeneity \citep[e.g.][]{vaneylen2018asteroseismic}.
Planets located inside the radius valley may have a different composition or could be undergoing the final stages of atmospheric loss by thermally-driven mechanisms and hence may be important targets for further studies \citep[e.g.][]{owen2017evaporation, gupta2019sculpting, petigura2020two}. The valley's location as a function of orbital period can be used to distinguish between thermal mass-loss models, which exhibit a negative slope as a function of orbital period, and late gas-poor formation models which have the opposite slope \citep[e.g.][]{vaneylen2019orbital,cloutier2020evolution,vaneylen2021masses}. Within thermal mass-loss models, photoevaporation and core-powered mass loss models predict a different dependence of the valley's location on stellar mass and age \citep[e.g.][]{rogers2021photoevaporation}.

Observationally, these valley characteristics have been challenging to reliably ascertain. 
A deficit of planets with sizes around 1.5-2 Earth radii ($R_\oplus$) was first observed by \cite{fulton2017california} in a sample of 2025 planets, with stellar radii determined spectroscopically as part of the California-\textit{Kepler} survey (CKS). These planets were about a factor of two rarer than planets both smaller and larger. Independently, \cite{vaneylen2018asteroseismic} (\citetalias{vaneylen2018asteroseismic} hereafter) analysed a subset of this sample (117 planets), incorporating higher-precision stellar parameters using asteroseismology and refitting transit light curves to achieve a median uncertainty on planet sizes of 3.3\%. This study revealed the valley's slope as a function of orbital period for the first time, and suggested the radius valley may be very deep or even entirely empty.The tension between the valley's views of \cite{fulton2017california} and \citetalias{vaneylen2018asteroseismic} was further exacerbated when the precision of stellar parameters of the former study were further improved by \citet{fulton2018california} (\citetalias{fulton2018california} hereafter). Despite improving stellar uncertainties from 11\% to 3\% by incorporating Gaia parallaxes, the valley remained partially filled in, with its depth largely unchanged.

\cite{petigura2020two} investigated the discrepancy in the valley's depth between \citetalias{vaneylen2018asteroseismic} and \citetalias{fulton2018california} and concluded it is unlikely to be caused by differing sample sizes or differing values or uncertainties in stellar radii. The study argued that the $6.9\%$ dispersion in planetary radii is instead primarily caused by a discrepancy in the ratio of planet to stellar radii ($R_p/R_\star$)) determined from the transit fits. \citetalias{fulton2018california} used radius ratios from \cite{mullally2015planetary}, which fitted \textit{Kepler} 30-minute 'long cadence' observations, whereas \citetalias{vaneylen2018asteroseismic} used \textit{Kepler} 1-minute 'short cadence' observations, also used for orbital eccentricity determination and described in \cite{vaneylen2015eccentricity} and \cite{vaneylen2019orbital}.

Here, we seek to refit planet transits for the full subset of \citetalias{fulton2018california} for which short cadence observations are available. This increases the sample of planets relevant for the radius valley for which short cadence transit fits are used from 60 in \citetalias{vaneylen2018asteroseismic} to 431 here. Furthermore, we will apply the methods to determine the valley's location and slope used by \citetalias{vaneylen2018asteroseismic}, notably the use of support vector machines, to this larger sample, and we expand to other dimensions such as stellar mass and age.

In Section~\ref{sect:methods}, we describe the sample and methodology used to analyse the radius valley. In Section~\ref{sect:results}, we present the results of this analysis, such as revised planetary sizes, the depth of the valley, and its dependence on parameters such as the orbital period and stellar mass and age. These findings are compared to other observational studies and theoretical models in Section~\ref{sect:discussion}. Finally, we provide conclusions in Section~\ref{sect:conclusion}.

\section{Methods} \label{sect:methods}
\subsection{Sample selection}
We use the sample of planets for which stellar parameters are available from \citetalias{fulton2018california} as a starting point. To focus on the radius valley, we limit the sample to planets with radii $1 \leq R_p/R_{\earth} \leq 4$ and orbital periods $1 \leq P/\text{days} \leq 100$, resulting in a sample of 1272 planets (for comparison, applying the same period and radius cuts to the sample studied by \citetalias{vaneylen2018asteroseismic} leaves 74 planets). As \textit{Kepler} 1-minute short cadence observations may yield superior precision \citep{petigura2020two}, we further limit our sample to those planets for which at least 6 months of \textit{Kepler} short cadence data are available.

To avoid issues with transit fitting related to transit timing variations (TTVs), we also remove planets with known TTVs based on  the catalogue by \citet{holczer2016transit}. We further exclude KOI-1576.03, as we find that the short cadence data suggested an orbital period different to the one recorded in the archive. Furthermore, we exclude any planets that are classified as potential false positives in \citet{petigura2017california}. The results in a  total sample size of 431 planets, 60 of which  have parameters previously analysed by \citetalias{vaneylen2018asteroseismic} and 371 which have not (a further 14 planets in \citetalias{vaneylen2018asteroseismic} have TTVs and are not reanalysed here).

\subsection{Data reduction}
The 1-minute \textit{Kepler} short cadence Pre-search Data Conditioning SAP \citep[PDCSAP,][]{stumpe2012kepler, smith2012kepler} light curves of these targets are downloaded from the NASA Mikulski Archive for Space Telescopes (MAST) database using the \texttt{lightkurve} package \citep{2018ascl.soft12013L}, which incorporates \texttt{astroquery} \citep{astroquery2019} and \texttt{astropy} \citep{astropy2013, astropy2018} dependencies. We only retain data within 0.2 days before the estimated ingress and after the estimated egress of the transits of the planets of interest, using the transit durations and mid-times in \citet{mullally2015planetary} as the expected transit locations. For multi-planet systems, we only retain transits of planets that are within our sample. We remove data outliers that lie beyond 6$\sigma$ from the median after masking the transits. We then flatten the transits by dividing the data points with the slope obtained by performing linear regression on the data points immediately before ingress and after egress, to remove long-term systematic trends present in the transits. We then again remove data outliers with $\sigma=5$ to further clean the data.

\subsection{Stellar multiplicity}
Around 46\% of solar-type stars have at least one stellar companion \citep{raghavan2010survey}. When a planet orbits a single star, the transit depth $\delta$ is approximately given by
\begin{equation}
    \delta = \frac{\Delta F}{F_{\text{tot}}} \approx \frac{R_p^2}{R_{\star}^2}
    \label{eq:trans_depth_rcf}
\end{equation}
where $F_{\text{tot}}$ is the total stellar flux, $\Delta F$ is the change in stellar flux, and $R_p$ and $R_{\star}$ are the planetary and stellar radius respectively. However, in a multi-stellar system, the total flux is the sum of fluxes of all stars in the system, but the change in flux during transit is only relative to the star(s) which the planet transits \citep{furlan2017kepler}. Therefore it is important to take into account the effect of nearby stars on the light curve flux. 

\citet{furlan2017kepler} compiled a catalogue of \textit{Kepler} Objects of Interest (KOI) observations with adaptive optics, speckle interferometry, lucky imaging, and imaging from space with the Hubble Space Telescope. The typical point spread function (PSF) widths and sensitivities ($\Delta m$) are different for every observation method, target and bandpass, hence whether stellar companions are detected is dependent on the above factors. For example, \citet{furlan2017kepler} were able to detect a median $\Delta m \sim 8\text{mag}$ with Keck in the \textit{K} band at a separation of $\sim$0.5'', but only at $\sim$2.5'' at Lick in the \textit{J} or \textit{H} bands. About 30\% of KOIs observed in \citet{furlan2017kepler} have at least one companion detected within 4'' \citep{furlan2017kepler}, and given a mean distance of 616pc for the 431 planets in our sample computed from distances reported in \citet{mathur2017revised}, corresponds to $\sim$2464AU.

Here, we adopt the `radius correction factor' (RCF), given in \citet{furlan2017kepler} as
\begin{equation}
    \text{RCF} = \frac{R_{p, \text{corr}}}{R_{p, \text{uncorr}}}
    \label{eq:rcf_rp}
\end{equation}
and multiply the normalised \textit{Kepler} light curve fluxes by RCF\textsuperscript{2}, and subtract $\left(\text{RCF}^2 -1\right)$ to re-normalise, to obtain the corrected light curve reflecting the transit of one planet orbiting around one star. 137 of the 431 planets in our sample (32\%) have RCF measurements from \citet{furlan2017kepler}.

\subsection{Transit fitting} \label{subsect:trans_fit}
We use the \texttt{exoplanet} package \citep{foremanmackey2021exoplanet} to generate a transit light curve model with quadratic stellar limb darkening, and then run a Hamiltonian Monte Carlo (HMC) algorithm implemented in \texttt{PyMC3} \citep{salvatier2016probabilistic} to perform fitting and determine orbital parameter posteriors. We also implement a Gaussian Process (GP) model \citep{rasmussen2006gaussian} to account for correlated noise in the light curves. However, for Kepler-65 and Kepler-21 A, we do not fit for a GP model due to convergence constraints. The parameters fitted for each planet are orbital period ($P$), transit mid-time ($t_0$), ratio between planetary and stellar radii ($R_p/R_{\star}$), impact parameter ($b$), eccentricity ($e$), argument of periapsis ($\omega$), and stellar density ($\rho_{\star}$). For each light curve, we further include two quadratic stellar limb darkening parameters ($u_0$ and $u_1$) for the host star, with bounds $0 < u_0, u_1 < 1$ and implemented with the \citet{kipping2013efficient} reparameterisation in \texttt{exoplanet}, the transit jitter ($\sigma_{\text{lc}}$), and two parameters describing the GP contribution ($\sigma_{\text{gp}}$, $\rho_{\text{gp}}$).

We initialise the HMC chains by using values presented in the \textit{Kepler} Q1-16 dataset \citep{mullally2015planetary} for $P$, $t_0$, $R_p/R_\star$, and $b$. We set the system to begin with near-circular orbits, with $e = 0.01$ and $\omega = 0.01$ rad. We take initial stellar densities from \citet{fulton2018california}. We use the Exoplanet Characterization ToolKit (ExoCTK) \citep{exoctk2021} to estimate the initial $u_0$ and $u_1$, which takes the stellar temperature, surface gravity, and metallicity, which we use values from the \textit{Kepler} Q1-16 dataset \citep{mullally2015planetary}, as inputs.

We apply Gaussian priors to $P$, $t_0$, $u_0$, $u_1$, and $\rho_{\star}$, using the initial guesses as the mean, and $\sigma_P = 2 \times 10^{-5}$ days, $\sigma_{t_0} = 10^{-3}$ days, $\sigma_u = 0.2$, and the $\rho_{\star}$ uncertainty from \citet{fulton2018california} if available, and \citet{mullally2015planetary} otherwise. A beta distribution prior according to \citet{vaneylen2019orbital} is placed on $e$, which is
\begin{equation}
    \text{PDF}(e, \alpha, \beta) \propto e^{\alpha} (1-e)^{\beta}
    \label{eq:beta}
\end{equation}
with $\alpha = 1.58$ and $\beta = 4.4$ for system with only one transiting planet, and $\alpha=1.52$ and $\beta = 29$ for a multi-transiting-planet system.

\section{Results} \label{sect:results}
\subsection{Revised planet parameters} \label{subsect:params}
\begin{figure}
    \centering
    \includegraphics[width=\columnwidth]{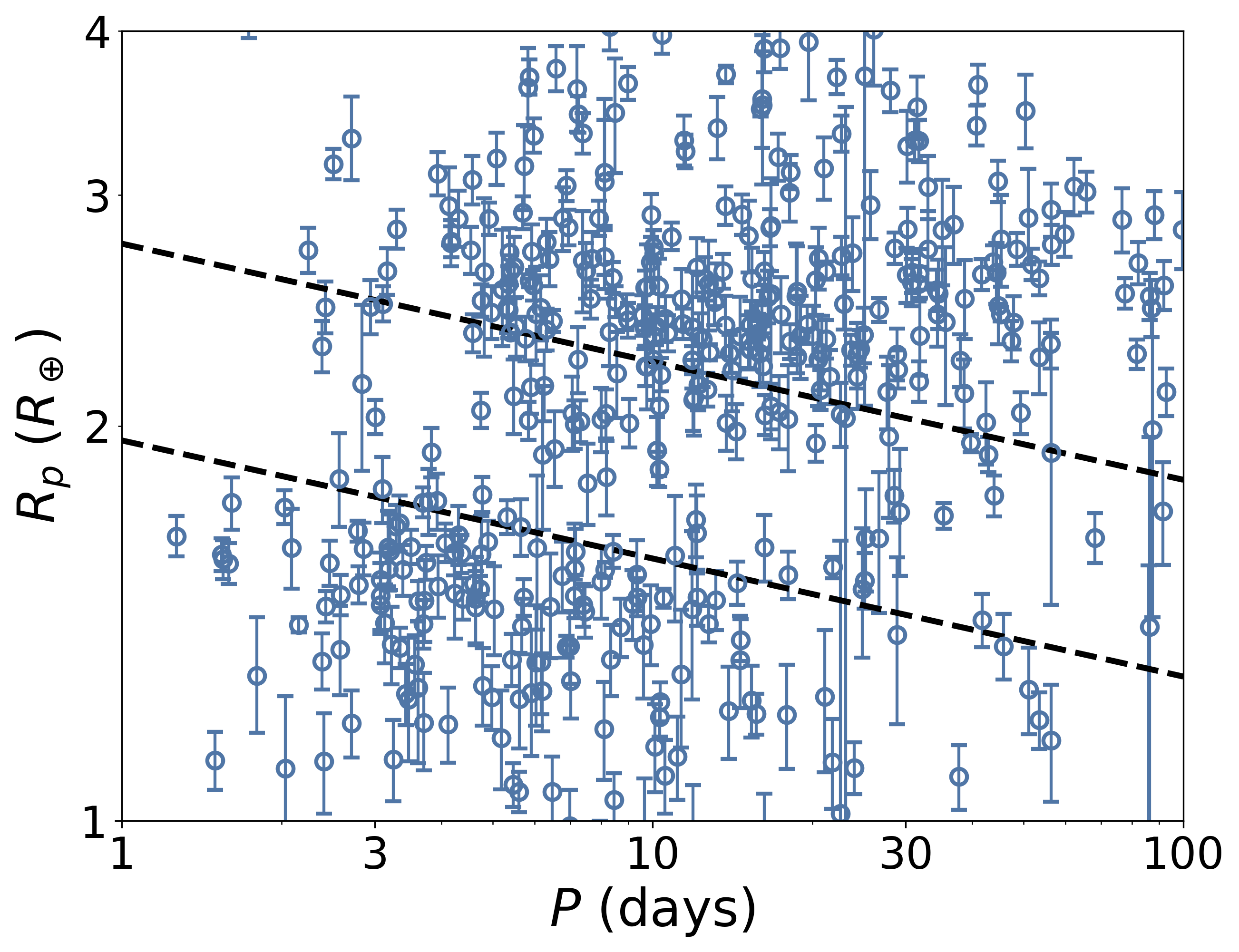}
    \caption{Radius-period plot of all 431 planets refitted in this work. The black dotted lines indicate the upper and lower boundaries of the radius valley defined in \citetalias{vaneylen2018asteroseismic}.}
    \label{fig:rp_plot_new_full}
\end{figure}

We report the updated orbital periods ($P$), planetary-to-stellar-radii ratio ($R_p/R_{\star}$), planetary radii ($R_p$), the number of transits in the fitted light curve ($N_{\text{tr}}$), and their uncertainties of the 431 planets fitted in this sample in Table~\ref{tab:planet_params}. The full list of parameters are provided in Appendix~\ref{appendix}. We convert our $R_p/R_{\star}$ to $R_p$ using the updated stellar parameters available: values used in \citetalias{vaneylen2018asteroseismic} from asteroseismology \citep[i.e. taken from][]{huber2013fundamental, silvaaguirre2015ages, lundkvist2016hot} if the planets are included in the \citetalias{vaneylen2018asteroseismic} samples, and \citetalias{fulton2018california} otherwise. Full homogeneity is lost by using stellar radii from two sources. To investigate the consequences of this, we compute the difference, $\delta_{R_p}$, between the planetary radii obtained by converting $R_p/R_{\star}$ to $R_p$ using $R_{\star}$ from \citetalias{fulton2018california} and \citetalias{vaneylen2018asteroseismic}, and found the mean $\delta$, $\bar{\delta} = 0.03 \pm 0.11$, hence $\delta = 0$ (no difference) is well within $1 \sigma$, and we conclude that there is no substantial drawbacks of using multiple sources. This sample of 431 planets with updated parameters is plotted on the radius-orbital period plot as shown in Figure~\ref{fig:rp_plot_new_full}.

We present the typical uncertainties of $R_p/R_{\star}$ and $R_p$ of planets fitted in this work, compared with \citetalias{fulton2018california} and \citetalias{vaneylen2018asteroseismic} in Table~\ref{tab:typical_err}. For our newly fitted results, we find that $R_p/R_{\star}$ has a mean uncertainty of 4.76\%, and a median uncertainty of 3.44\%. This is smaller when compared to the mean and median uncertainties of 8.73\% and 4.07\% respectively for the \citetalias{fulton2018california} sample. For $R_p$, we find a mean and median uncertainty of 6.09\% and 4.70\% in our work, again smaller compared to 10.00\% and 5.22\% for \citetalias{fulton2018california}. However, they are larger than that of \citetalias{vaneylen2018asteroseismic}, possibly due to \citetalias{vaneylen2018asteroseismic} analysing brighter stars, hence the light curves are less noisy. This can be seen by comparing the mean photometric flux error of the transit light curves fitted: 1007 parts per million (ppm) for this work, and 271~ppm for \citetalias{vaneylen2018asteroseismic}. 

We select the planets in \citetalias{fulton2018california} that are in common with the planets in our sample, and examine the change in $R_p$. We find that 217 planets have a larger $R_p$ after refitting, and smaller for 214 planets. Of the planets whose sizes have increased, the mean change is 9.79\%, and 8.69\% for planets with reduced sizes. Considering all 431 planets, our new results change $R_p$ by a mean and median of only 0.62\% and 0.02\% respectively, indicating that our refitting results do not systematically alter the planet sizes. We observe that 120 planets (28\%) have a revised $R_p$ $> 2\sigma$ away from their corresponding values from \citetalias{fulton2018california}, and 62 planets (14\%) $> 3\sigma$ away.

\begin{table*}
    \renewcommand*{\arraystretch}{1.35}
    \centering
    \caption{Table showing estimates of the orbital periods ($P$), planetary-to-stellar-radii ratio ($R_p/R_{\star}$), planetary radii ($R_p$), the number of transits in the fitted light curve ($N_{\text{tr}}$) of 431 planets refitted in this work. The source of $R_{\star}$ to convert $R_p/R_{\star}$ to $R_p$ is listed in the References column: (1) \citet{fulton2018california}, (2) \citet{vaneylen2018asteroseismic}.`Flag' refers to whether the planet passes the filter checks and is included in the smaller subset for further analyses (1 for True and 0 for False). The complete list of parameters are provided in Appendix~\ref{appendix}. Only the first 10 planets are shown here; the full table is available online in a machine-readable format.}
    \begin{tabular}{llllllll}
    \hline 
KOI & Kepler name & $P$ (days) & $R_p/R_{\star}$ & $R_p$ ($R_{\oplus}$) & $R_{\star}$ Source & $N_{\text{tr}}$ & Flag \\ 
\hline 
K00041.01 & Kepler-100 c & $12.815893 \pm 0.000008$ & $0.0138 \pm 0.0002$ & $2.28_{-0.03}^{+0.03}$ & (2) & 93 & 1 \\ 
K00041.02 & Kepler-100 b & $6.887062 \pm 0.000007$ & $0.0082 \pm 0.0001$ & $1.36_{-0.03}^{+0.03}$ & (2) & 173 & 1 \\ 
K00041.03 & Kepler-100 d & $35.333093 \pm 0.000019$ & $0.0104 \pm 0.0002$ & $1.71_{-0.04}^{+0.04}$ & (2) & 35 & 1 \\ 
K00046.02 & Kepler-101 c & $6.029792 \pm 0.000020$ & $0.0073 \pm 0.0007$ & $1.32_{-0.14}^{+0.14}$ & (1) & 50 & 0 \\ 
K00049.01 & Kepler-461 b & $8.313784 \pm 0.000015$ & $0.0287 \pm 0.0010$ & $4.03_{-0.17}^{+0.17}$ & (1) & 34 & 1 \\ 
K00069.01 & Kepler-93 b & $4.726739 \pm 0.000001$ & $0.0151 \pm 0.0001$ & $1.50_{-0.04}^{+0.04}$ & (2) & 275 & 1 \\ 
K00070.01 & Kepler-20 A c & $10.854089 \pm 0.000003$ & $0.0290 \pm 0.0002$ & $2.79_{-0.07}^{+0.07}$ & (1) & 116 & 1 \\ 
K00070.02 & Kepler-20 A b & $3.696115 \pm 0.000001$ & $0.0182 \pm 0.0002$ & $1.75_{-0.04}^{+0.05}$ & (1) & 336 & 1 \\ 
K00070.03 & Kepler-20 A d & $77.611598 \pm 0.000019$ & $0.0263 \pm 0.0003$ & $2.52_{-0.07}^{+0.07}$ & (1) & 15 & 1 \\ 
K00070.05 & Kepler-20 A f & $19.577627 \pm 0.000020$ & $0.0091 \pm 0.0004$ & $0.88_{-0.04}^{+0.04}$ & (1) & 62 & 1 \\ 
... & ... & ... & ... & ... & ... & ... & ... \\ 
\hline
    \end{tabular}
    \label{tab:planet_params}
\end{table*}

\begin{table}
    \centering
    \caption{Mean and median values radii determined in this work, \citetalias{fulton2018california}, and \citetalias{vaneylen2018asteroseismic}. Values are listed as percentages (\%).}
    \begin{tabular}{ccccc}
    \hline
        \multicolumn{2}{c}{Parameter} & This work & \citetalias{fulton2018california} & \citetalias{vaneylen2018asteroseismic} \\
    \hline
        \multicolumn{2}{c}{Sample Size} & 431 & 1901 & 117 \\
    \hline
        \multirow{2}{*}{$R_p/R_{\star}$} & Mean & 4.76 & 8.73 & 3.20 \\
        & Median & 3.44 & 4.07 & 2.44 \\
        \hline
        \multirow{2}{*}{$R_{\star}$} & Mean & 3.23 & 3.17 & 2.49 \\
        & Median & 2.78 & 2.74 & 2.20 \\
        \hline
        \multirow{2}{*}{$R_p$} & Mean & 6.09 & 10.00 & 3.96 \\
        & Median & 4.70 & 5.22 & 3.36 \\
    \hline
    \end{tabular}
    
    \label{tab:typical_err}
\end{table}

\subsection{Radius valley dependence on orbital period} \label{subsect: radvalley_pos}
To obtain the most precise planetary sample, we apply the following conservative cuts to our sample for subsequent analyses in the rest of this paper:
\begin{enumerate}
    \item \textit{Precision on $R_p$}: We exclude planets with a planet radius precision $\sigma_{R_p} > 10\%$.
    \item \textit{Radius correction factor (RCF)}: We exclude planets with an RCF $> 5\%$, as reported in \citet{furlan2017kepler}. RCFs may themselves be uncertain due to observational challenges in detecting nearby stellar companions and measuring their brightness \citep{furlan2017kepler}. Of the 137 planets with RCF measurements from \citet{furlan2017kepler}, 18 (13\%) have RCF $> 5\%$. The remaining 294 planets do not have RCF measurements currently available.
    \item \textit{Number of transits}: We exclude planets with fewer than 3 transits in the \textit{Kepler} short cadence data to limit the risk that correlated noise during an individual transit strongly affects the resulting fit.
\end{enumerate}
After implementing these filters, 375 planets remain in our sample which we will use throughout the remainder of this work.

\begin{figure*}
    \centering
    \begin{tabular}{cc}
    \includegraphics[width=0.99\columnwidth]{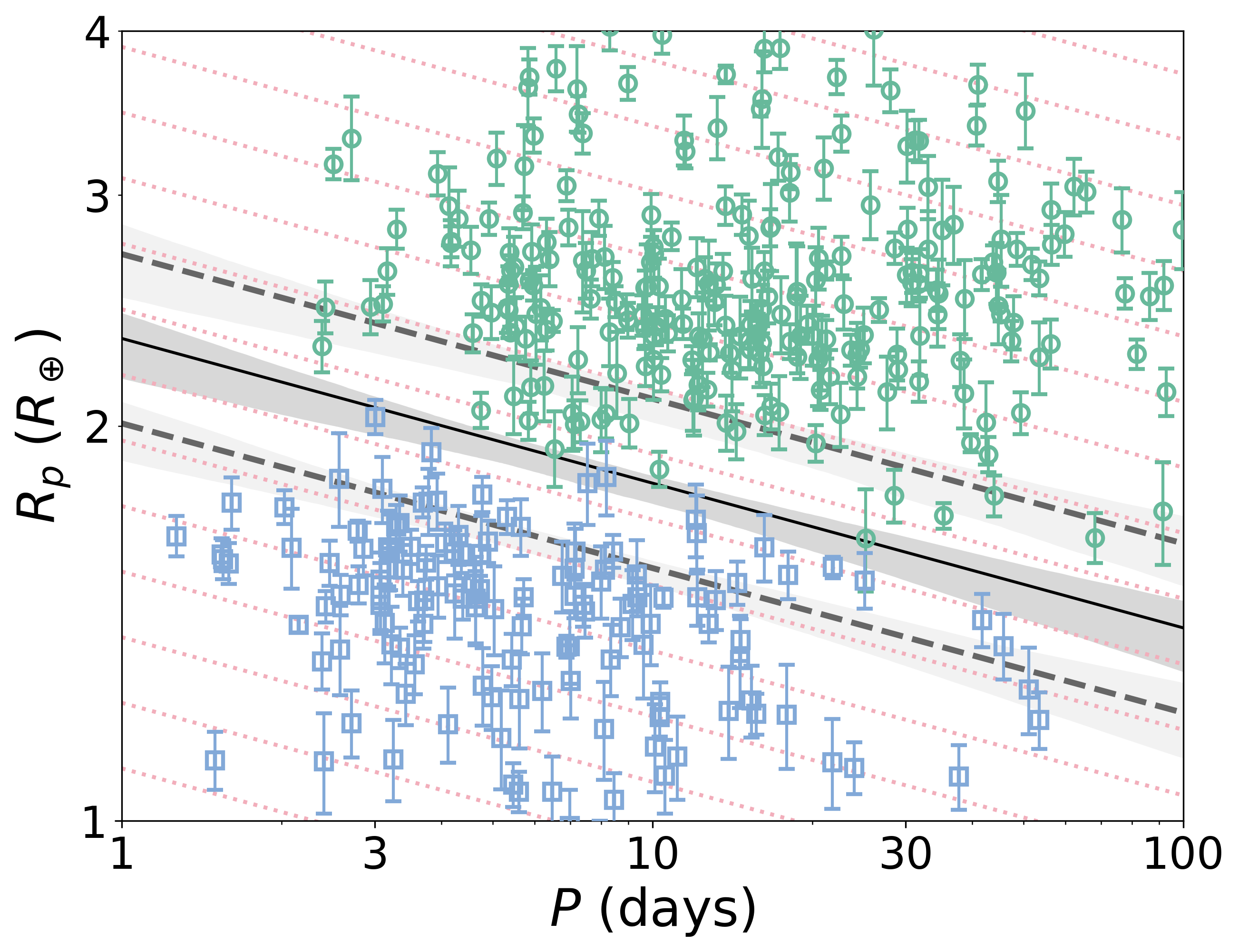} &
    \includegraphics[width=0.99\columnwidth]{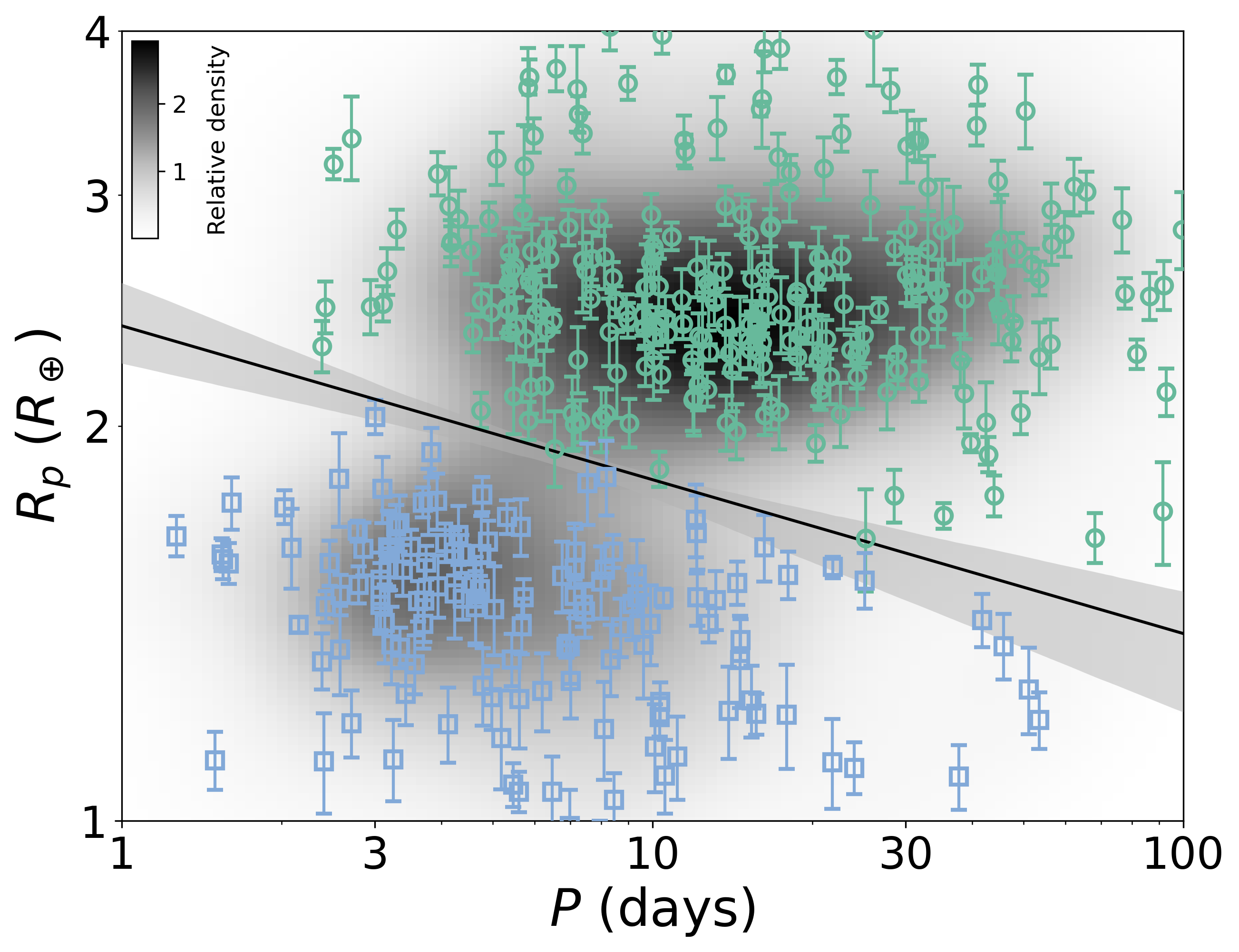}
    \end{tabular}
    \caption{\textit{Left:} Radius valley location determined with the support-vector machine (SVM), with $m=-0.11 \pm 0.02$, and $c = 0.36_{-0.03}^{+0.02}$, indicated by the black solid line. The dashed lines represent the median boundaries passing through the supporting vectors determining the position of the solid line. We define the area between the two lines as the radius valley region. The green and blue points show planets above and below the radius valley respectively. The grey shaded regions represent the $\pm 1\sigma$ uncertainties of the lines determined using the bootstrap method. The red dotted lines show the plot divided into multiple orbital-period dependent bins, which is then used for plotting the adjusted histogram in Fig.~\ref{fig:tilthist}. \textit{Right}: Radius valley position determined by fitting a line through the region where the kernel density estimate is minimum. With this method, we obtain $m=-0.12_{-0.05}^{+0.03}$, and $c=0.37_{-0.03}^{+0.0}$, indicated by the black solid line with $\pm 1 \sigma$ uncertainty shaded in grey.}
    \label{fig:rp_location}
\end{figure*}

We first use this sample to investigate the location of the radius valley as a function of orbital period. Following the procedure outlined in \citet{vaneylen2018asteroseismic}, we calculate the position of the radius valley by determining the hyperplane of maximum separation. We perform this with a linear support-vector machine (SVM). To initialise our model, we initially classify our sample into two groups, `above' and `below' the radius valley on the radius-orbital period plane, by applying a Gaussian Mixture Model with two components in the $R_p$-$P$ plane. Since the orders of magnitude of $R_p$ and $P$ are different, we divide $P$ by 5 before applying the above clustering algorithm to allow the model to separate the planetary population into two groups above and below the radius valley. Otherwise, the population would be clustered in a way dominated by the difference in period. Following \citet{vaneylen2018asteroseismic} and \citet{david2021evolution}, we select an SVM penalty parameter of $C=10$ for the hyperplane, to minimise misclassification of data points above or below the radius valley, but still allow the hyperplane location to be determined by a sufficient number of data points. To determine accurate uncertainties on the location of the valley, we then perform a bootstrap by generating 1000 new sample sets on which we repeat the above procedure. Each bootstrap sample is generated by generating a new sample of the same size from the original sample, allowing replacement. Each bootstrapped sample is then categorised into two groups with a Gaussian Mixture Model, and the SVM procedure is repeated. Reporting the median value and taking the 16\textsuperscript{th} and 84\textsuperscript{th} percentiles as the upper and lower uncertainties, we find
\begin{equation}
    \log_{10}{\left(R_p/R_{\earth}\right)} = m\log_{10}{\left(P/{\text{days}}\right)} + c
    \label{eq:logr_logp}
\end{equation}
with $m=-0.11 \pm 0.02$, and $c=0.37^{+0.02}_{-0.03}$. The location of the radius valley is plotted in Fig.~\ref{fig:rp_location}.

\begin{table*}
    \renewcommand*{\arraystretch}{1.5}
    \centering
    \caption{Dependencies of the radius valley in $n$ dimensions ($m_i$), given by the equation $\log_{10}{\left(R_p/R_{\earth} \right)} = \sum_{i=1}^{n} m_i x_i$, with different parameter combinations $x_i$. The methods used to obtain the equation of the radius valley hyperplanes are also given, where SVM and KDE stand for the support-vector machine and fitting the minima of the kernel density estimates respectively.}
    \begin{tabular}{cccccccc}
    \hline
        Dimensions & $\log_{10}{\left(P/\text{days}\right)}$ & $\log_{10}{\left(S/S_{\earth} \right)}$ & $\log_{10}{\left(M/M_{\sun} \right)}$ & $\log_{10}{\left(\text{Age/Gyr} \right)}$ & [Fe/H] & Intercept & Method \\
    \hline
        \multirow{6}{*}{2} & $-0.11_{-0.02}^{+0.02}$ & & & & & $0.37_{-0.03}^{+0.02}$ & SVM \\
        & $-0.12_{-0.05}^{+0.03}$ & & & & & $0.37_{-0.03}^{+0.05}$ & KDE \\
        &  & $0.07_{-0.01}^{+0.02}$ &  &  &  & $0.11_{-0.04}^{+0.03}$ & SVM \\
        &  &  & $0.23_{-0.08}^{+0.09}$ &  &  & $0.27_{-0.01}^{+0.01}$ & SVM \\
        &  &  &  & $0.02_{-0.02}^{+0.01}$ &  & $0.26_{-0.01}^{+0.01}$ & SVM \\
        &  &  &  &  & $0.06_{-0.08}^{+0.06}$ & $0.26_{-0.01}^{+0.01}$ & SVM \\
    \hline
        \multirow{4}{*}{3} & $-0.09_{-0.03}^{+0.02}$ &  & $0.21_{-0.07}^{+0.06}$ &  &  & $0.35_{-0.03}^{+0.02}$ & SVM \\
        &  & $0.07_{-0.02}^{+0.02}$ & $-0.01_{-0.09}^{+0.07}$ &  &  & $0.11_{-0.05}^{+0.04}$ & SVM \\
        & $-0.10_{-0.02}^{+0.02}$ &  &  & $0.03_{-0.03}^{+0.02}$ &  & $0.34_{-0.02}^{+0.03}$ & SVM \\
        & $-0.10_{-0.03}^{+0.03}$ &  &  &  & $0.03_{-0.04}^{+0.03}$ & $0.36_{-0.03}^{+0.02}$ & SVM \\
    \hline
        4 & $-0.096_{-0.027}^{+0.023}$ &  & $0.231_{-0.064}^{+0.053}$ & $0.033_{-0.025}^{+0.017}$ &  & $0.339_{-0.018}^{+0.026}$ & SVM \\
    \hline
    \end{tabular}
    \label{tab:svm_values}
\end{table*}

We also implement the method adopted by \citet{petigura2022california}, which involves computing the planet density in the radius-period plane with a Gaussian kernel density estimate (KDE), and fitting the planet radius at the KDE minima between $\log_{10}{\left(P/\text{days}\right)}=0.5-1.5$ (i.e. $P \approx 3.16-31.6~\text{days}$) and $\log_{10}{\left(R_p/R_{\earth}\right)}=0.15-0.35$ (i.e. $R_p \approx 1.41-2.24R_{\earth}$) according to equation~\ref{eq:logr_logp}, and performing bootstrap with 1000 sample sets to find the uncertainties. The result is illustrated in Figure~\ref{fig:rp_location}. We use a KDE bandwidth of 0.467 in $\log_{10}{P}$ and 0.075 in $\log_{10}{R_p}$, which is based on the bandwidth used by \citet{petigura2022california}, but scaled up linearly based on the ratios between the two sample sizes. We note that this method fits a narrower period range compared to the SVM method, which fits for the full $P=1-100~\text{days}$. With this method, we obtain $m = -0.12_{-0.05}^{+0.03}$ and $c=0.37_{-0.03}^{+0.05}$.  We find this value to be slightly steeper than that determined with the SVM, however the two values are well within $1\sigma$ of each other. Here, we do not correct for detection completeness, and we note that this method is highly sensitive to the choice of the KDE bandwidth, and using a bandwidth five times larger results with a slope approximately twice as steep. 

\begin{figure}
    \centering
    \includegraphics[width=\linewidth]{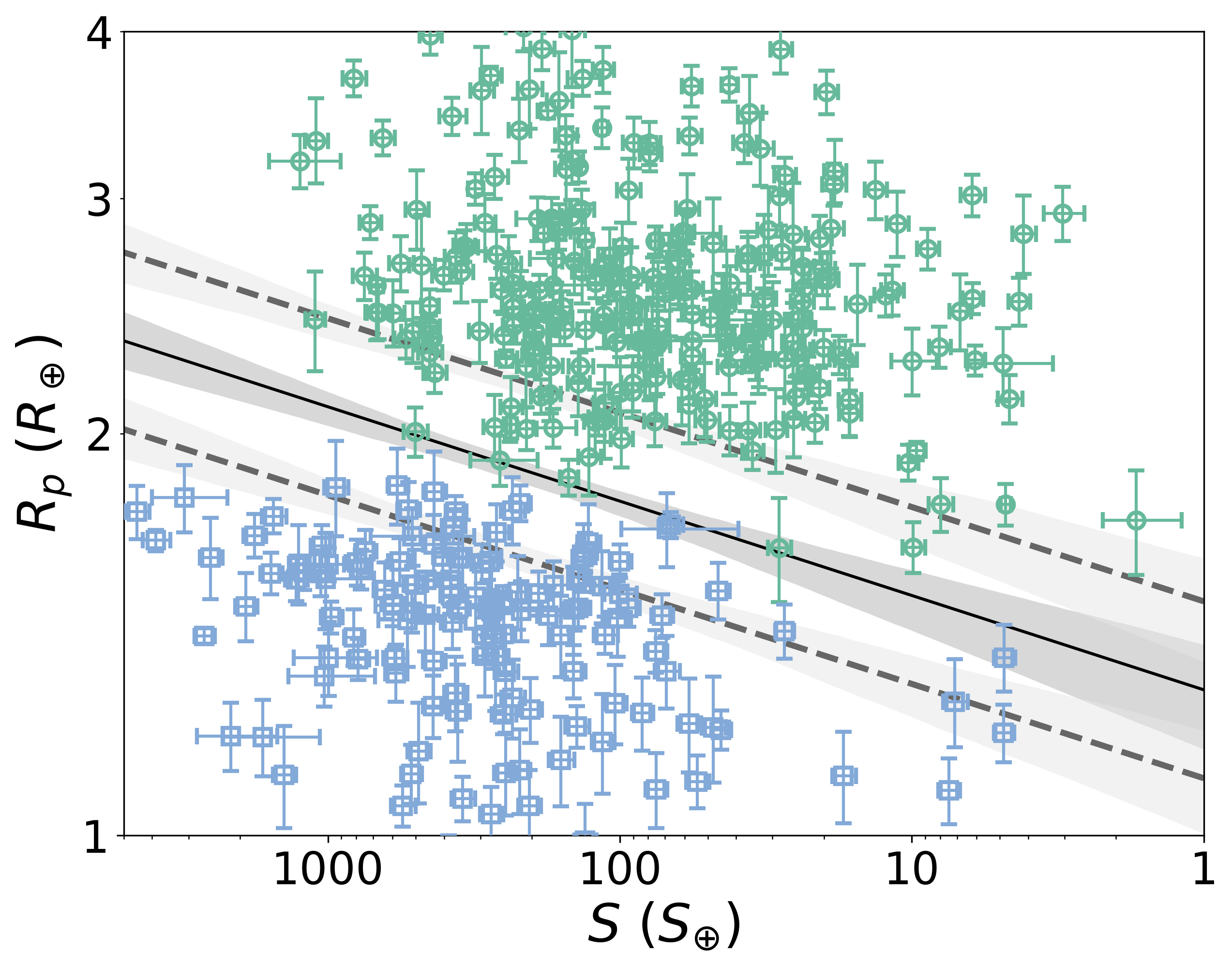}
    \caption{Same as Figure~\ref{fig:rp_location} left, but as a function of incident flux $S$ instead of orbital period $P$.}
    \label{fig:svm_RS}
\end{figure}

Some studies have opted to study the radius valley location as a function of incident flux $S$, in addition to, or instead of, $P$ \citep[e.g.][]{rogers2021photoevaporation, petigura2022california}. We calculate $S$ according to the formula
\begin{equation}
    S = \frac{L_{\star}}{4\pi a^2}
    \label{eq:S}
\end{equation}
where $L_{\star}$ is the host star's luminosity. $L_{\star}$ can itself be calculated using
\begin{equation}
    L_{\star} = 4\pi R_{\star}^2 \sigma_{\text{sb}} T_{\text{eff}}^4
\end{equation}
where $R_{\star}$ is the stellar radius, $\sigma_{\text{sb}}$ is the Stefan-Boltzmann constant, and $T_{\text{eff}}$ is the effective temperature of the star. Finally, $a$ is the orbital semi-major axis, which is given by
\begin{equation}
    \left(\frac{a}{R_{\star}}\right)^3 = \frac{GP^2 \rho_{\star}}{3\pi} \frac{\left(1+e\sin{\omega}\right)^3}{\left(1-e^2 \right)^{1.5}}
    \label{eq:kepler_3rd}
\end{equation}
according to Kepler's third law of planetary motion \citep[e.g.][]{vaneylen2015eccentricity, petigura2020two}. Here, $G$ is the gravitational constant, $P$ is the orbital period, $\rho_{\star}$ is the stellar density, $e$ and $\omega$ are the orbital eccentricity and argument of periapsis respectively. As before, we obtain the stellar properties ($R_{\star}$, $\rho_{\star}$, $T_{\text{eff}}$) from \citetalias{vaneylen2018asteroseismic} when available, and from \citetalias{fulton2018california} otherwise. $P$, $e$, and $\omega$ are obtained from the transit fitting results of this work. Fitting the valley with the SVM, we obtain
\begin{equation}
    \log_{10}{R_p/R_{\earth}} = m\log_{10}{S/S_{\earth}} + c
\end{equation}
with $m = 0.07_{-0.01}^{+0.02}$, and $c = 0.11_{-0.04}^{+0.03}$. The location of the radius valley in terms of $S$ is shown in Figure~\ref{fig:svm_RS}.

\subsection{Depth of the radius valley} \label{subsect:results_deepness}
\begin{figure}
    \centering
    \includegraphics[width=\columnwidth]{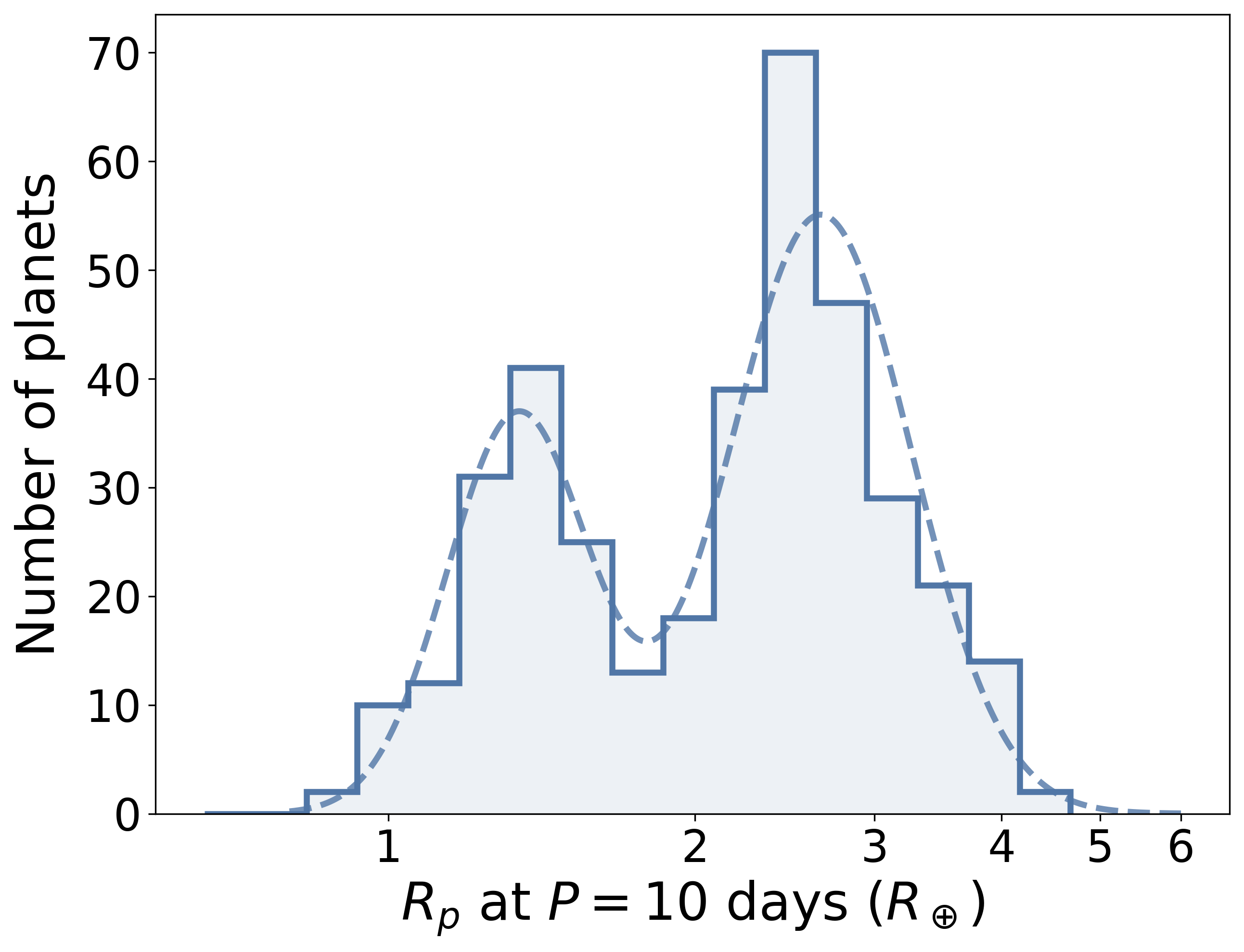}
    \caption{Histogram of planet radii, adjusted to equivalent radii at $P = 10$ days, according to the radius valley slope calculated in Section~\ref{subsect: radvalley_pos} with the SVM. Note that completeness corrections are not performed. Here, $E_{\text{avg}} =  2.98_{-0.47}^{+0.60}$.}
    \label{fig:tilthist}
\end{figure}

We investigate the depth of the radius valley. As we find the position of the radius valley is dependent on orbital period, we divide the radius valley into multiple tilted bins (as shown in Fig.~\ref{fig:rp_location} left) and plot an adjusted histogram in logarithmic scale. We shift planets along the slope of the radius valley obtained with the SVM method in Section~\ref{subsect: radvalley_pos}, i.e. $m=-0.11$, and plot a histogram of `expected' planetary radii at an orbital period of 10 days, shown in Fig.~\ref{fig:tilthist}. We choose to fit the histogram with a Gaussian Mixture Model of two clusters, as opposed to a Gaussian kernel density estimate, as the former is independent of the sizes and locations of the histogram bins, as well as the Gaussian bandwidth. Also, with the Gaussian Mixture Model, we are able to force the planets to fall into two groups only, matching the bimodal distribution of small planets.

Here, we propose the metric $E$, defined as
\begin{equation}
    E_{\text{SN}} = N_{\text{sub-Neptune,peak}}/{N_{\text{valley}}}
    \label{eq:E_SN}
\end{equation}
and 
\begin{equation}
    E_{\text{SE}} = N_{\text{super-Earth,peak}}/{N_{\text{valley}}}
    \label{eq:E_SE}
\end{equation}
to compare the number of planets inside the valley and the peak number outside the valley. A higher $E$ indicates a deeper radius valley.  $N_{\text{sub-Neptune,peak}}$ and $N_{\text{super-Earth,peak}}$ are the number of planets at the sub-Neptune and super-Earth Gaussian peaks respectively, and $N_{\text{valley}}$ is the number of planets at the lowest point between the two Gaussian peaks. As $N_{\text{sub-Neptune, peak}}$, $N_{\text{super-Earth, peak}}$, and $N_{\text{valley}}$ are determined directly from the curve resulting from the Gaussian Mixture Model, this $E$ metric is also independent of the histogram bin sizes or locations. To calculate the uncertainties of $E$, we again perform a bootstrap with 1000 sample sets, where each bootstrap sample is generated by generating a new sample of the same size from the original, allowing replacements. We also replace the radius of each selected planet by randomly drawing from a normal distribution with the reported planet radius as the mean, and the uncertainties on the radius as the variance. We report the median of this bootstrap distribution as our results, and the 84\textsuperscript{th} and 16\textsuperscript{th} percentiles as our $\pm 1 \sigma$ uncertainties. We test this metric on known planetary samples on \citetalias{fulton2018california} and \citetalias{vaneylen2018asteroseismic}, which we know the difference in the radius valley depth, and find that their corresponding $E$ value differ, hence demonstrating the reliability of such metric. The difference in the depth of the radius valley is further discussed in Section~\ref{subsect:discussion_deepness}. We find that for our new short cadence results, the ratios are $E_\text{SN} = 3.59_{-0.62}^{+0.77}$ and $E_\text{SE} = 2.40_{-0.41}^{+0.61}$. Averaging the two numbers gives us $E_\text{avg} = 2.98_{-0.47}^{+0.60}$.

\subsection{Radius valley dependence on stellar mass} \label{subsect:ms_results}
\begin{figure}
    \centering
    \includegraphics[width=\columnwidth]{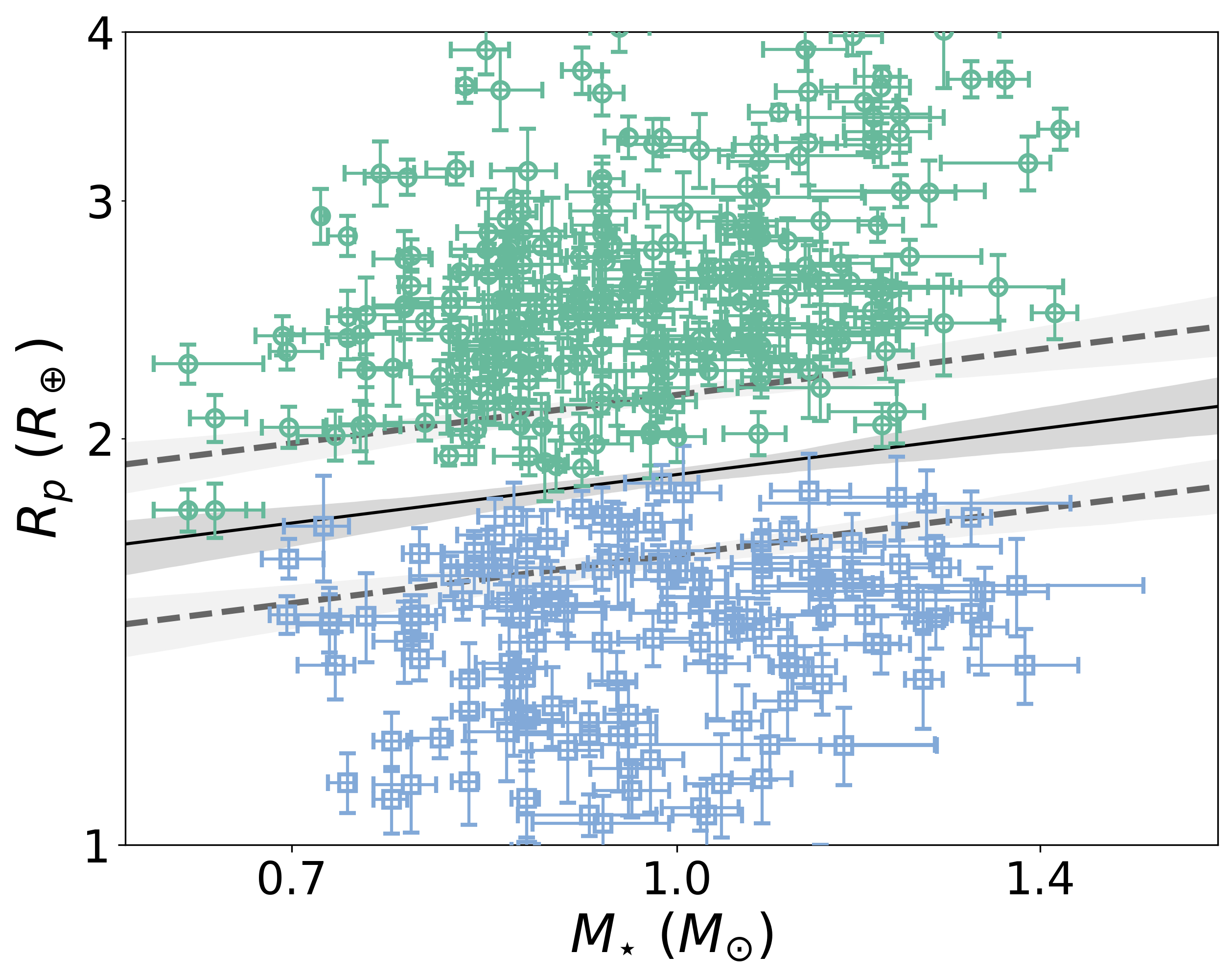}
    \caption{Plot of planetary radius against mass of host star. The solid line shows the location of the radius valley determined with the SVM, with slope $m = 0.23_{-0.08}^{+0.09}$, and $c = 0.27 \pm 0.01$. The dashed lines show the boundaries of the radius valley, given by the same $m$, and $c_\text{upper} = 0.33 \pm 0.01$, $c_\text{lower} = 0.22 \pm 0.01$. The shaded regions depict the $\pm 1\sigma$ uncertainties of the lines determined with bootstrapping.}
    \label{fig:svm_RM}
\end{figure}

We investigate the radius valley location as a function of stellar mass. We first implement a 2-dimensional SVM, using the same method as in Section~\ref{subsect: radvalley_pos}. The result is shown in Figure~\ref{fig:svm_RM}. We find $\mathrm{d}\log{R_p}/\mathrm{d}\log{M_{\star}} = 0.23_{-0.08}^{+0.09}$, and the intercept $c = 0.27 \pm 0.01$.

\begin{figure*}
    \centering
    \begin{tabular}{cc}
        \includegraphics[width=0.99\columnwidth]{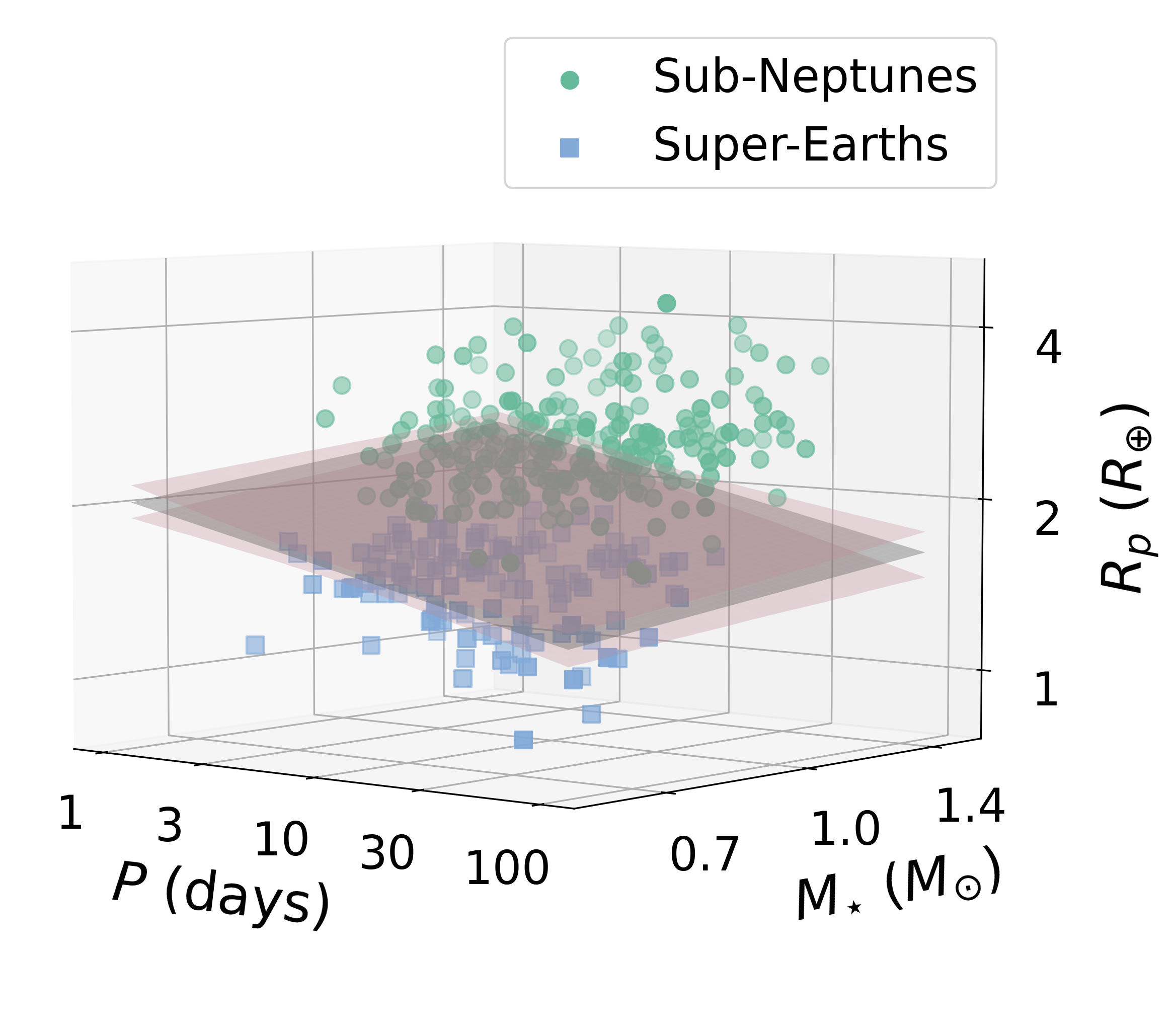} & \includegraphics[width=0.99\columnwidth]{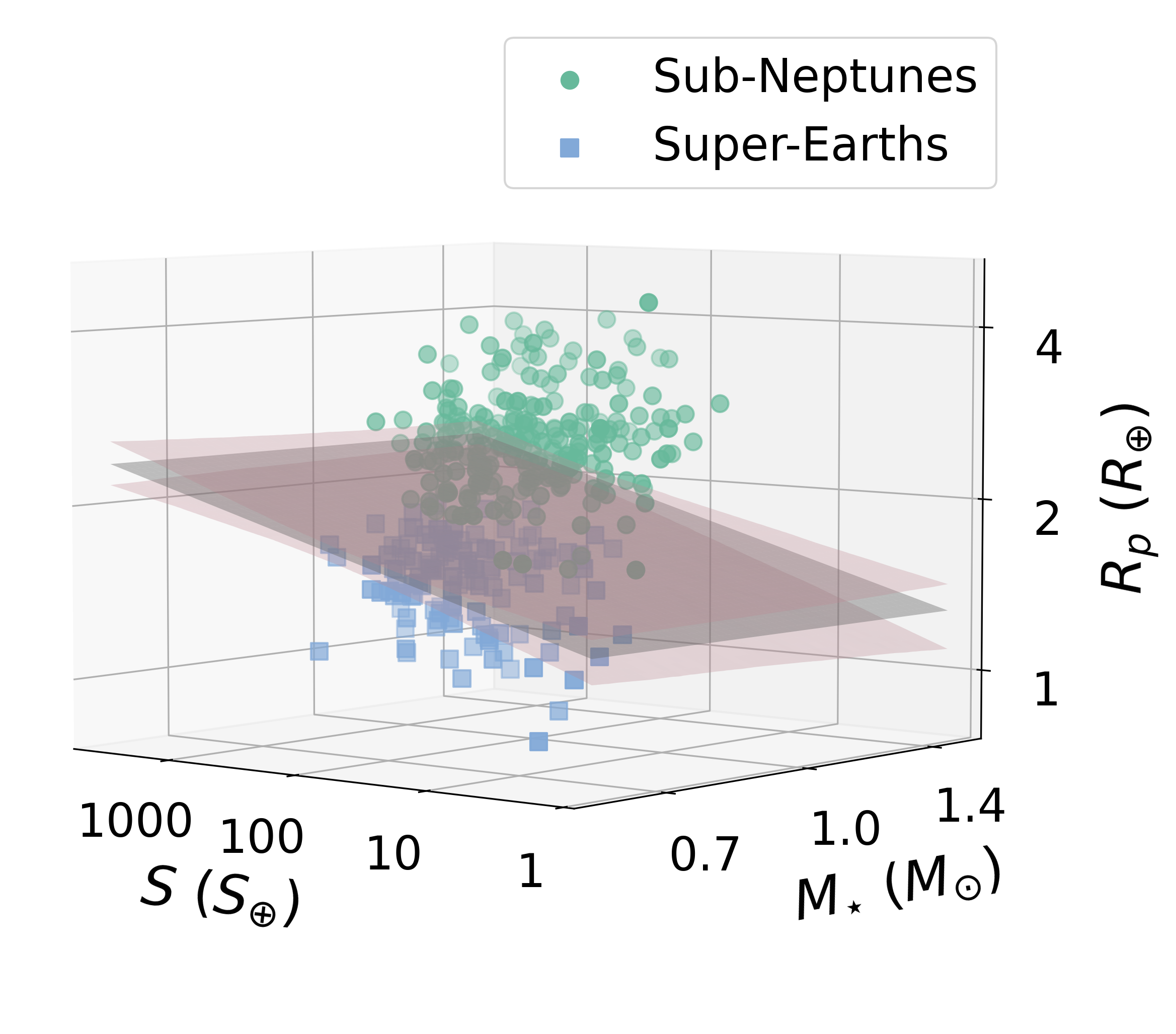}
    \end{tabular}
    \caption{Plot of planet radius against orbital period and mass of host star (left), and planet radius against incident flux and stellar mass (right). The grey plane shows the radius valley location determined with the SVM in 3 dimensions, with the $\pm 1\sigma$ uncertainties shown in pink. The uncertainties on individual planet parameters are not displayed in the interest of clarity.}
    \label{fig:svm_3d}
\end{figure*}

\citet{rogers2021photoevaporation} suggested the degeneracy between the photoevaporation and core-powered mass loss scenarios could be broken with an analysis of the radius valley in 3 dimensions. Hence, we implement an SVM in 3 dimensions: planet radius $R_p$, orbital period $P$, and mass of the host star $M_{\star}$, to fit the radius valley in the form of a plane. We perform bootstrapping with 1000 sample sets as per previous. We obtain the relation
\begin{equation}
    \log_{10}{\left(R_p/R_{\earth}\right)} = A\log_{10}{\left(P/\text{days}\right)} + B\log_{10}{\left(M_{\star}/M_{\sun}\right)} + C
    \label{eq:p_ms_rp_relation}
\end{equation}
with $A=-0.09_{-0.03}^{+0.02}$, $B=0.21_{-0.07}^{+0.06}$, $C=0.35_{-0.02}^{+0.02}$. An illustration of the SVM plane is shown in Figure~\ref{fig:svm_3d}.

We also investigate the radius valley location in the $R_p$--$S$--$M_{\star}$ space. Figure~\ref{fig:svm_3d} right shows the radius valley location in this space. We find
\begin{equation}
    \log_{10}{\left(R_p/R_{\earth}\right)} = A\log_{10}{\left(S/S_{\earth}\right)} + B\log_{10}{\left(M_{\star}/M_{\sun}\right)} + C
    \label{eq:s_ms_rp_relation}    
\end{equation}
with $A = 0.07 \pm 0.02$, $B = -0.01_{-0.09}^{+0.07}$, and $C = 0.11_{-0.05}^{+0.04}$.

\subsection{Radius valley dependence on stellar age} \label{subsect:results_age}

We investigate the location of the radius valley as a function of stellar age. We obtain the stellar ages from \citetalias{fulton2018california}. Kepler-174 does not have stellar age data from this source, hence we omit Kepler-174 b and Kepler-174 c in our analysis with stellar age. 

\begin{figure*}
    \centering
    \begin{tabular}{cc}
    \includegraphics[width=0.99\columnwidth]{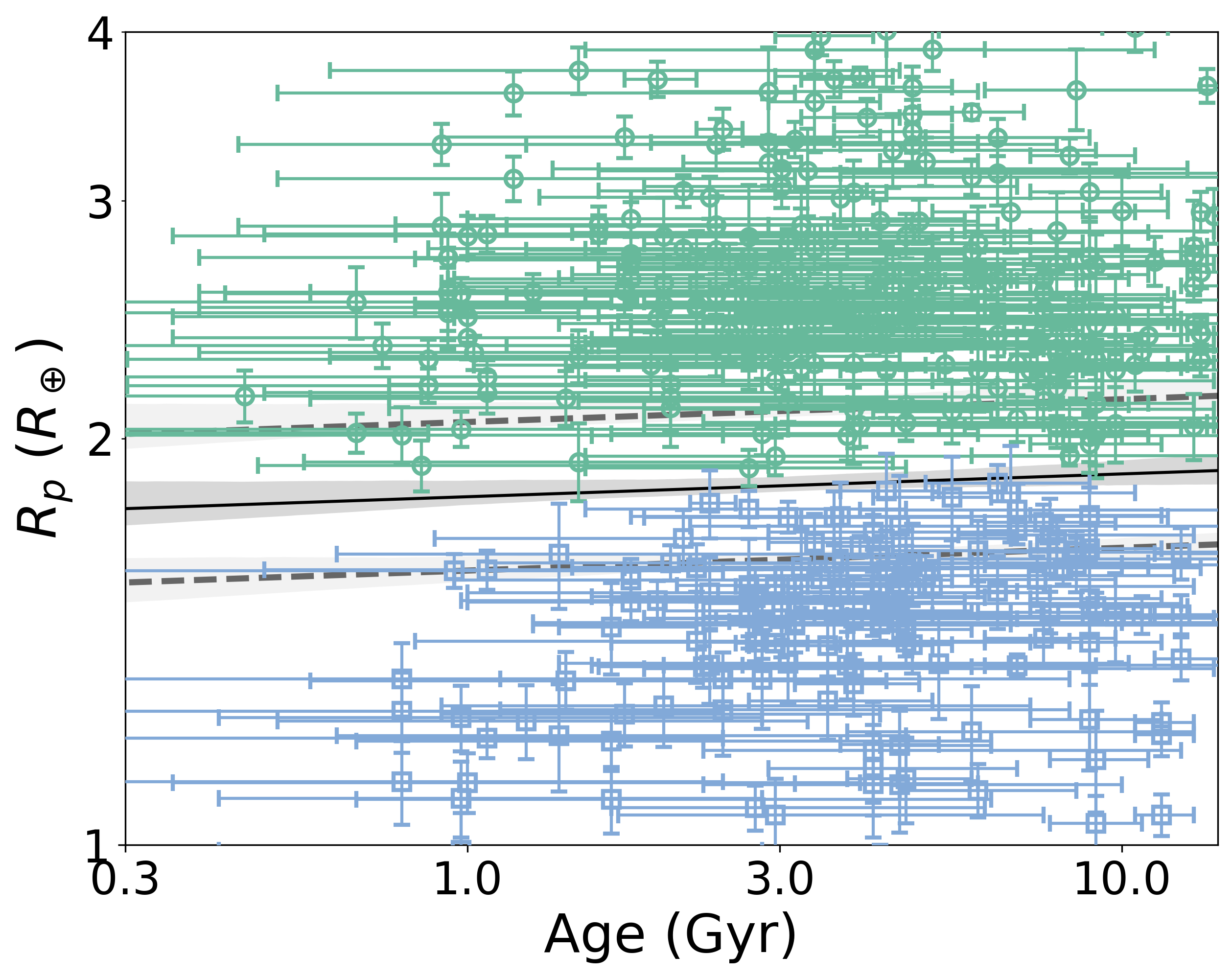} & 
    \includegraphics[width=0.97\columnwidth]{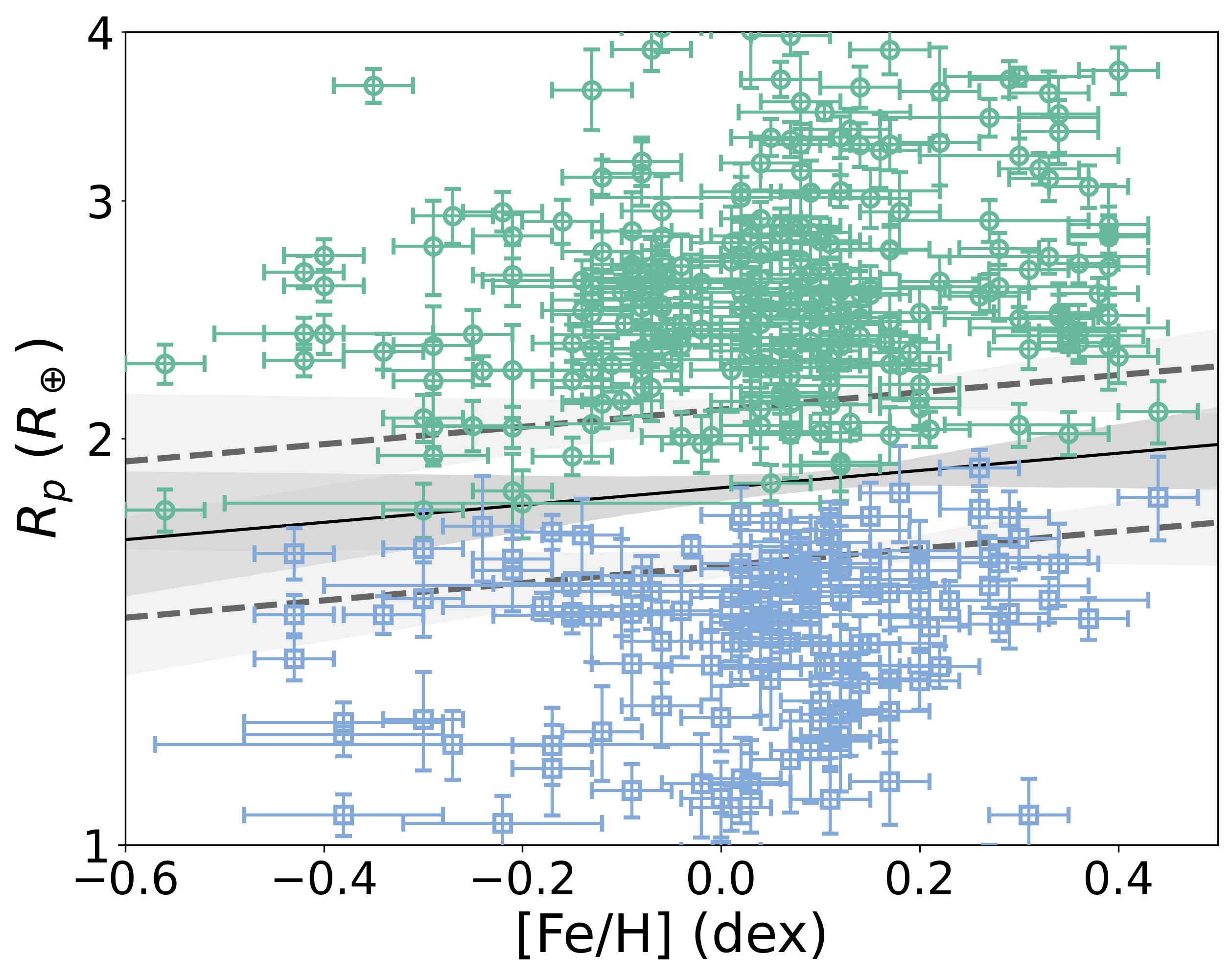}
    \end{tabular}
    \caption{Same as Figure~\ref{fig:svm_RM}, but for stellar age (left, $m = 0.01_{-0.02}^{+0.01}$, $c = 0.26_{-0.01}^{+0.01}$), and stellar metallicity [Fe/H] (right, $m = 0.06_{-0.08}^{+0.06}$, $c = 0.26_{-0.01}^{+0.01}$).}
    \label{fig:svm_other}
\end{figure*}

Fitting the valley with the SVM, we obtain
\begin{equation}
    \log_{10}{\left(R_p/R_{\earth}\right)} = m \log_{10}{\left(\text{Age/Gyr}\right)} + c
\end{equation}
with $m = 0.02_{-0.02}^{+0.01}$, $c = 0.26_{-0.01}^{+0.01}$, showing no significant correlation with the radius valley location in 2-dimensional space. The result is displayed in Figure~\ref{fig:svm_other} (left panel).

\begin{figure*}
    \centering
    \begin{tabular}{cc}
        \includegraphics[width=0.99\columnwidth]{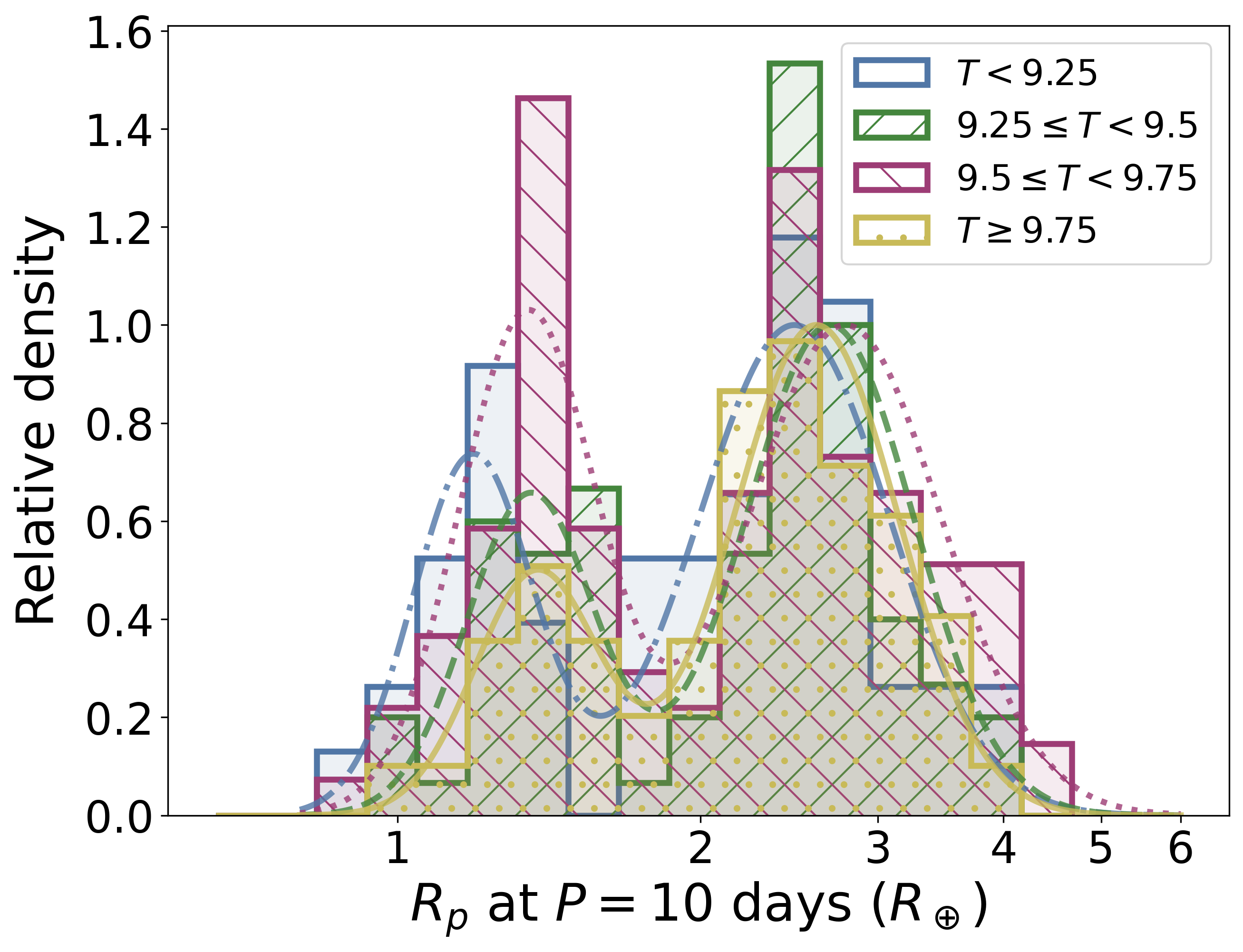} &
        \includegraphics[width=0.99\columnwidth]{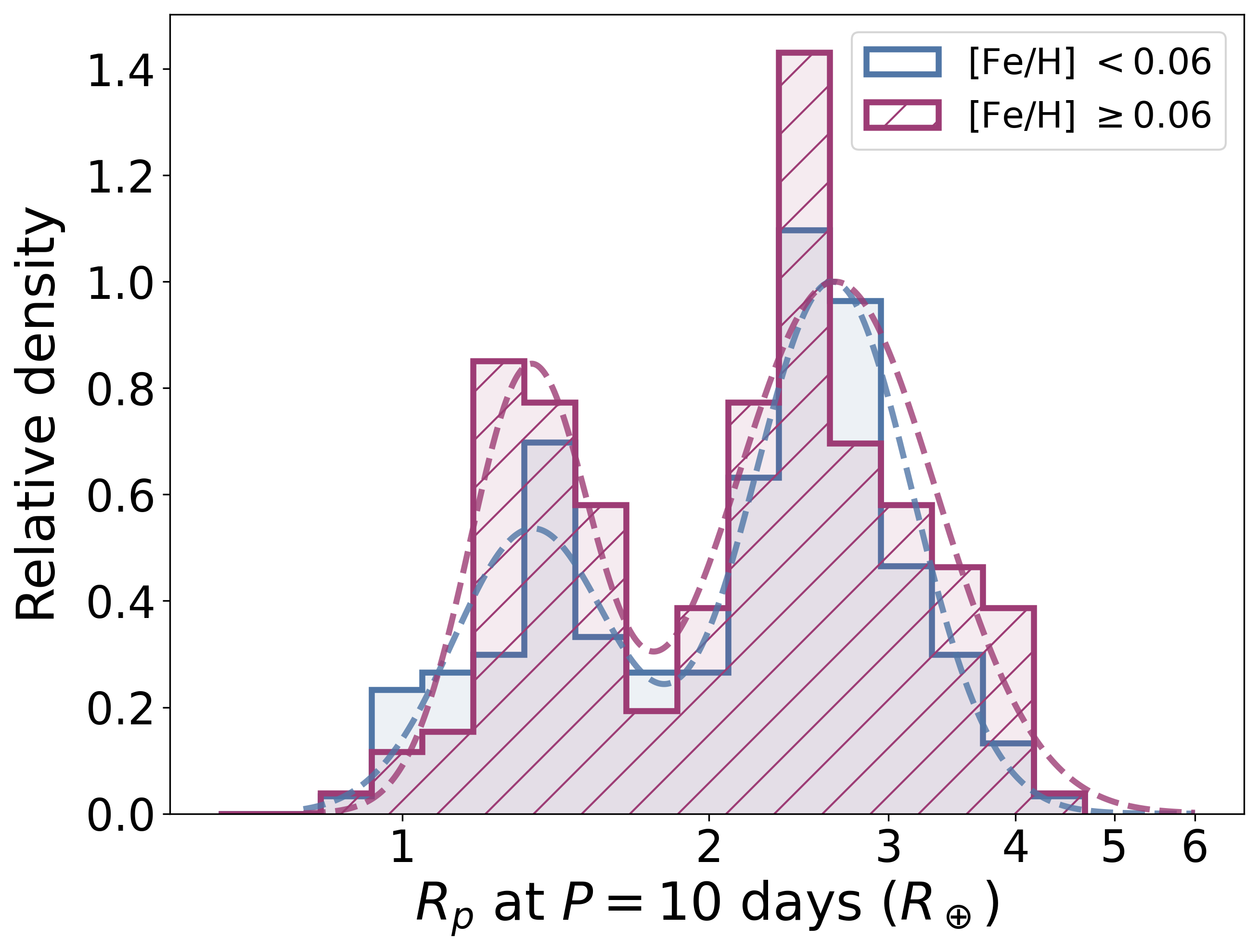}
    \end{tabular}
    \caption{Same as Figure~\ref{fig:tilthist}, but separated into different stellar ages (left), and metallicities (right). Here, $T = \log_{10}{\left(\text{Age/yr}\right)}$. The histograms are normalised such that the relative density of the sub-Neptune peak equals to unity.}
    \label{fig:tilthist_agefeh}
\end{figure*}

We further investigate whether the radius valley \emph{depth} may be a function of stellar age. To do so we generate histograms, similar to Figure~\ref{fig:tilthist}, split into different stellar age subsamples. Figure~\ref{fig:tilthist_agefeh} (left panel) shows that for older stars, the radius valley location shifts to higher $R_p$, and the radius valley becomes shallower. The change in the $E$ metric is reported in Table~\ref{tab:tilthist_age}. These findings suggest that the radius valley has a dependence on the age of the host stars. 

\begin{figure*}
    \centering
    \begin{tabular}{cc}
        \includegraphics[width=0.99\columnwidth]{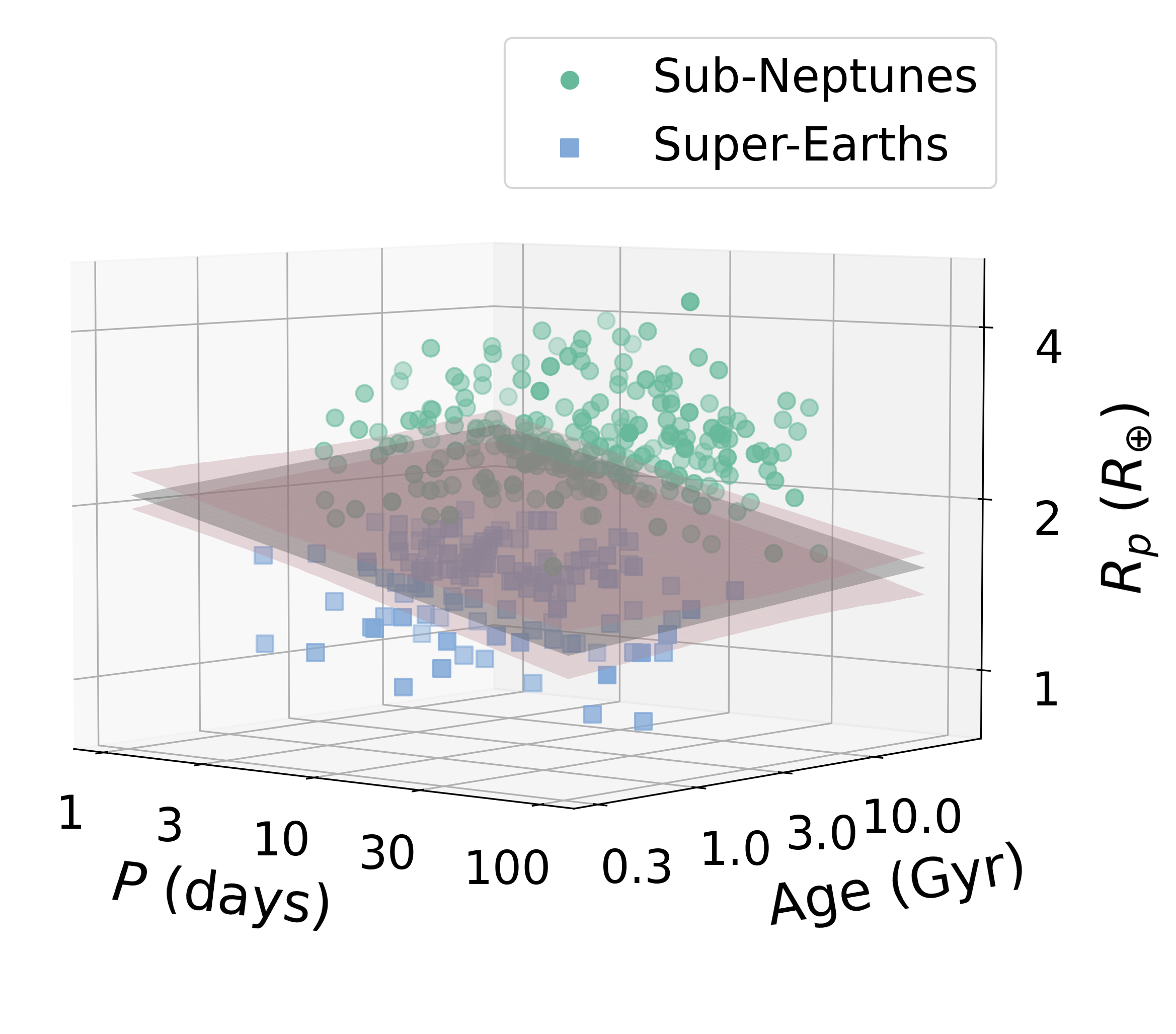} & \includegraphics[width=0.99\columnwidth]{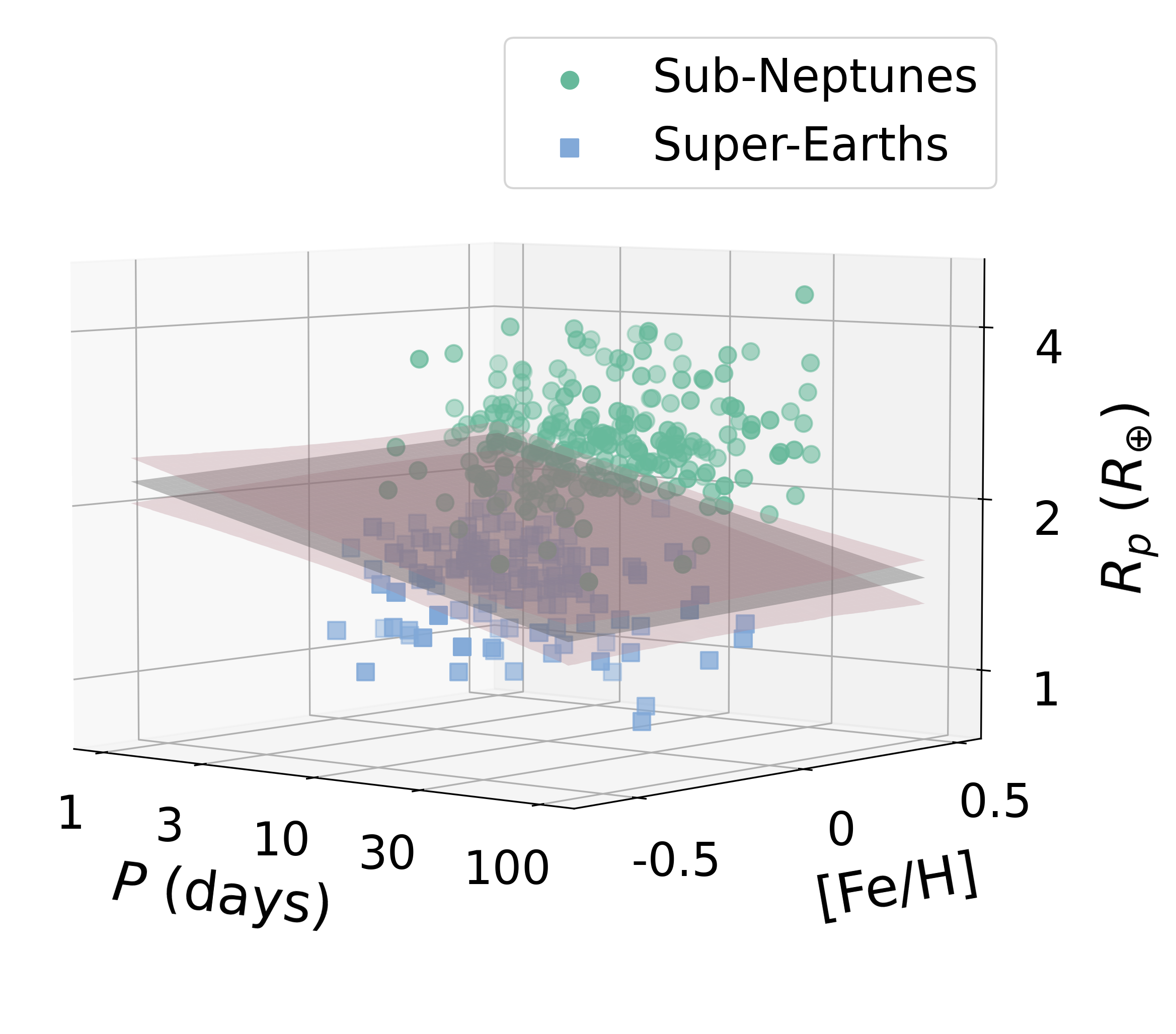}
    \end{tabular}
    \caption{Same as Figure~\ref{fig:svm_3d}, but in $R_p$--$P$--age space (left) and $R_p$--$P$--[Fe/H] space (right).}
    \label{fig:svm_3d_agefeh}
\end{figure*}

We further plot the radius valley in $R_p$--$P$--age space, and fit the valley with the SVM, as shown in Figure~\ref{fig:svm_3d_agefeh} (left panel). We find
\begin{equation}
    \log_{10}{\left(R_p/R_{\earth}\right)} = A \log_{10}{\left(P/\text{days}\right)} + B \log_{10}{\left(\text{Age/Gyr}\right)} + C
\end{equation}
with $A = -0.10 \pm 0.02$, $B = 0.03_{-0.03}^{+0.02}$, and $C = 0.34_{-0.02}^{+0.03}$.

\begin{table}
    \centering
    \caption{$E$ values of the radius valley for different ages of the host stars. $E_{\text{avg}} = (E_{\text{SN}} + E_{\text{SE}})/2$.}
    \begin{tabular}{cccccc}
        \hline
        $\log_{10}{\left(\text{Age/yr}\right)}$ & Age (Gyr) & $N_{\text{planets}}$ & $E_{\text{SN}}$ & $E_{\text{SE}}$ & $E_{\text{avg}}$ \\
        \hline
        $<9.25$ & $<1.78$ & 53 & 4.93 & 3.64 & 4.28 \\
        $9.25-9.5$ & $1.78-3.16$ & 95 & 4.69 & 3.08 & 3.89 \\
        $9.5-9.75$ & $3.16-5.62$ & 114 & 3.22 & 3.32 & 3.27 \\
        $>9.75$ & $>5.62$ & 111 & 2.21 & 4.40 & 3.30 \\
        \hline
    \end{tabular}
    \label{tab:tilthist_age}
\end{table}

\begin{figure}
    \centering
    \includegraphics[width=\linewidth]{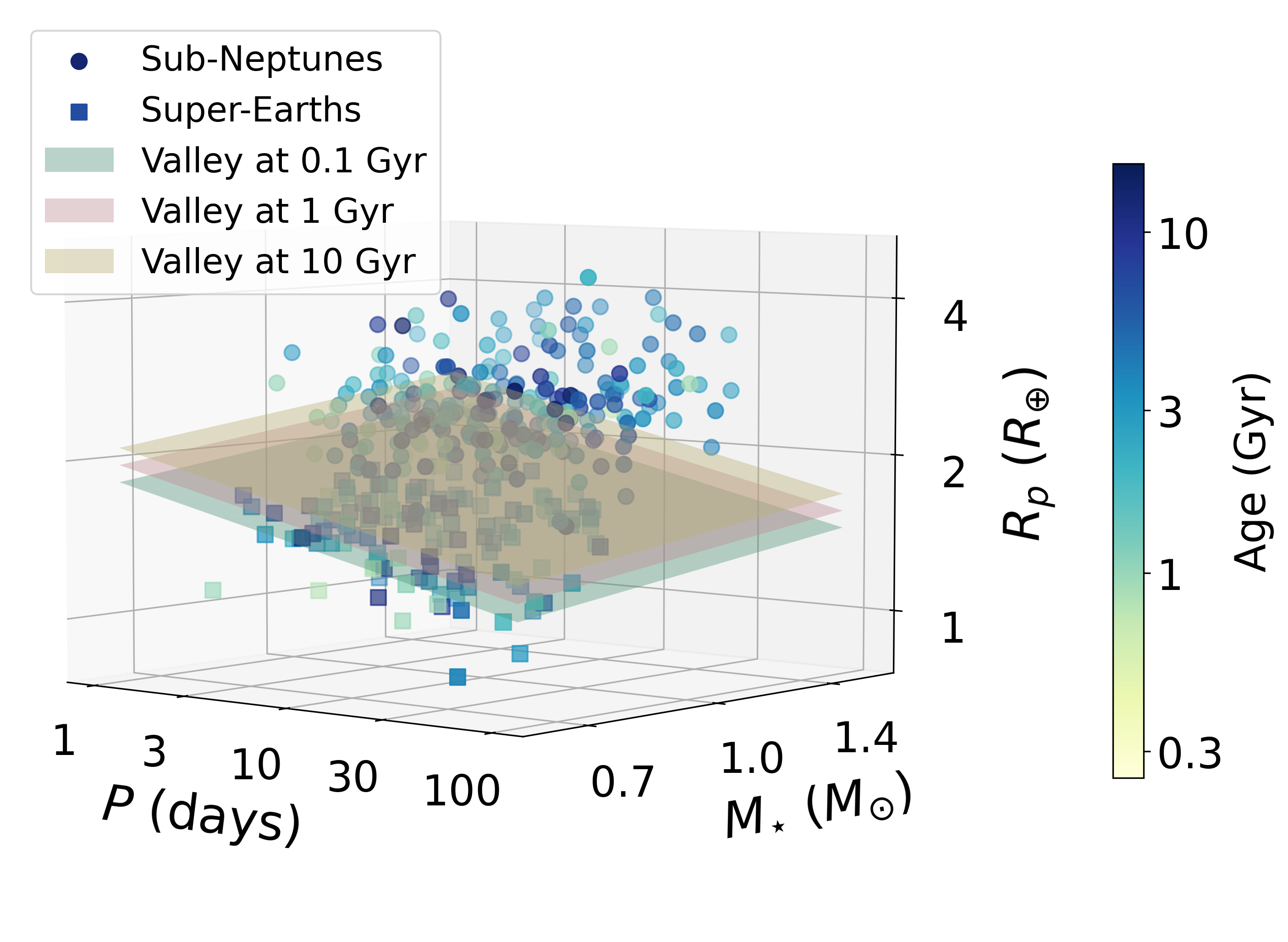}
    \caption{Radius valley location in terms of orbital period, stellar mass, and stellar age. The colour bar represents the age of the planetary host stars, and planes of different colours indicate the radius valley location for different stellar ages.}
    \label{fig:svm_4d}
\end{figure}

We can also combine $R_p$, $P$, $M_{\star}$, and stellar age, and determine the radius valley with a 4-dimensional SVM, as shown in Figure~\ref{fig:svm_4d}. The resulting equation representing the radius valley is in the form of a 4-dimensional hyperplane
\begin{multline}
    \log_{10}{\left(R_p/R_{\earth} \right)} = A\log_{10}{\left(P/\text{days}\right)} + B\log_{10}{\left(M_{\star}/M_{\sun}\right)} \\
    + C\log_{10}{\left(\text{Age/Gyr} \right)} + D
    \label{eq:svm_4d}
\end{multline}
with $A = -0.096_{-0.027}^{+0.023}$, $B = 0.231_{-0.064}^{+0.053}$, $C = 0.033^{+0.017}_{-0.025}$, $D = 0.339_{-0.018}^{+0.026}$. These results imply there is strong evidence the radius valley location is dependent on $P$ and $M_{\star}$, and weak evidence for its dependence on stellar age ($>1 \sigma$). These values are also consistent within $1 \sigma$ with their corresponding dependencies in two and three dimensions (see Table~\ref{tab:svm_values}).

\subsection{Radius valley dependence on stellar metallicity} \label{subsect:results_feh}

We perform a similar analysis in terms of the stellar metallicity. As for age, we obtain the stellar metallicity from \citetalias{vaneylen2018asteroseismic} if available, and \citetalias{fulton2018california} otherwise. We find
\begin{equation}
    \log_{10}{\left(R_p/R_{\earth}\right)} = m\text{[Fe/H]} + c
\end{equation}
with $m = 0.06_{-0.08}^{+0.06}$, $c = 0.26_{-0.01}^{+0.01}$, again displaying no significant correlation with the radius valley location in 2-dimensional space.

We divide the planet population into two groups, based on the median [Fe/H] $= 0.06$. The adjusted histograms in Figure~\ref{fig:tilthist_agefeh} (right panel) show that the super-Earth peak is lower for metal-poor stars. The $E$ values are reported in Table~\ref{tab:tilthist_feh}.

\begin{table}
    \centering
    \caption{$E$ values of the radius valley for different stellar metallicities. $E_{\text{avg}} = (E_{\text{SN}} + E_{\text{SE}})/2$.}
    \begin{tabular}{ccccc}
        \hline
        [Fe/H] & $N_{\text{planets}}$ & $E_{\text{SN}}$ & $E_{\text{SE}}$ & $E_{\text{avg}}$ \\
        \hline
        $<0.06$ & 181 & 4.10 & 2.20 & 3.15 \\
        $\geq 0.06$ & 194 & 3.28 & 2.78 & 3.03 \\
        \hline
    \end{tabular}
    \label{tab:tilthist_feh}
\end{table}

We perform a similar SVM analysis in $R_p$--$P$--[Fe/H] space (shown in Figure~\ref{fig:svm_3d_agefeh} right panel), and find
\begin{equation}
    \log_{10}{\left(R_p/R_{\earth}\right)} = A \log_{10}{\left(P/\text{days}\right)} + B \text{[Fe/H]} + C
\end{equation}
with $A = -0.10 \pm 0.03$, $B = 0.03_{-0.04}^{+0.03}$, and $C = 0.36_{-0.03}^{+0.02}$. These values imply that we find no evidence that the radius valley location depends on stellar metallicity.

\section{Discussion} \label{sect:discussion}

\subsection{$R_p$-$P$ relation suggests a thermally-driven mass loss model} \label{subsect:slope comparison}

As presented in Section~\ref{subsect: radvalley_pos}, we observe the radius valley scales as $m = \mathrm{d}\log{R_p}/\mathrm{d}\log{P} = -0.11 \pm 0.02$. This negative period-dependence is a robust finding which remains roughly similar even when other parameters are included in the fit (see Table~\ref{tab:svm_values}). 

\begin{table*}
    \renewcommand*{\arraystretch}{1.25}
    \centering
    \caption{Slope of the radius valley on the radius-period plane from various sources.}
    \begin{tabular}{cccc}
        \hline
        & Source & $m = \text{d}\log{R_p}/\text{d}\log{P}$ & Stellar type \\
        \hline
        \multirow{8}{*}{Observations} & This work & $-0.11_{-0.02}^{+0.02}$ & FGK \\
        & \citet{vaneylen2018asteroseismic} & $-0.09_{-0.04}^{+0.02}$ & FGK \\
        & \citet{martinez2019spectroscopic} & $-0.11_{-0.02}^{+0.02}$ & FGK \\
        & \citet{macdonald2019examining} & $-0.319_{-0.116}^{+0.088}$ & FGK \\
        & \citet{cloutier2020evolution} & $0.058_{-0.022}^{+0.022}$ & M \\
        & \citet{vaneylen2021masses} & $-0.11_{-0.04}^{+0.05}$ & M \\
        & \citet{petigura2022california} & $-0.11_{-0.02}^{+0.02}$ & FGKM \\
        & \citet{luque2022density} & $-0.02_{-0.05}^{+0.05}$ & M \\
        \hline
        & Source & $m = \text{d}\log{R_p}/\text{d}\log{P}$ & Model \\
        \hline
        \multirow{6}{*}{Theory} & \citet{owen2017evaporation} & $-0.25 \leq m \leq -0.16$ & Photoevaporation \\
        & \multirow{2}{*}{\citet{lopez2018how}} & -0.09 & Photoevaporation\\
        & & 0.11 & Gas-poor formation \\
        & \citet{gupta2019sculpting} & $-0.11$ & Core-powered mass loss \\
        & \multirow{2}{*}{\citet{rogers2021photoevaporation}} & $-0.16$ & Photoevaporation \\
        & & $-0.11$ & Core-powered mass loss \\
        \hline
    \end{tabular}
    \label{tab:rp_values}
\end{table*}

Different theoretical mechanisms to create the radius valley result in a different slope as a function of orbital period. For example, \citet{lopez2018how} predicted that if the rocky planets are core remnants of sub-Neptunes with evaporated atmospheres, the radius valley location should decrease with increasing orbital period, with $m = -0.09$, whereas if those rocky planets were formed after disk dissipation (i.e., late gas-poor formation), the radius valley location tends to larger planetary radii at longer orbital periods, with $m=0.11$. Similarly, \citet{owen2017evaporation} predicted a negative period-radius valley slope for a photoevaporation model, with $-0.25 \leq m \leq -0.16$ depending on the photoevaporation efficiency. If the radius valley is thermally driven but powered by the core rather than photoevaporation, the slope would be similarly negative, with e.g., \citet{gupta2019sculpting} predicting that $m \approx -0.11$ in this case. Theoretically predicted slopes for different formation mechanisms are summarised in Table~\ref{tab:rp_values}.

Our observed negative slope is consistent with thermally driven mass-loss models but inconsistent with late gas-poor formation models. We can also compare our observed slope with other observational studies (see again Table~\ref{tab:rp_values}). 
The period-radius slope was first observed by  \citet{vaneylen2018asteroseismic}, who used the SVM approach that we adopted here and who found $m = -0.09^{+0.02}_{-0.04}$. A different approach was followed by \citet{martinez2019spectroscopic}, who divided their planetary sample into 10 bins with equal number of planets, determined the minimum radius in each bin, and fitted a linear relationship to obtain equation~\ref{eq:logr_logp}. These two approaches led to a consistent result, with $m=-0.11 \pm 0.02$. \citet{macdonald2019examining} adopted machine learning approaches, and report $m=-0.319^{+0.088}_{-0.116}$. The above studies all focus on samples of FGK stars, where various approaches to model the valley's location appear to result in negative slopes with consistent magnitude, matching thermally-driven atmospheric loss models.

For smaller and cooler (M type) stars, \citet{cloutier2020evolution} found a positive slope ($m=0.058 \pm 0.022$) using a method similar to \citet{martinez2019spectroscopic}, suggesting for these stars the valley may be the result of gas-poor formation rather than being thermally driven. \cite{vaneylen2021masses} used the SVM approach to measure the M dwarf valley and found a negative slope instead, of $-0.11_{-0.04}^{+0.05}$. \citet{luque2022density} used the \texttt{gapfit} package \citep{loyd2020current} and found $m = -0.02 \pm 0.05$. A recent study by \citet{petigura2022california} also included M type stars in addition to FGK stars, and they found $m = -0.11 \pm 0.02$ for this sample. Our sample does not include M type stars but does span a mass range from about 0.6 to 1.4~$M_\star$. 

\begin{figure*}
    \centering
    \begin{tabular}{cc}
    \includegraphics[width=0.99\columnwidth]{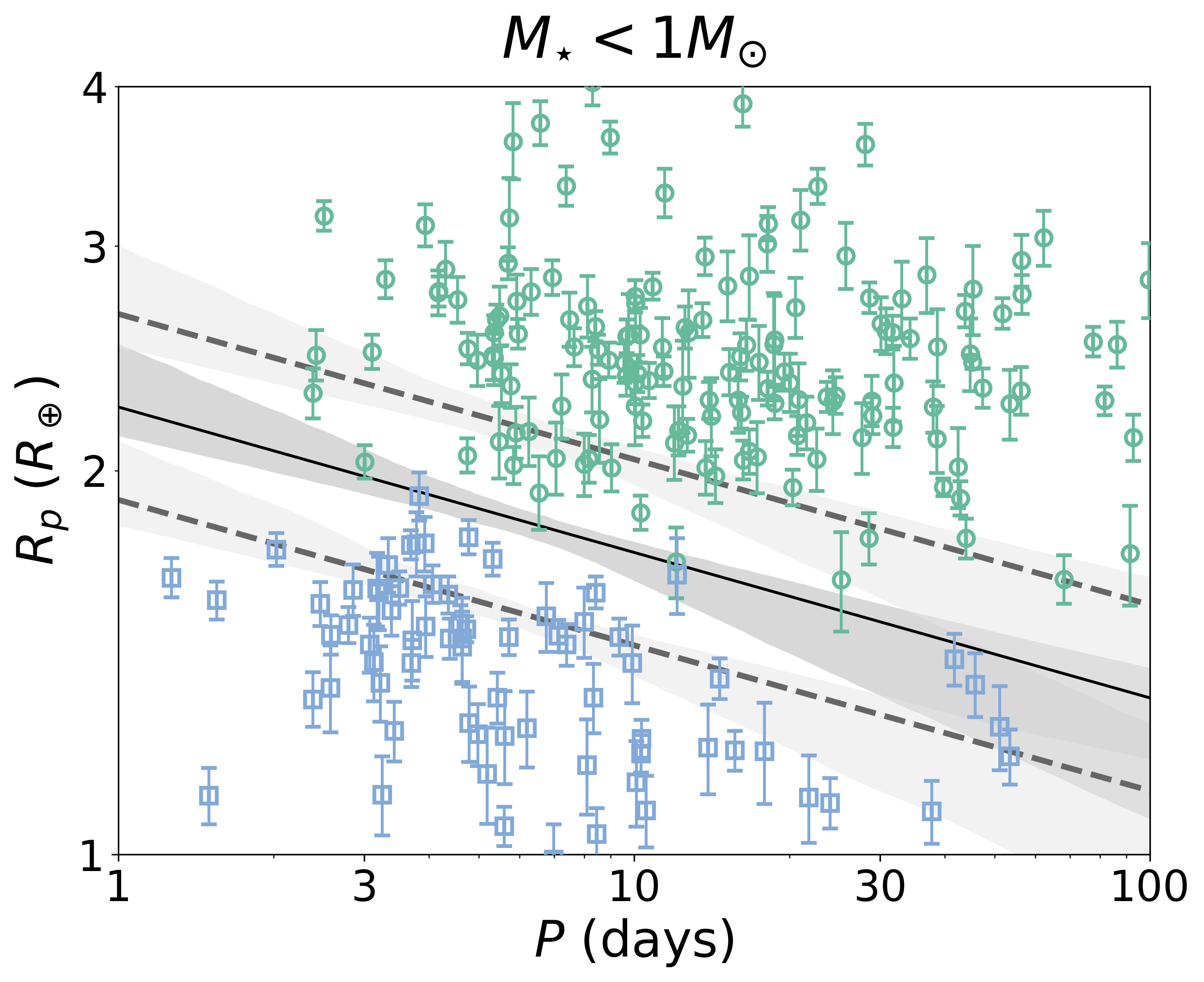} &
    \includegraphics[width=0.99\columnwidth]{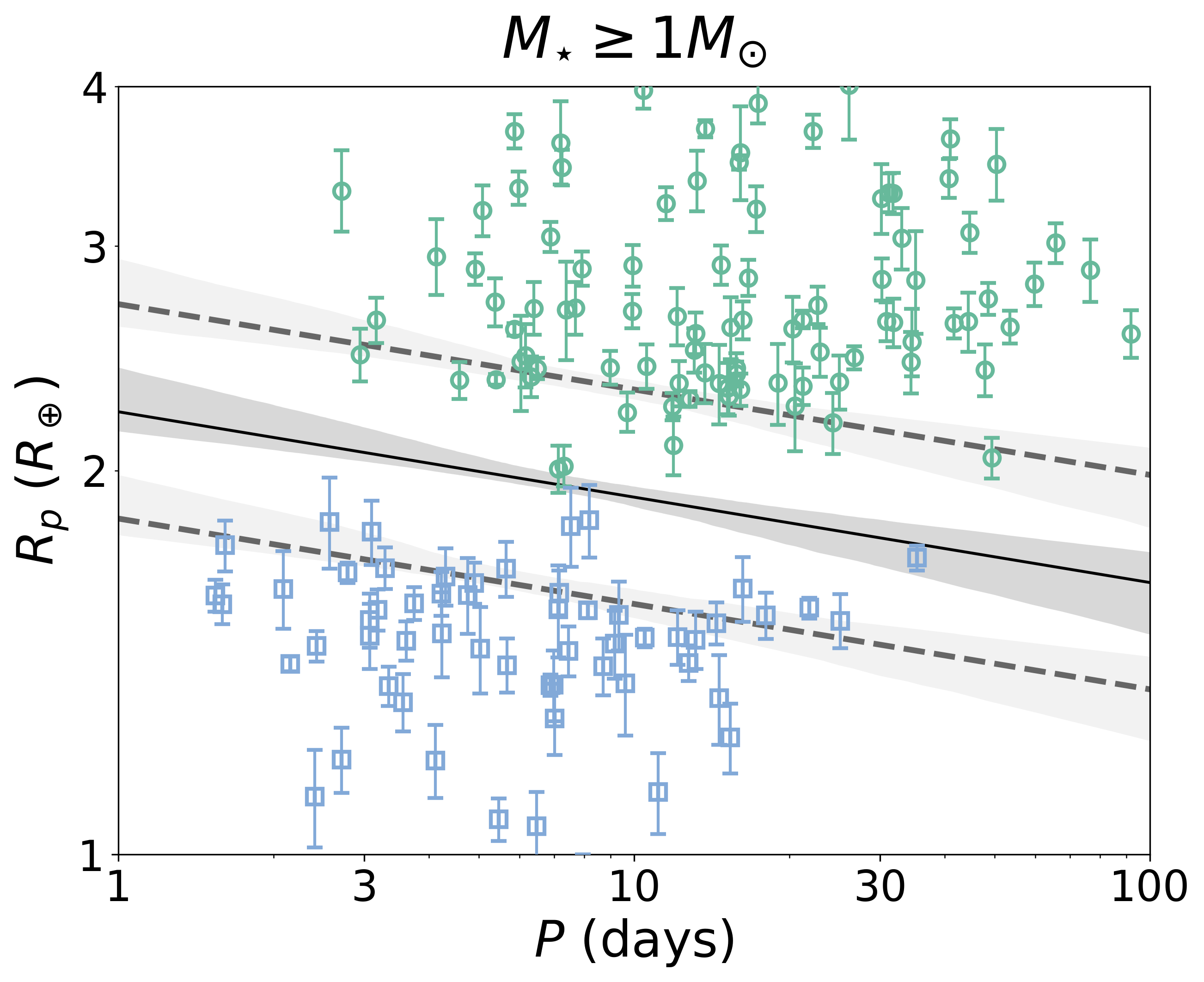}
    \end{tabular}
    \caption{Radius valley position for planets with host star mass $M_{\star} < 1M_{\odot}$ (left), and $M_{\star} \geq 1M_{\odot}$ (right). For $M_{\star} < 1M_{\odot}$, $m = -0.11_{-0.07}^{+0.02}$ and $c = 0.35_{-0.02}^{+0.05}$; for $M_{\star} \geq 1M_{\odot}$, $m = -0.07_{-0.04}^{+0.02}$ and $c = 0.35_{-0.02}^{+0.03}$. The green and blue points show planets above and below the radius valley respectively. The grey shaded region represents the $\pm 1 \sigma$ uncertainty in the radius valley position determined with bootstrapping.}
    \label{fig:svm_RP_stellar}
\end{figure*}

To investigate whether the slope of $m$ changes with stellar mass within our sample, we split our planetary sample into two groups: $M_{\star} \geq 1M_{\sun}$, and $M_{\star} < 1M_{\sun}$. We determine the $R_p-P$ relation separately for these two groups with the same methods as above. We find for $M_{\star} \geq 1M_{\sun}$, $m = -0.07_{-0.04}^{+0.02}$ and $c=0.35_{-0.02}^{+0.03}$; for $M_{\star} < 1M_{\sun}$, $m = -0.11_{-0.07}^{+0.02}$ and $c = 0.35_{-0.02}^{+0.05}$. These results are shown in Figure~\ref{fig:svm_RP_stellar}. The two values are in agreement within $1\sigma$, suggesting that within our sample the radius valley location as a function of orbital period is inconsistent with the gas-poor formation scenario.

We can also look at the slope of the valley as a function of incident flux ($S$) rather than orbital period. By Kepler's third law (as shown in equation~\ref{eq:kepler_3rd}), planets at longer orbital periods are located further away from the planet, thus the incident flux $S$ is lower for planets with larger star-planet distances as shown in equation~\ref{eq:S}. Hence, we expect for a thermally-driven planetary mass loss scenario, the radius valley location tends to larger planetary radii for higher $S$. We observe this positive relationship in this work, in agreement with other previous observations as shown in Table~\ref{tab:rs_values}, and consistent with thermally-driven mass loss models which is also shown in radius-period space.

\begin{table*}
    \renewcommand*{\arraystretch}{1.25}
    \centering
    \caption{Same as Table~\ref{tab:rp_values}, but for the radius valley slope on the radius-incident-flux plane.}
    \begin{tabular}{cccc}
        \hline
        & Source & $m = \text{d}\log{R_p}/\text{d}\log{S}$ & Stellar type \\
        \hline
        \multirow{5}{*}{Observations} & This work & $0.07_{-0.01}^{+0.02}$ & FGK \\
        & \citet{martinez2019spectroscopic} & $0.12 \pm 0.02$ & FGK \\
        & \citet{cloutier2020evolution} & $-0.060 \pm 0.025$ & M \\
        & \citet{petigura2022california} & $0.06 \pm 0.01$ & FGKM \\
        & \citet{luque2022density} & $0.02 \pm 0.02$ & M \\
        \hline
    \end{tabular}
    \label{tab:rs_values}
\end{table*}

\subsection{$R_p$-$M_{\star}$ relation supports a thermally-driven mass loss model} \label{subsect:discuss_mstar}

As presented in Section~\ref{subsect:ms_results}, we find that in two dimensions, $m = \mathrm{d}\log{R_p}/\mathrm{d}\log{M_{\star}} = 0.23_{-0.08}^{+0.09}$. 

A stellar mass dependence has been predicted by radius valley models. Both thermally driven mass-loss models predict a similar dependence of the valley on stellar mass. For example, \citet{rogers2021photoevaporation} predicted $m = 0.29$ and $m = 0.32$ for photoevaporation
\citep{owen2017evaporation} and core-powered mass loss models \citep{gupta2019sculpting, gupta2020signatures} respectively. Our results are consistent with both sets of models within $1\sigma$.

A stellar mass dependence was observed by  \citet{berger2020gaia2}, who find $m=0.26^{+0.21}_{-0.16}$ by fitting the minima of the 2-dimensional KDE in $R_p-M_{\star}$ space. A recent study by \citet{petigura2022california}, similarly following a binning approach and incorporating data from Data Release 2 (DR2) of the California-\textit{Kepler} Survey (CKS) for cooler stars, estimated $m = 0.18_{-0.07}^{+0.08}$. It is therefore reassuring to see that despite the different method adopted here, the slope derived in this work is consistent with both of these studies within $1 \sigma$. For lower mass stars, \citet{luque2022density} found $m = 0.08 \pm 0.12$; this may be inconsistent with our results at $1\sigma$, however the stellar mass range they studied is significantly lower than that in our sample with no overlaps. The results are summarised in Table~\ref{tab:rm_values}.

\begin{table*}
    \renewcommand*{\arraystretch}{1.25}
    \centering
    \caption{Same as Table~\ref{tab:rp_values}, but for the radius valley slope on the radius-stellar-mass plane.}
    \begin{tabular}{cccc}
        \hline
        & Source & $m = \text{d}\log{R_p}/\text{d}\log{M_{\star}}$ & Stellar type \\
        \hline
        \multirow{4}{*}{Observations} & This work & $0.23_{-0.08}^{+0.09}$ & FGK \\
        & \citet{berger2020gaia2} & $0.26_{-0.16}^{+0.21}$ & FGKM \\
        & \citet{petigura2022california} & $0.18_{-0.07}^{+0.08}$ & FGKM \\
        & \citet{luque2022density} & $0.08_{-0.12}^{+0.12}$ & M \\
        \hline
        & Source & $m = \text{d}\log{R_p}/\text{d}\log{M_{\star}}$ & Model \\
        \hline
        \multirow{3}{*}{Theory} & \citet{gupta2020signatures} & 0.33 & Core-powered mass loss \\
        & \multirow{2}{*}{\citet{rogers2021photoevaporation}} & 0.29 & Photoevaporation \\
        & & 0.32 & Core-powered mass loss \\
        \hline
    \end{tabular}
    \label{tab:rm_values}
\end{table*}

\begin{figure}
    \centering
    \includegraphics[width=\linewidth]{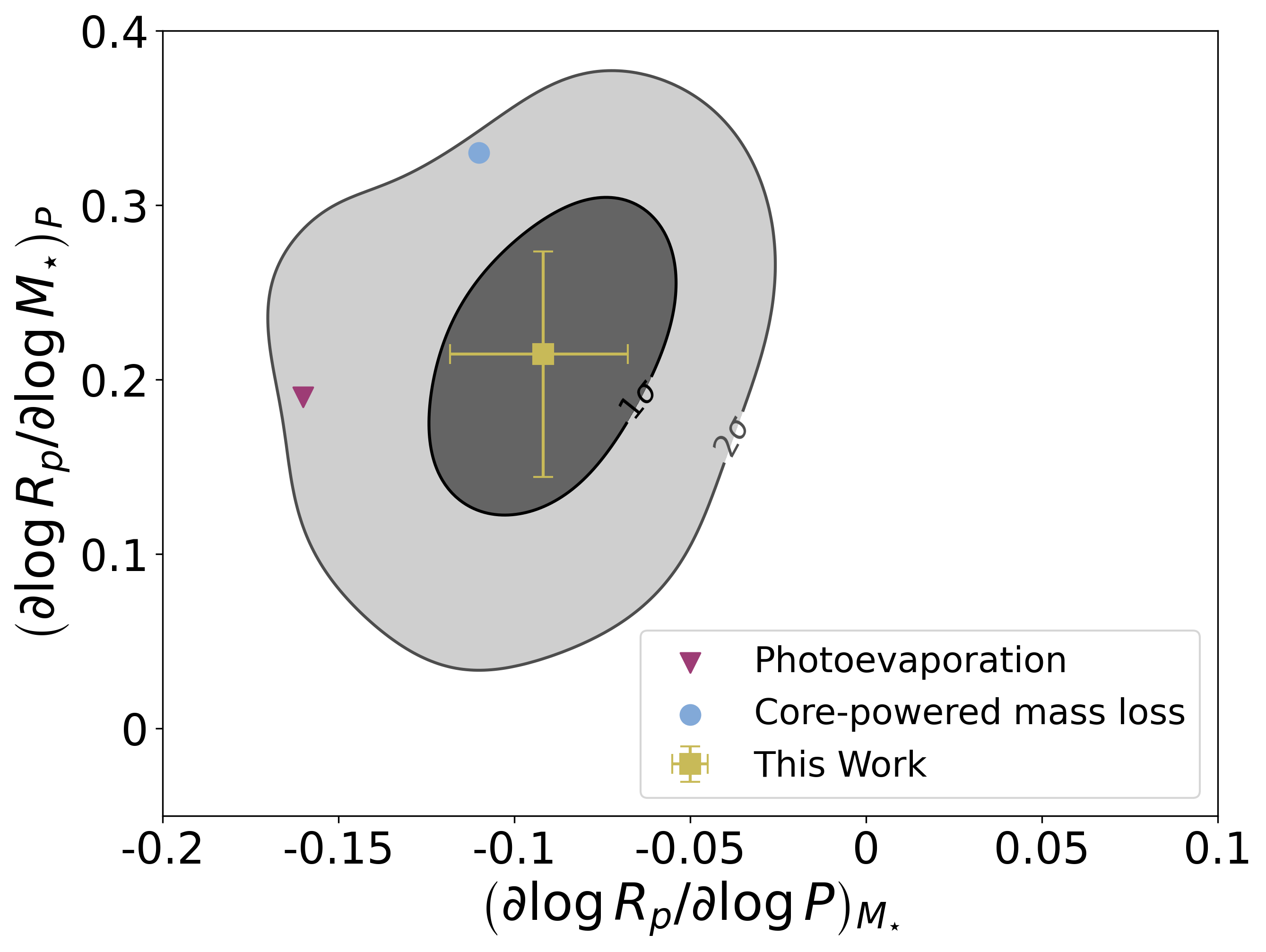}
    \caption{Posterior distributions of the radius valley location dependence with respect to orbital period at constant stellar mass $\left( \partial\log{R_p}/\partial\log{P}\right)_{M_{\star}}$, and stellar mass at constant orbital period $\left(\partial\log{R_p}/\partial\log{M_{\star}} \right)_P$. The dark and light-coloured shades represent the $1\sigma$ and $2\sigma$ uncertainties respectively. The theoretical models of photoevaporation and core-powered mass loss are taken from \citet{vaneylen2021masses}.}
    \label{fig:rpm_contour}
\end{figure}

When extending our analysis to 3 dimensions as a function of $P$ and $M_{\star}$, we obtain $A = \left(\partial\log{R_p}/\partial\log{P}\right)_{M_{\star}} = -0.09_{-0.03}^{+0.02}$, $B = \left(\partial\log{R_p}/\partial\log{M_{\star}}\right)_{P} = 0.21_{-0.08}^{+0.06}$ from determining the radius valley location in the $R_p$--$P$--$M_{\star}$ space. Note that this is different to the total derivative $\mathrm{d}\log{R_p}/\mathrm{d}\log{M_{\star}}$ in two dimensions (shown in Table~\ref{tab:svm_values}). Based on the models of photoevaporation \citep{owen2017evaporation, owen2019effects, mordasini2020planetary} and core-powered mass loss \citep{gupta2019sculpting, gupta2020signatures}, \citet{vaneylen2021masses} predicted $B=0.19$ for a photoevaporation model, and $B=0.33$ for a core-powered mass-loss model. Our resulting posterior distribution of $A$ and $B$ determined from the bootstrapping presented in Section~\ref{subsect:ms_results}, as shown in Figure~\ref{fig:rpm_contour}, is consistent with both the photoevaporation and core-powered mass loss cases at $2\sigma$, hence we are unable to distinguish between the two models in this particular parameter space.

\begin{figure}
    \centering
    \includegraphics[width=\linewidth]{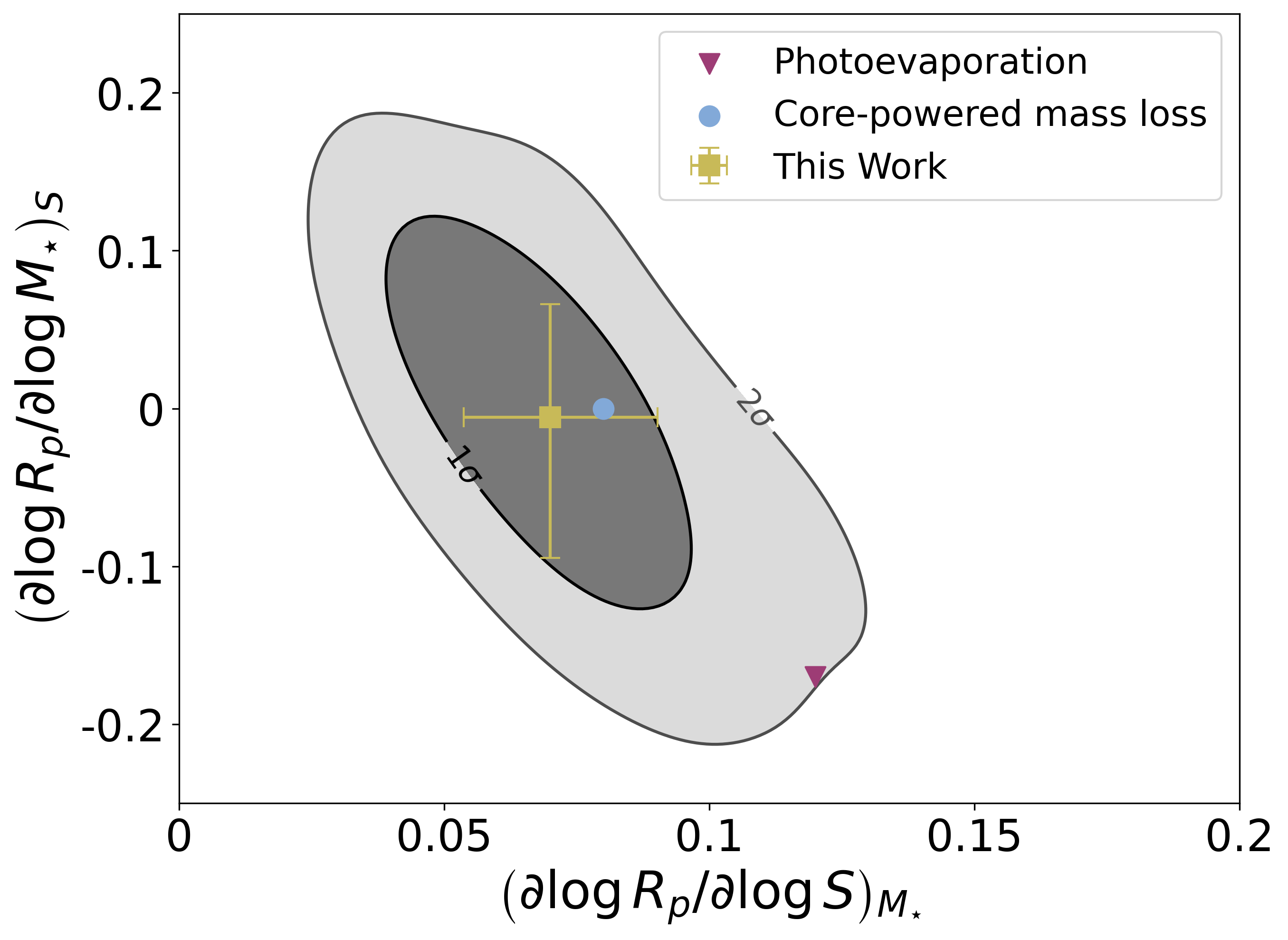}
    \caption{Same as Figure~\ref{fig:rpm_contour}, but for $\left( \partial\log{R_p}/\partial\log{S}\right)_{M_{\star}}$ and $\left( \partial\log{R_p}/\partial\log{M_{\star}}\right)_{S}$.}
    \label{fig:rsm_contour}
\end{figure}

\citet{rogers2021photoevaporation} proposed an analysis of the radius valley in $R_p$--$S$--$M_{\star}$ space that could distinguish between the two different thermally-driven mass-loss mechanisms. Using theoretical models, they predicted the radius valley scales as a function of $S$ and $M_{\star}$ as equation~\ref{eq:p_ms_rp_relation}, with $A = \left(\partial\log{R_p}/\partial\log{S}\right)_{M_{\star}} \simeq 0.12$ and $B = \left(\partial\log{R_p}/\partial\log{M_{\star}}\right)_{S} \simeq -0.17$ for a photoevaporation model, and $A \simeq 0.08$ and $B \simeq 0.00$ for a core-powered mass loss model. Again, we plot the posterior distributions of $A$ and $B$ as shown in Figure~\ref{fig:rsm_contour}, and observe that our results are consistent with the core-powered mass loss case well within $1\sigma$. For the photoevaporation scenario, our values overlap with the theoretical predictions at the edge of the 2$\sigma$ confidence interval. \citet{rogers2021photoevaporation} also measured the planet density of the California-\textit{Kepler} Survey \citep[CKS,][]{fulton2018california}, and the \textit{Gaia}-\textit{Kepler} Survey \citep[GKS,][]{berger2020gaia2}, in $R_p$--$S$--$M_{\star}$ space. They found for the CKS data, $A = 0.13_{-0.05}^{+0.03}$, $B = -0.21_{-0.39}^{+0.33}$, and for the GKS data, $A = 0.10_{-0.02}^{+0.03}$, $B = -0.03_{-0.12}^{+0.10}$. Our results are in agreement with both the CKS and GKS values, and our measurements have smaller uncertainties. 

There are some caveats to this comparison between our observation results and theoretical models. Firstly, the thermally-driven mass loss models predict the slope of the bottom of the valley \citep{vaneylen2018asteroseismic, rogers2021photoevaporation}, whereas our SVM finds the slope for the middle of the radius valley. Some studies have suggested a different planet size dependence with orbital period for super-Earths and sub-Neptunes \citep[e.g.][]{petigura2022california}, hence these two slopes may not be equal. Since the radius valley is not completely empty, the bottom of the radius valley is not clearly defined, and there would be challenges locating and fitting the bottom of the radius valley. As a result, our observed values may not be fully comparable with theoretical model values. Furthermore, the method of extracting the radius valley is prone to transit biases, which we do not correct for in this work. \citet{rogers2021photoevaporation} showed that even when modelling synthetic transit surveys based on evolving planets with theoretical models, the resulting posteriors may not be fully consistent with the theoretically predicted slope. Further work, such as generating synthetic surveys from both photoevaporation and core-powered mass loss models based on conditions similar to that of our sample in a method similar to that performed in \citet{rogers2021photoevaporation}, and fitting the valley with the same method as in this work, or analysing more planets around stars in a larger mass range, is required to compare our observations to theoretical models in a homogeneous way.

\subsection{Deeper radius valley suggests a homogeneous initial planetary core composition} \label{subsect:discussion_deepness}

\begin{figure*}
    \centering
    \begin{tabular}{ccc}
    \includegraphics[width=0.32\linewidth]{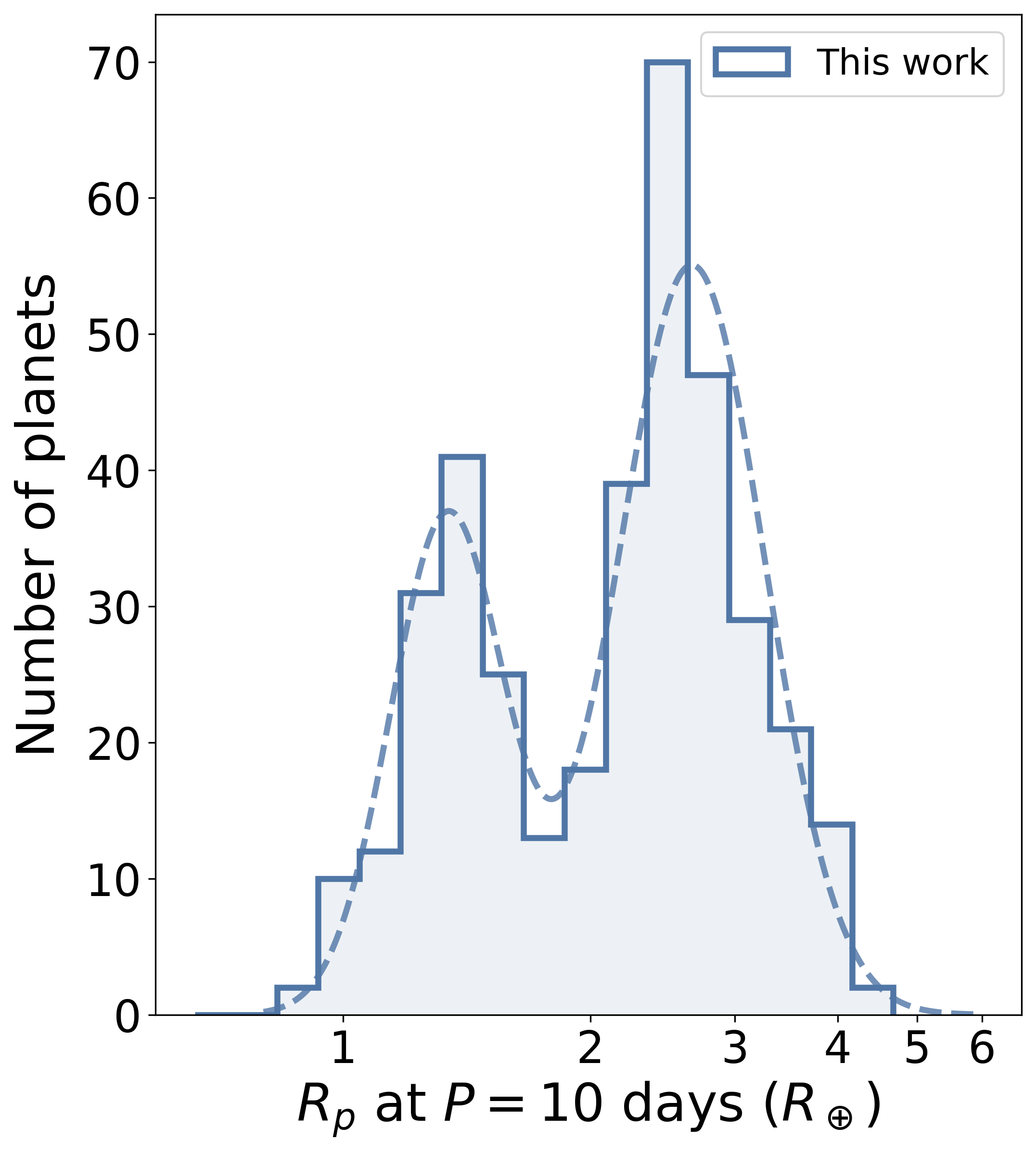} &
    \includegraphics[width=0.32\linewidth]{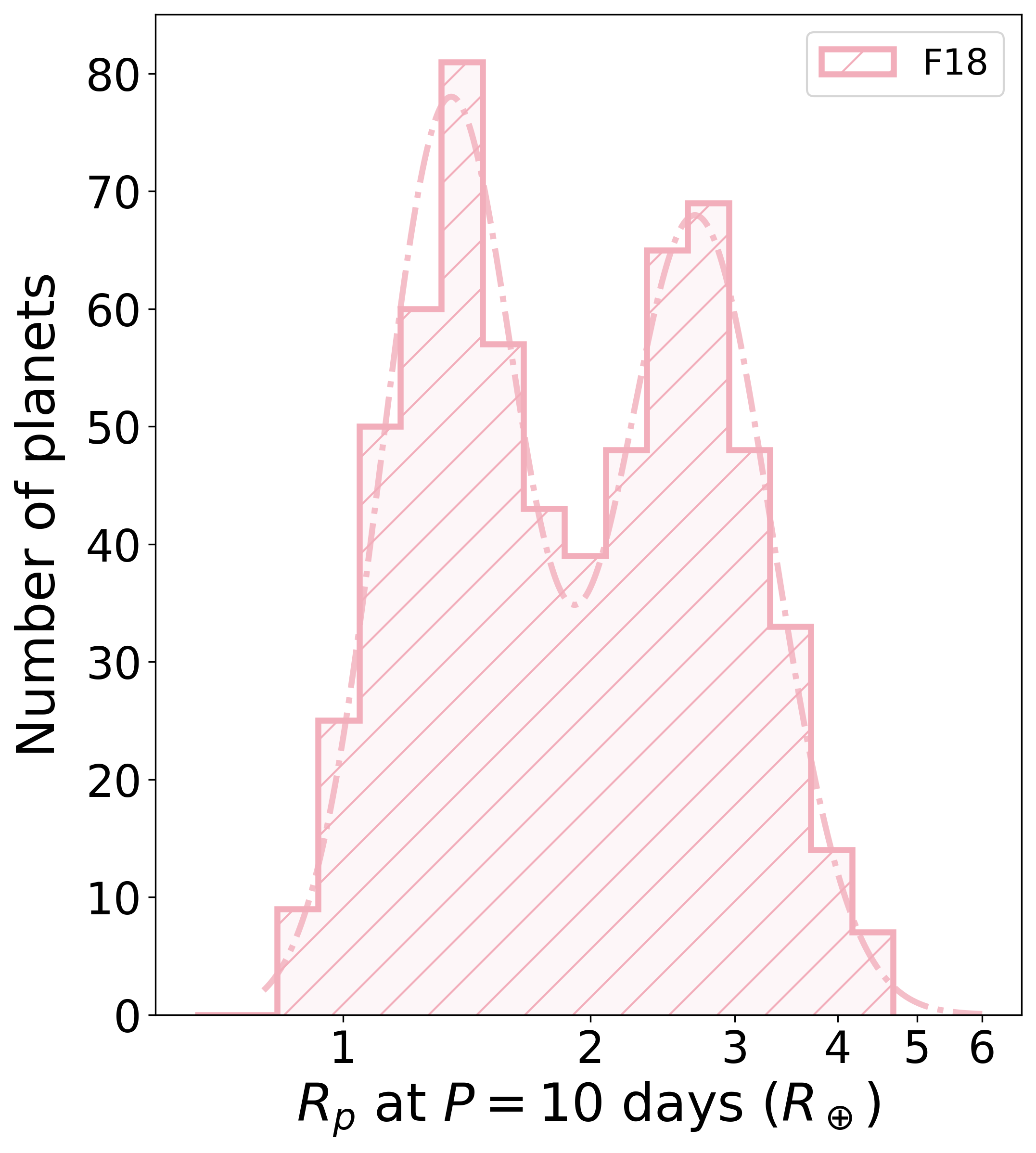} &
    \includegraphics[width=0.32\linewidth]{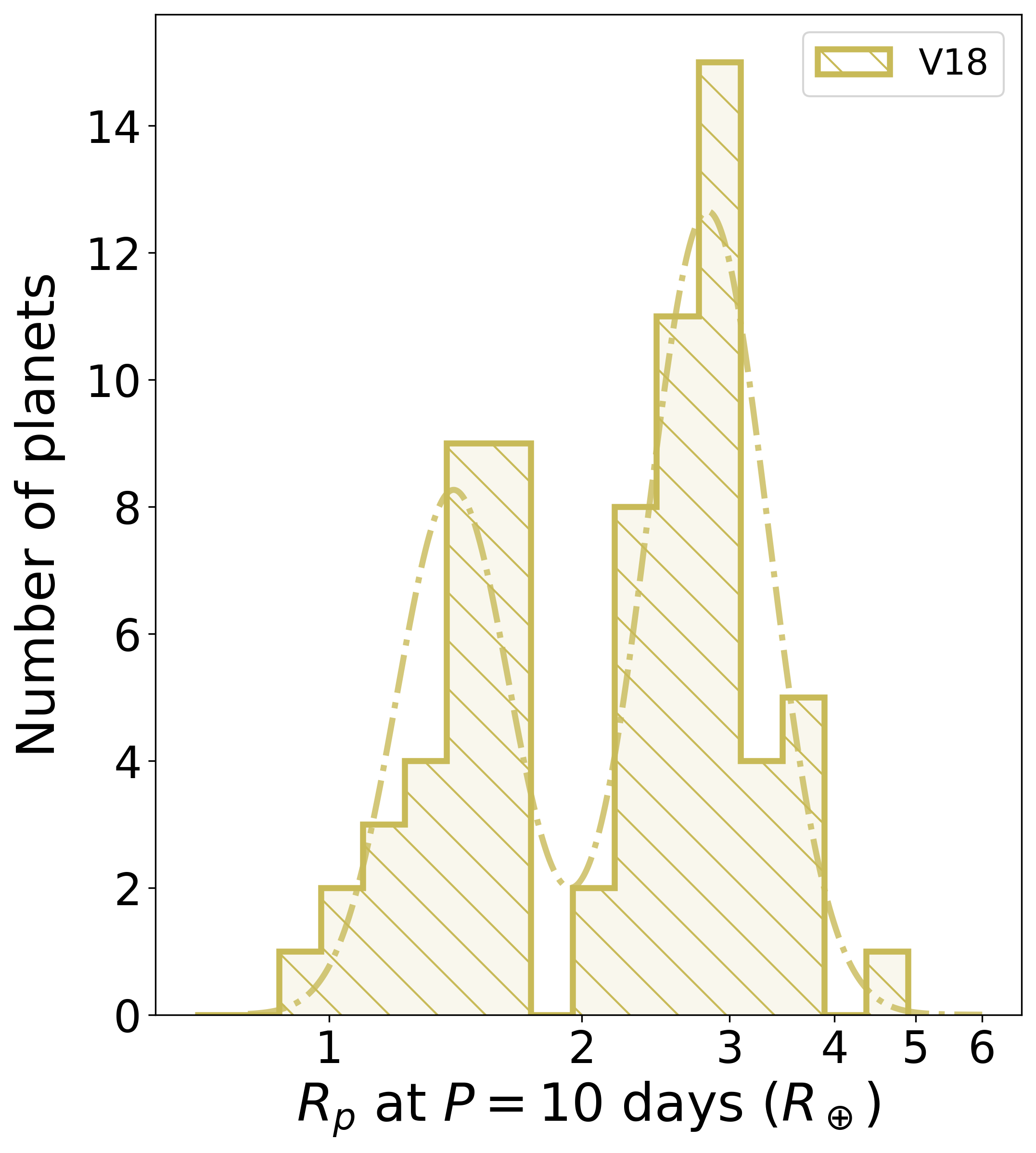}
    \end{tabular}
    \caption{Histogram of planet radii, adjusted to $P = 10$ days for the planetary population in this work (\textit{left}, identical to Figure~\ref{fig:tilthist}), \citetalias{fulton2018california} (\textit{centre}), and \citetalias{vaneylen2018asteroseismic} (\textit{right}). Planets in this work and \citetalias{fulton2018california} are shifted according to the slope defined in Section~\ref{subsect: radvalley_pos} with the SVM (i.e. $-0.11_{-0.02}^{+0.02}$), whereas planets in \citetalias{vaneylen2018asteroseismic} are shifted according to the slope found in \citetalias{vaneylen2018asteroseismic} (i.e. $m=-0.09_{-0.04}^{+0.02}$). The $E$ metrics defining the average peak-to-valley ratio are $2.98_{-0.47}^{+0.60}$, $2.14_{-0.21}^{+0.26}$, and $6.05_{-2.14}^{+3.87}$ respectively.}
    \label{fig:tilthist_compare}
\end{figure*}

We now turn to the depth of the radius valley. Using the previously defined depth metric ($E$, equations~\ref{eq:E_SN} and~\ref{eq:E_SE}), we find a valley depth of $E_{\text{avg}} = 2.98_{-0.47}^{+0.60}$ (see Section~\ref{subsect:results_deepness}). We can compare this depth to the valley observed by \citetalias{fulton2018california}. Shifting the planets along the slope calculated in Section~\ref{subsect: radvalley_pos}, and applying the same metric to their filtered sample of 907 planets, we calculate $E_{\text{SN}} = 1.99_{-0.23}^{+0.26}$, $E_{\text{SE}} = 2.28_{-0.27}^{+0.31}$, giving $E_{\text{avg}} = 2.14_{-0.21}^{+0.26}$ for that sample. For \citetalias{vaneylen2018asteroseismic}, we shift the planets according to the slope obtained in their study, i.e. $m=-0.09_{-0.04}^{+0.02}$, and we find $E_{\text{SN}} = 7.11_{-2.49}^{+4.70}$, $E_{\text{SE}} = 4.75_{-1.70}^{+3.42}$, giving $E_{\text{avg}} = 6.05_{-2.14}^{+3.87}$. These values imply that compared to \citetalias{fulton2018california}, we observe a deeper radius valley. On the other hand, the radius valley appears less deep than observed by \citetalias{vaneylen2018asteroseismic} for a smaller sample. This finding is visualised in 
Figure~\ref{fig:tilthist_compare}, which shows the adjusted histograms of the sample studied here next to the \citetalias{fulton2018california} and the \citetalias{vaneylen2018asteroseismic} samples.

\begin{figure*}
    \centering
    \begin{tabular}{cc}
        \includegraphics[width=0.99\columnwidth]{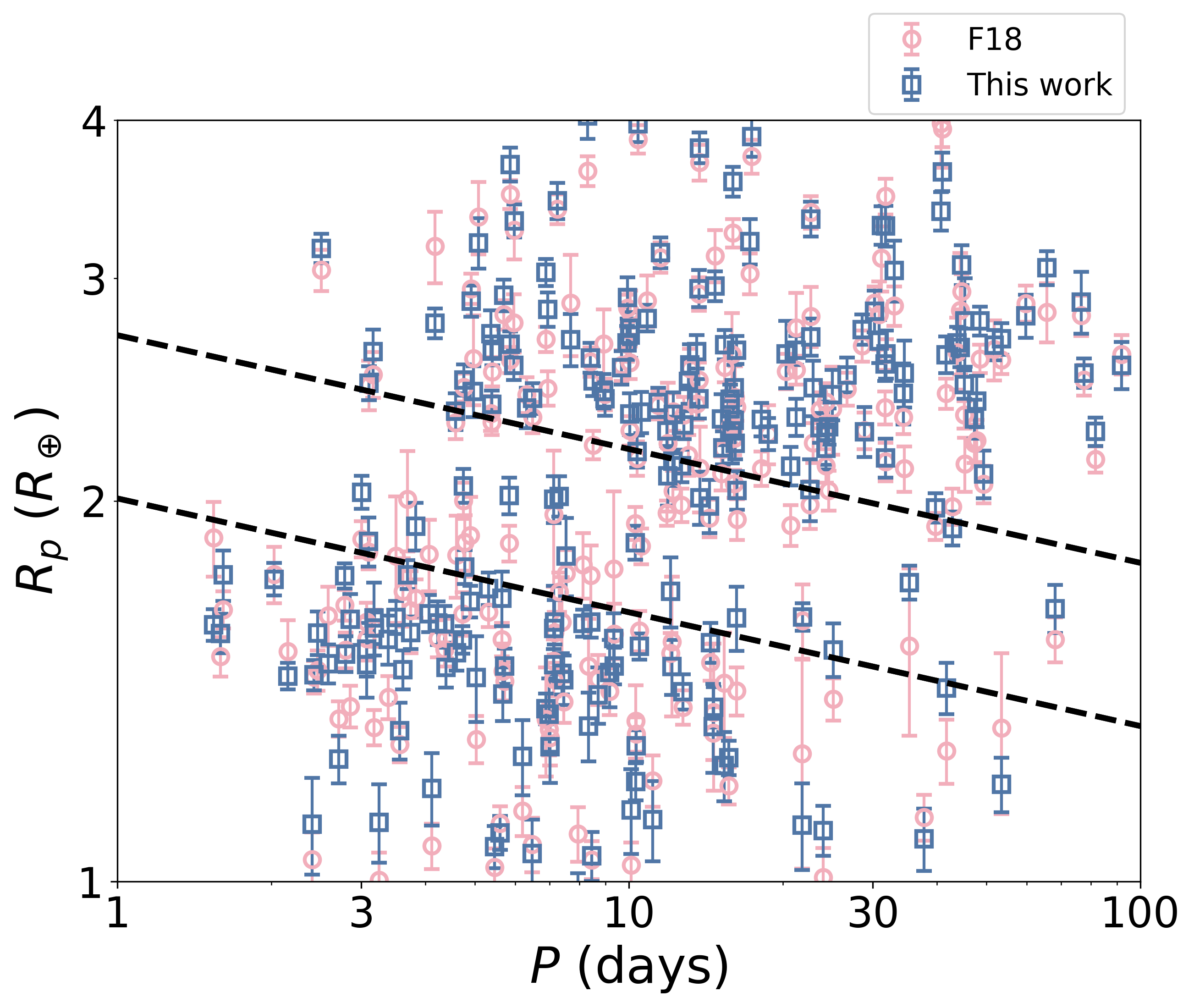} & \includegraphics[width=0.99\columnwidth]{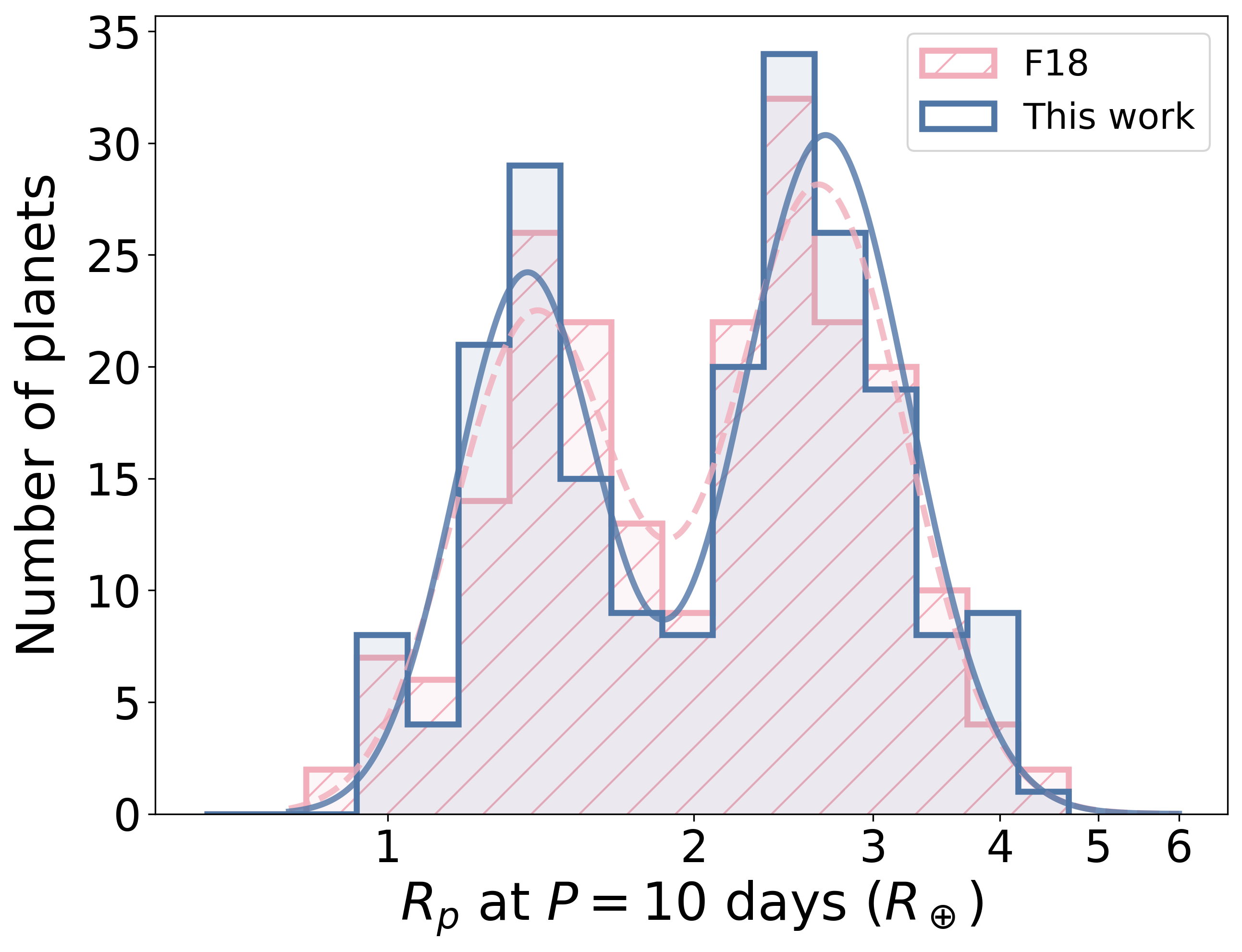}
    \end{tabular}
    \caption{\textit{Left:} Radius-period plot of the 211 planets common to this work (blue) and \citetalias{fulton2018california} (red), using the same stellar radii to calculate planetary radii. \textit{Right:} Same as Figure~\ref{fig:tilthist_compare}, but the planetary population is filtered to the 211 common planets in both samples. Planets in both samples are adjusted to equivalent radii at $P=10$ days according to the slope calculated in Section~\ref{subsect: radvalley_pos} with the SVM. The red and blue histograms are produced with the parameters obtained from \citetalias{fulton2018california} and this work respectively. The histograms and fitting with the Gaussian Mixture Model show that the observed valley is deeper in this work.}
    \label{fig:tilthist_newf18}
\end{figure*}

To investigate the reason for observing a deeper valley than \citetalias{fulton2018california}, we compare the 211 planets common in both our sample and the filtered sample of \citetalias{fulton2018california}. To investigate the role of transit fitting, we convert all our $R_p/R_{\star}$ into $R_p$ using $R_{\star}$ from \citetalias{fulton2018california} (even when \citetalias{vaneylen2018asteroseismic} values are available). The results are shown in Figure~\ref{fig:tilthist_newf18}, which compares the same planets with the same stellar parameters but different transit fitting. We observe that in this case, the $R_p$ of 56 (27\%) and 24 (11\%) planets change by $>2\sigma$ and $> 3\sigma$ respectively, compared to the values reported in \citetalias{fulton2018california}. We find for this common planetary sample, for our planetary parameters, $E_{\text{SN}} = 3.63_{-0.70}^{+1.02}$, $E_{\text{SE}} = 2.95_{-0.63}^{+0.86}$, giving $E_{\text{avg}} = 3.29_{-0.59}^{+0.91}$, whereas for parameters from \citetalias{fulton2018california}, $E_{\text{SN}} = 2.40_{-0.47}^{+0.59}$, $E_{\text{SE}} = 1.93_{-0.39}^{+0.49}$, giving $E_{\text{avg}} = 2.18_{-0.42}^{+0.49}$. These findings suggest that our updated transit fittings are directly responsible for deepening (although not fully emptying) the radius valley.

A deeper radius valley is associated with a more homogeneous planet core composition. For example, in photoevaporation models the radius valley position is dependent on the mass of the planet core ($M_c$), and the density of a $1M_{\earth}$ core of a particular core composition ($\rho_{M_{\earth}}$), as
\begin{equation}
    R_{\text{valley}} \propto \rho_{M_{\earth}}^{-1/3} M_c^{1/4}.
    \label{eq:rad_valley_owen2017}
\end{equation}
Hence, if $R_{\text{valley}}$ is known, and the planets' mean masses are known, the planetary core compositions could be deduced \citep{owen2017evaporation}.

Using the above relation, if the planetary cores were icy at formation, the radius valley would be located at a higher planetary radius than if the cores were rocky/terrestrial at formation. Hence, if the planetary cores are of mixed composition, a superposition of the two models would be predicted, and we would expect the radius valley to be smeared and less distinct, as each type of planet would have its own `radius valley' at a different location \citep{owen2017evaporation}. Our deep radius valley found in this work implies the opposite case, where the planetary cores are more similar in composition. In this scenario, planets inside the valley may have a different (e.g. icy) composition.

\citet{owen2017evaporation} compared their models to observations, and found that the planet compositions are more likely to be Earth-like (i.e. rocky), but that the apparent shallowness of the valley  suggested a wide distribution of iron fractions $(f_{\text{Fe}})$ in their cores, as planets with a single value iron fraction ($f_{\text{Fe}}=0.5$) produces a deeper valley compared to planets with a uniform distribution ($f_{\text{Fe}} \in [0,1]$). Comparing our finding of a deeper valley to models in \citet{owen2017evaporation} would indicate that the planet compositions are more likely to have similar iron fractions with a narrower spread.

Similarly, in the core-powered mass loss model, the location of the radius valley scales as
\begin{equation}
    R_{\text{valley}} \propto \rho_c^{-4/9}
\end{equation}
where $\rho_c$ is the planet core density \citep{gupta2019sculpting}. The same reasoning as the photoevaporation case then applies: given the larger $\rho_c$ for icy cores, planets with homogeneous icy cores will produce a radius valley at a larger planetary radii compared to rocky/terrestrial cores, implying that the radius valley would be smeared if planetary cores are of inhomogeneous compositions. Our deep radius valley supports the opposite case, i.e., a similar planetary core composition.

\begin{figure*}
    \centering
    \begin{tabular}{cc}
    \includegraphics[width=0.99\columnwidth]{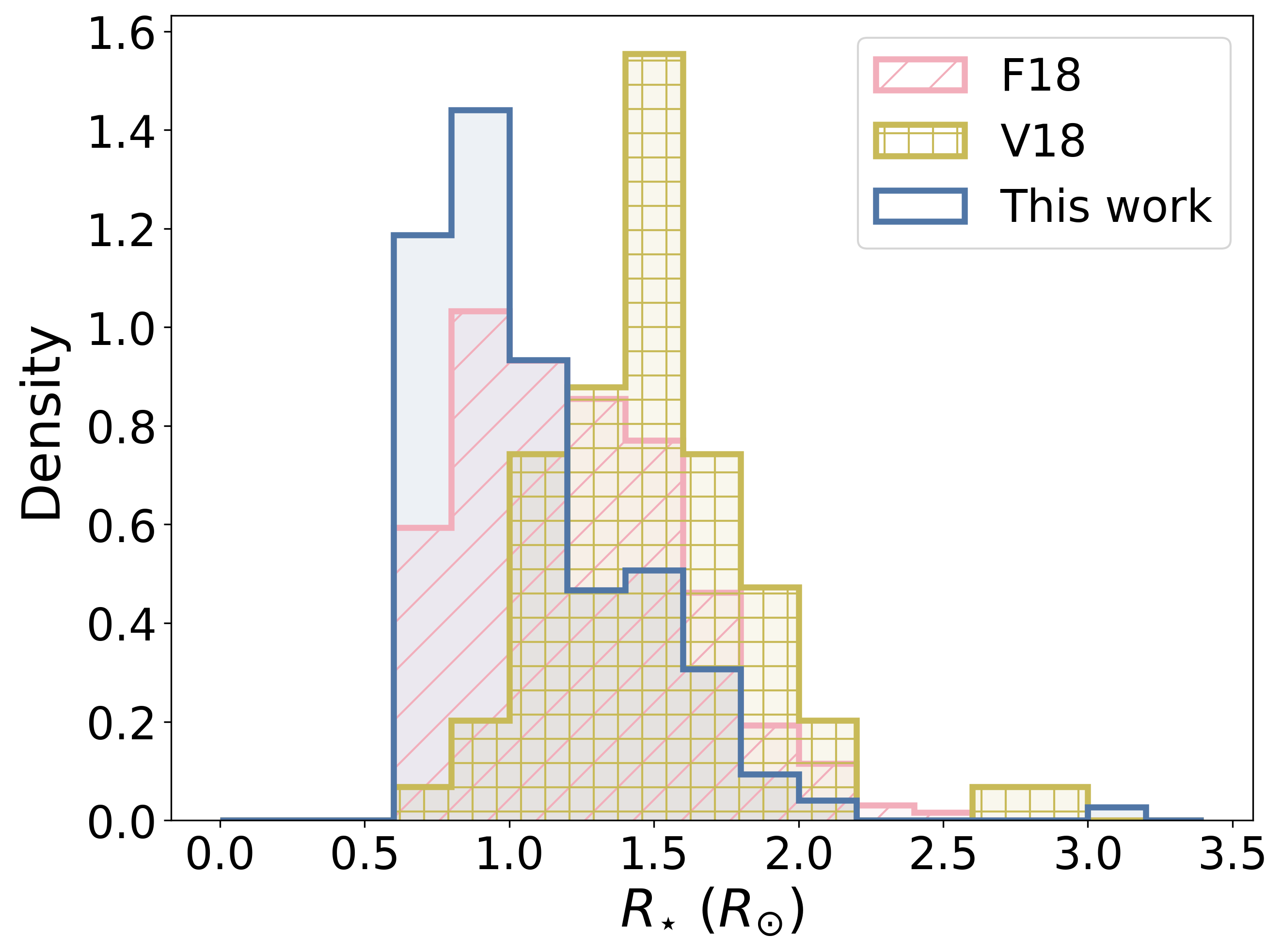} &
    \includegraphics[width=0.99\columnwidth]{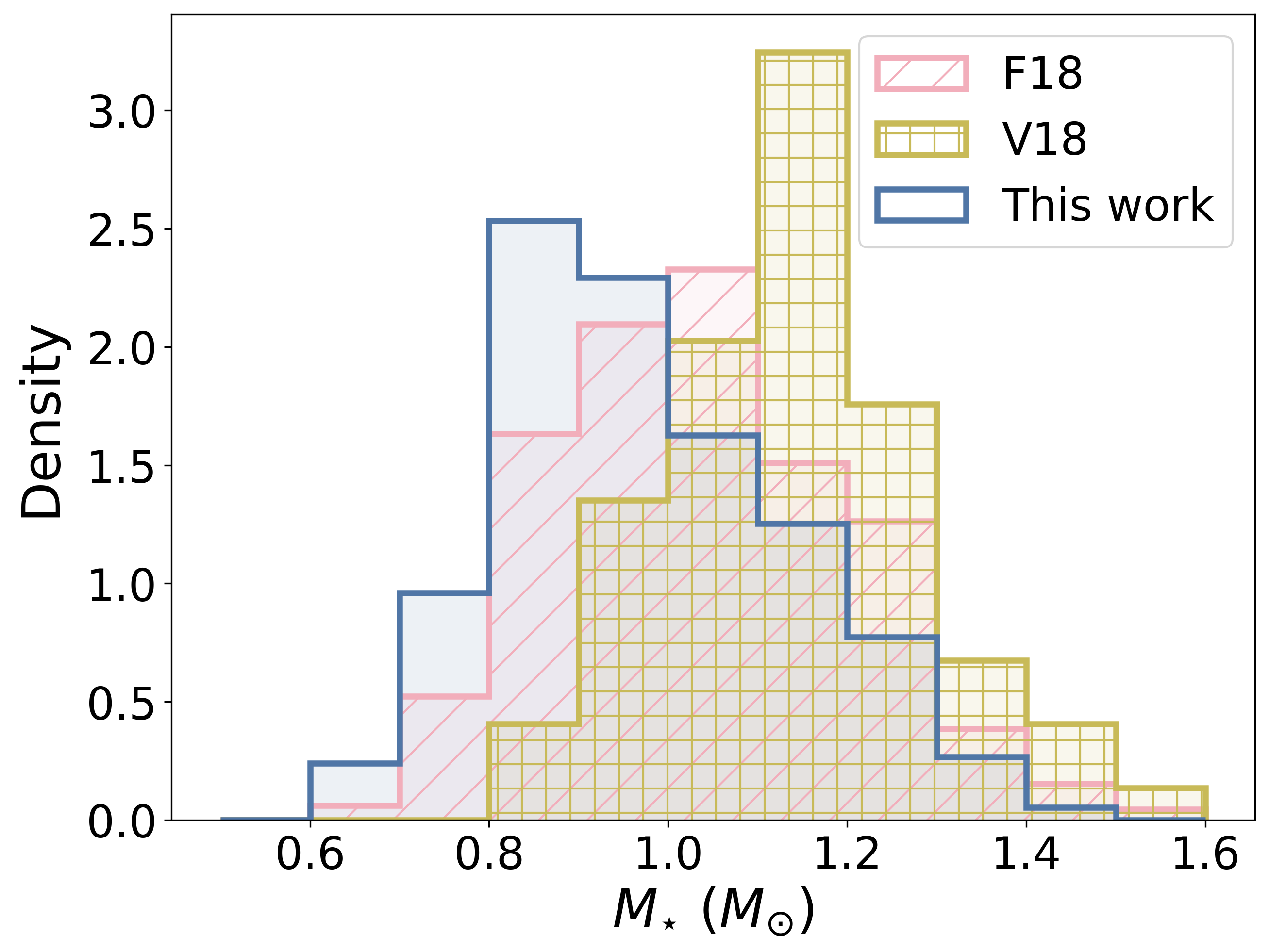} \\
    \includegraphics[width=0.99\columnwidth]{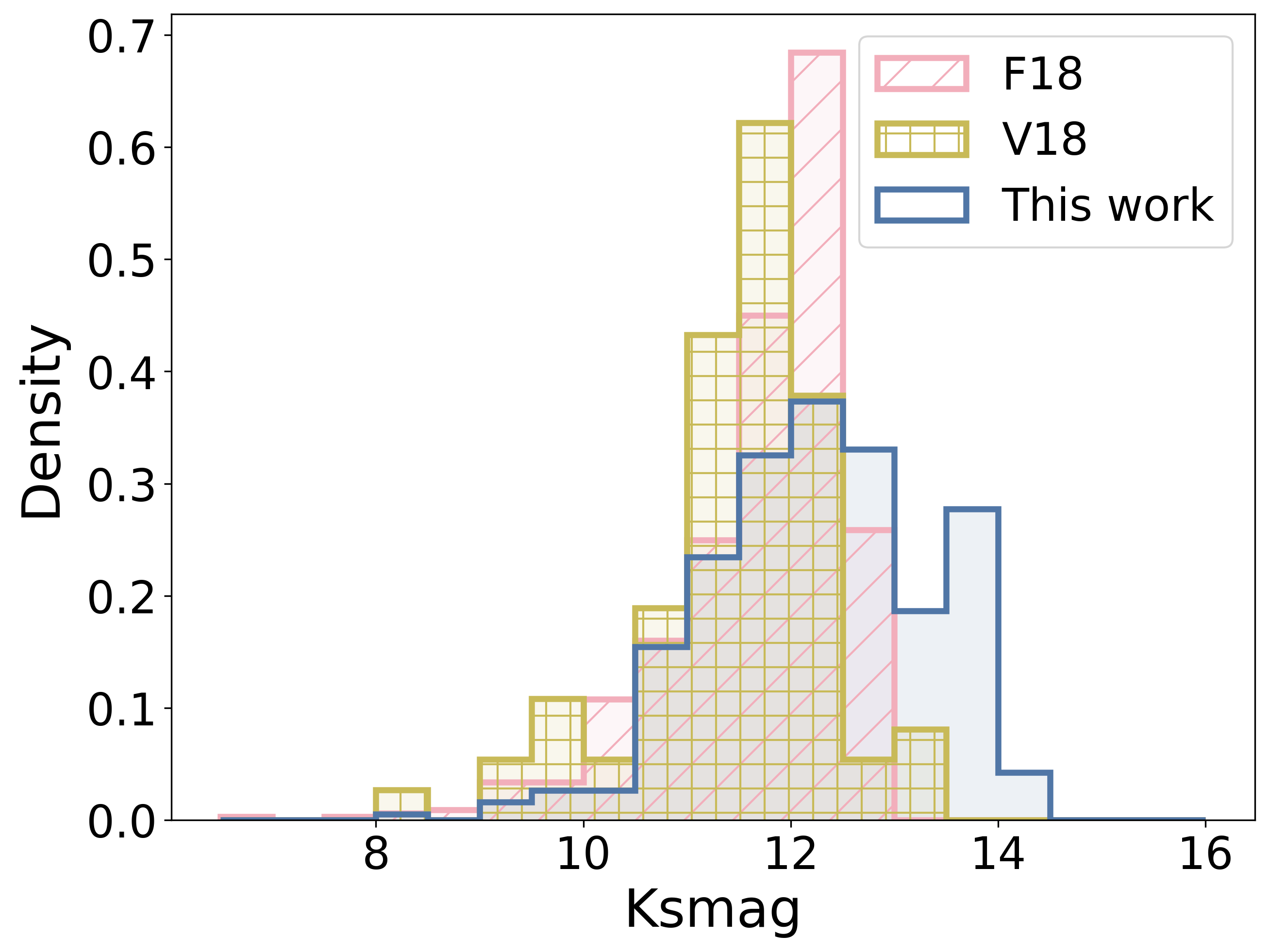} &
    \includegraphics[width=0.99\columnwidth]{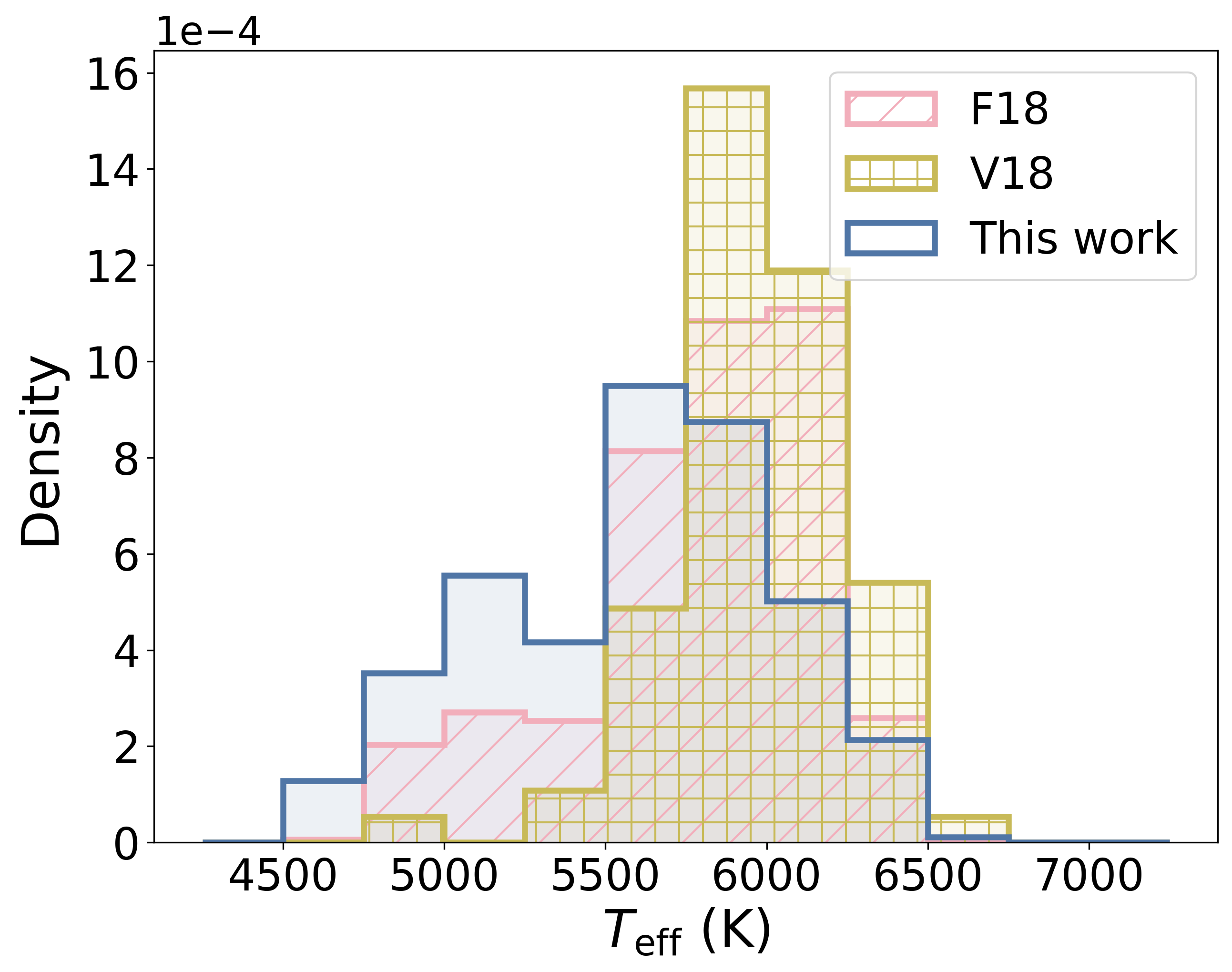}
    \end{tabular}
    \caption{Distribution of host stars parameters in our work, compared with \citetalias{fulton2018california} and \citetalias{vaneylen2018asteroseismic}. The properties shown here, from top left to bottom right, are stellar radius $R_{\star}$, mass $M_{\star}$, \textit{Kepler} magnitude Ksmag, and effective temperature $T_{\text{eff}}$. For systems with multiple transiting planets, stars are counted multiple times.} 
    \label{fig:hist_params}
\end{figure*}

Figure~\ref{fig:hist_params} shows the stellar parameter distributions for the planet host stars in the three planet samples, and the mean and median values are listed in Table~\ref{tab:stellar_params_mean}. We notice a similar stellar parameter range between this work and \citetalias{fulton2018california}, however the stars in \citetalias{vaneylen2018asteroseismic} are brighter, have a larger mean radius and mass, and higher effective temperature. This is likely due to \citetalias{vaneylen2018asteroseismic} selecting stars which display strong asteroseismic signals, which usually are brighter and larger stars. This observation may indicate that the radius valley of such stars are emptier, however the details are left for future studies.

Despite our new results revealing that the radius valley deepens by refitting planets with 1-minute short cadence light curves, it is still uncertain whether the difference between results from this work and \citetalias{fulton2018california} is solely due to the cadence in transit data used, as different methods are used in the transit fitting process. 
\citet{mullally2015planetary} fitted planets using the method described in \citet{rowe2014validation}, which first fits a multi-planet transit model to the light curves, with fixed limb darkening parameters from \citet{claret2011gravity}, and subsequently fitting for each planet in a system independently by removing photometric contributions of other planets based on the parameters from the multi-planet fit. In our work, we fit planets in multi-planet systems simultaneously, such that each system shares the same stellar parameters including limb-darkening parameters and stellar density. \citet{mullally2015planetary} assumed a circular orbit when performing the transit fits. On the contrary, we leave orbital eccentricity $e$ as a free parameter, and place a prior on $e$ based on the expected distribution from \citet{vaneylen2019orbital} and the stellar density $\rho_{\star}$ from \citetalias{fulton2018california}. However, most of the planets in our sample have near-circular orbits, with over 85\% of planets having $e < 0.1$. Therefore planetary orbital eccentricity is not sufficient to explain the difference between the two results. The possible presence of TTVs also do not contribute to the discrepancy as planets with known TTVs are excluded in our like-for-like planet comparisons. In fact, when fitting transits using identical methods, precisions in $R_p/R_{\star}$ obtained from fitting transit light curves of shorter cadences has been found to substantially improve, compared to 30-minute cadence light curves (Camero, Ho \& Van Eylen, in prep.). We therefore expect the photometry cadence to contribute significantly to the difference in the views of the radius valley. Further work, such as refitting the long cadence data of the same planet population with identical transit fitting methods, is needed to further investigate the effect of light curve cadence on planet parameter estimates and the radius valley. We leave such considerations for future studies.

\begin{table*}
    \centering
    \caption{Average values of stellar properties of the host stars in the planetary sample used in this work, compared with \citetalias{fulton2018california} and \citetalias{vaneylen2018asteroseismic}. $\bar{x}$ and $\tilde{x}$ represent the mean and median values of the parameter $x$ respectively.}
    \begin{tabular}{ccccccccc}
    Sample & $\bar{R_{\star}}$ ($R_{\sun}$) & $\tilde{R_{\star}}$ ($R_{\sun}$) & $\bar{M_{\star}}$ ($M_{\sun}$) & $\tilde{M_{\star}}$ ($M_{\sun}$) & $\bar{\text{Ksmag}}$ & $\tilde{\text{Ksmag}}$ & $\bar{T_{\text{eff}}}$ (K) & $\tilde{T_{\text{eff}}}$ (K) \\
    \hline
    This work & 1.08 & 0.99 & 0.98 & 0.95 & 12.27 & 12.31 & 5587 & 5630 \\
    \citetalias{fulton2018california} & 1.23 & 1.19 & 1.04 & 1.03 & 11.72 & 11.97 & 5788 & 5860 \\
    \citetalias{vaneylen2018asteroseismic} & 1.50 & 1.45 & 1.14 & 1.14 & 11.49 & 11.57 & 5980 & 5952
    \end{tabular}
    \label{tab:stellar_params_mean}
\end{table*}

\subsection{Radius valley relation with stellar age consistent with core-powered mass loss model} \label{subsect:discussion_age}
In Section~\ref{subsect:results_age}, we present a positive relationship between the radius valley location and stellar age. Photoevaporation is predicted to occur in the first 100~Myr of the planet's formation \citep{owen2017evaporation}, well before observations are able to detect the evolution signals, whereas core-powered mass loss occurs throughout the main-sequence lifetime of the stars, on Gyr timescales \citep{ginzburg2016super,gupta2019sculpting, gupta2020signatures}. Hence, in the photoevaporation case, the radius valley is expected to be located at a constant radius. On the other hand, in the core-powered mass loss case, the radius valley shifts to higher planet radii for older systems, as the atmospheres of planets with more massive cores are stripped off later in the evolution process than their less massive counterparts \citep[e.g.][]{david2021evolution, rogers2021unveiling}.

Our results reveal a weak positive radius valley dependence on the stellar age, which is consistent with the core-powered mass loss scenario, as is the observed radius valley dependence on stellar mass as discussed in Section~\ref{subsect:discuss_mstar}. However, a small age dependence does not preclude photoevaporation, since even in this scenario a subset of planets may still lose their atmospheres and evolve at Gyr timescales \citep{david2021evolution, rogers2021photoevaporation}, and we are unable to observe stars younger than 100 Myr and hence cannot rule out the possibility of a dominant photoevaporation effect on planets at the early stages of the stars' lifetime. Also, stellar age measurements are highly uncertain; the mean percentage uncertainty in stellar age for our sample is 54\%, hence there is also a probability that some stars are younger than observed.

Table~\ref{tab:valley_planets} lists the 50 planets located inside the radius valley in our sample. To do so, we here defined the new radius valley region as the area bounded by the two lines passing through the supporting vectors in the 4D SVM model in Section~\ref{subsect:results_age}, given by equation~\ref{eq:svm_4d} with $A = -0.096$, $B = 0.231$, $C=0.033$, $D{_\text{lower}} = 0.272$ for the lower line, and $D_{\text{upper}} = 0.405$ for the upper line. These planets are potentially interesting for future characterisation study as their atmospheres and interiors may provide additional insights regarding formation and evolution mechanisms.

\begin{table*}
    \renewcommand*{\arraystretch}{1.25}
    \centering
    \caption{List of planets inside the radius valley as defined by this work. The coordinates (RA and Dec) are taken from \textit{Kepler} Q1-17 Data Release 25 catalogue \citep{thompson2018planetary}, except for K02533.03, where data is taken from \textit{Kepler} Q1-16 catalogue \citep{mullally2015planetary}.}
    \begin{tabular}{cccccccc}
\hline 
KOI & Kepler name & $P$ (days) & $t_0$ (BJD-2454833) & $R_p/R_{\star}$ & $R_p$ ($R_{\earth}$) & RA (deg) & Dec (deg) \\ 
\hline 
K00049.01 & Kepler-461 b & $8.313784 \pm 0.000015$ & $175.9915 \pm 0.0008$ & $0.0287 \pm 0.0010$ & $4.03_{-0.17}^{+0.17}$ & 292.24902 & 46.164822 \\ 
K00070.03 & Kepler-20 A d & $77.611598 \pm 0.000019$ & $164.7274 \pm 0.0005$ & $0.0263 \pm 0.0003$ & $2.52_{-0.07}^{+0.07}$ & 287.698 & 42.338718 \\ 
K00092.01 & KOI-92.01 & $65.704594 \pm 0.000018$ & $137.4419 \pm 0.0005$ & $0.0263 \pm 0.0006$ & $3.02_{-0.11}^{+0.11}$ & 283.37482 & 43.788219 \\ 
K00094.02 & Kepler-89 A c & $10.423684 \pm 0.000005$ & $138.0092 \pm 0.0005$ & $0.0266 \pm 0.0006$ & $3.97_{-0.13}^{+0.13}$ & 297.33307 & 41.891121 \\ 
K00105.01 & Kepler-463 b & $8.981015 \pm 0.000002$ & $136.6493 \pm 0.0005$ & $0.0313 \pm 0.0004$ & $3.65_{-0.10}^{+0.11}$ & 298.93704 & 44.85791 \\ 
K00107.01 & Kepler-464 b & $7.256964 \pm 0.000011$ & $134.0232 \pm 0.0007$ & $0.0198 \pm 0.0004$ & $3.46_{-0.11}^{+0.11}$ & 294.83517 & 48.982361 \\ 
K00108.01 & Kepler-103 b & $15.965333 \pm 0.000013$ & $142.1780 \pm 0.0006$ & $0.0221 \pm 0.0002$ & $3.49_{-0.04}^{+0.04}$ & 288.98456 & 40.064529 \\ 
K00111.03 & Kepler-104 A d & $51.755294 \pm 0.000018$ & $271.0894 \pm 0.0005$ & $0.0232 \pm 0.0003$ & $2.66_{-0.07}^{+0.07}$ & 287.60461 & 42.166779 \\ 
K00122.01 & Kepler-95 b & $11.523073 \pm 0.000005$ & $131.9686 \pm 0.0004$ & $0.0205 \pm 0.0002$ & $3.24_{-0.10}^{+0.10}$ & 284.48245 & 44.398041 \\ 
K00157.02 & Kepler-11 d & $22.687159 \pm 0.000014$ & $148.4549 \pm 0.0006$ & $0.0282 \pm 0.0006$ & $3.34_{-0.10}^{+0.11}$ & 297.11511 & 41.909142 \\ 
K00174.01 & Kepler-482 b & $56.354185 \pm 0.000019$ & $144.8366 \pm 0.0007$ & $0.0339 \pm 0.0013$ & $2.92_{-0.14}^{+0.14}$ & 296.82291 & 48.107552 \\ 
K00238.01 & Kepler-123 b & $17.232309 \pm 0.000018$ & $135.0923 \pm 0.0008$ & $0.0223 \pm 0.0006$ & $3.21_{-0.13}^{+0.13}$ & 296.99863 & 42.78196 \\ 
K00285.01 & Kepler-92 b & $13.748833 \pm 0.000015$ & $179.2788 \pm 0.0007$ & $0.0198 \pm 0.0003$ & $3.71_{-0.06}^{+0.06}$ & 289.08606 & 41.562958 \\ 
K00317.01 & Kepler-521 b & $22.208119 \pm 0.000015$ & $206.3592 \pm 0.0006$ & $0.0207 \pm 0.0003$ & $3.69_{-0.11}^{+0.11}$ & 298.81638 & 43.998039 \\ 
K00351.03 & Kepler-90 d & $59.737034 \pm 0.000020$ & $158.9612 \pm 0.0009$ & $0.0217 \pm 0.0006$ & $2.80_{-0.11}^{+0.11}$ & 284.4335 & 49.305161 \\ 
K00351.04 & Kepler-90 e & $91.940461 \pm 0.000020$ & $134.2987 \pm 0.0010$ & $0.0198 \pm 0.0007$ & $2.56_{-0.11}^{+0.11}$ & 284.4335 & 49.305161 \\ 
K00386.01 & Kepler-146 b & $31.158789 \pm 0.000019$ & $173.9038 \pm 0.0009$ & $0.0297 \pm 0.0007$ & $3.30_{-0.12}^{+0.12}$ & 294.11075 & 38.710232 \\ 
K00386.02 & Kepler-146 c & $76.732517 \pm 0.000020$ & $200.6716 \pm 0.0010$ & $0.0258 \pm 0.0013$ & $2.87_{-0.16}^{+0.16}$ & 294.11075 & 38.710232 \\ 
K00408.01 & Kepler-150 c & $7.381981 \pm 0.000007$ & $173.0729 \pm 0.0008$ & $0.0355 \pm 0.0005$ & $3.34_{-0.12}^{+0.12}$ & 288.2341 & 40.520901 \\ 
K00416.01 & Kepler-152 b & $18.207957 \pm 0.000015$ & $185.8427 \pm 0.0006$ & $0.0383 \pm 0.0006$ & $3.12_{-0.09}^{+0.10}$ & 286.86548 & 41.989079 \\ 
K00435.05 & Kepler-154 c & $62.302788 \pm 0.000020$ & $179.0982 \pm 0.0010$ & $0.0266 \pm 0.0010$ & $3.04_{-0.15}^{+0.15}$ & 289.78052 & 49.89653 \\ 
K00509.02 & Kepler-171 c & $11.463477 \pm 0.000013$ & $137.3859 \pm 0.0008$ & $0.0328 \pm 0.0008$ & $3.30_{-0.14}^{+0.15}$ & 296.77191 & 41.755539 \\ 
K00510.04 & Kepler-172 e & $35.118523 \pm 0.000020$ & $152.1229 \pm 0.0010$ & $0.0248 \pm 0.0021$ & $2.82_{-0.26}^{+0.26}$ & 283.36841 & 41.821861 \\ 
K00555.02 & Kepler-598 c & $86.494779 \pm 0.000020$ & $181.8831 \pm 0.0009$ & $0.0272 \pm 0.0008$ & $2.51_{-0.10}^{+0.10}$ & 293.12341 & 40.934769 \\ 
K00665.01 & Kepler-207 d & $5.868083 \pm 0.000009$ & $170.3244 \pm 0.0009$ & $0.0202 \pm 0.0004$ & $3.69_{-0.11}^{+0.11}$ & 290.03052 & 42.16605 \\ 
K00707.02 & Kepler-33 f & $41.028059 \pm 0.000019$ & $172.5788 \pm 0.0009$ & $0.0207 \pm 0.0004$ & $3.64_{-0.13}^{+0.13}$ & 289.07755 & 46.005219 \\ 
K00707.03 & Kepler-33 e & $31.784774 \pm 0.000020$ & $135.8721 \pm 0.0009$ & $0.0188 \pm 0.0005$ & $3.30_{-0.12}^{+0.12}$ & 289.07755 & 46.005219 \\ 
K00708.01 & Kepler-216 c & $17.406653 \pm 0.000017$ & $171.0063 \pm 0.0008$ & $0.0230 \pm 0.0006$ & $3.88_{-0.14}^{+0.14}$ & 293.72806 & 46.12915 \\ 
K00711.01 & Kepler-218 c & $44.699505 \pm 0.000020$ & $174.8232 \pm 0.0008$ & $0.0272 \pm 0.0007$ & $3.07_{-0.11}^{+0.11}$ & 295.41281 & 46.266472 \\ 
K00800.02 & Kepler-234 c & $7.212030 \pm 0.000017$ & $172.8172 \pm 0.0009$ & $0.0283 \pm 0.0012$ & $3.61_{-0.26}^{+0.28}$ & 291.65353 & 38.494659 \\ 
K00834.02 & Kepler-238 d & $13.233546 \pm 0.000019$ & $140.3211 \pm 0.0010$ & $0.0197 \pm 0.0005$ & $3.37_{-0.18}^{+0.19}$ & 287.89713 & 40.637821 \\ 
K00834.05 & Kepler-238 f & $50.447315 \pm 0.000020$ & $178.4929 \pm 0.0010$ & $0.0203 \pm 0.0009$ & $3.48_{-0.22}^{+0.23}$ & 287.89713 & 40.637821 \\ 
K00881.01 & Kepler-712 b & $21.022471 \pm 0.000017$ & $207.6765 \pm 0.0007$ & $0.0391 \pm 0.0008$ & $3.14_{-0.17}^{+0.18}$ & 294.90973 & 42.935261 \\ 
K00907.04 & Kepler-251 e & $99.640965 \pm 0.000021$ & $198.6899 \pm 0.0009$ & $0.0320 \pm 0.0017$ & $2.82_{-0.19}^{+0.19}$ & 296.56622 & 44.105862 \\ 
K00921.02 & Kepler-253 d & $18.119898 \pm 0.000018$ & $182.6165 \pm 0.0008$ & $0.0346 \pm 0.0011$ & $3.01_{-0.15}^{+0.16}$ & 291.84198 & 44.858089 \\ 
K00934.01 & Kepler-254 b & $5.826654 \pm 0.000006$ & $173.0110 \pm 0.0009$ & $0.0368 \pm 0.0011$ & $3.62_{-0.24}^{+0.26}$ & 288.1647 & 45.816509 \\ 
K00941.01 & Kepler-257 c & $6.581482 \pm 0.000009$ & $174.7876 \pm 0.0009$ & $0.0429 \pm 0.0009$ & $3.75_{-0.15}^{+0.15}$ & 297.31598 & 46.023258 \\ 
K00954.02 & Kepler-259 c & $36.924954 \pm 0.000019$ & $174.2245 \pm 0.0010$ & $0.0291 \pm 0.0017$ & $2.85_{-0.19}^{+0.19}$ & 288.21194 & 46.615002 \\ 
K01001.01 & Kepler-264 b & $40.806846 \pm 0.000020$ & $155.7126 \pm 0.0009$ & $0.0159 \pm 0.0003$ & $3.39_{-0.12}^{+0.12}$ & 292.04462 & 37.37624 \\ 
K01198.01 & Kepler-275 c & $16.088329 \pm 0.000018$ & $139.4928 \pm 0.0010$ & $0.0232 \pm 0.0011$ & $3.55_{-0.29}^{+0.31}$ & 292.47971 & 38.514919 \\ 
K01215.02 & Kepler-277 c & $33.006310 \pm 0.000020$ & $145.3914 \pm 0.0010$ & $0.0161 \pm 0.0008$ & $3.04_{-0.17}^{+0.17}$ & 286.58316 & 39.077202 \\ 
K01270.01 & Kepler-57 b & $5.729326 \pm 0.000005$ & $138.5598 \pm 0.0009$ & $0.0356 \pm 0.0024$ & $3.16_{-0.24}^{+0.24}$ & 293.6413 & 44.65704 \\ 
K01486.02 & Kepler-302 b & $30.183689 \pm 0.000018$ & $146.6480 \pm 0.0009$ & $0.0286 \pm 0.0011$ & $3.27_{-0.20}^{+0.21}$ & 294.31699 & 43.629341 \\ 
K01563.04 & Kepler-305 d & $16.738655 \pm 0.000019$ & $359.4178 \pm 0.0009$ & $0.0338 \pm 0.0022$ & $2.84_{-0.22}^{+0.22}$ & 299.22433 & 40.343182 \\ 
K01598.01 & Kepler-310 c & $56.476167 \pm 0.000019$ & $143.8052 \pm 0.0008$ & $0.0301 \pm 0.0007$ & $2.75_{-0.10}^{+0.10}$ & 288.83936 & 46.98674 \\ 
K02051.01 & Kepler-355 c & $25.762459 \pm 0.000020$ & $147.7050 \pm 0.0009$ & $0.0230 \pm 0.0010$ & $2.95_{-0.17}^{+0.18}$ & 285.79947 & 42.811779 \\ 
K02390.01 & Kepler-1219 b & $16.104672 \pm 0.000020$ & $135.0156 \pm 0.0010$ & $0.0120 \pm 0.0011$ & $3.51_{-0.46}^{+0.46}$ & 297.21579 & 47.378521 \\ 
K02414.02 & Kepler-384 c & $45.348527 \pm 0.000020$ & $142.2527 \pm 0.0010$ & $0.0157 \pm 0.0012$ & $2.78_{-0.22}^{+0.22}$ & 286.02612 & 44.782871 \\ 
K02533.03 & KOI-2533.03 & $26.115290 \pm 0.000019$ & $145.5642 \pm 0.0010$ & $0.0122 \pm 0.0011$ & $4.01_{-0.38}^{+0.38}$ & 286.71564 & 48.645279 \\ 
K02639.01 & KOI-2639.01 & $25.108060 \pm 0.000020$ & $146.4407 \pm 0.0010$ & $0.0191 \pm 0.0083$ & $3.69_{-1.62}^{+1.63}$ & 285.36517 & 49.201561 \\ 
\hline
    \end{tabular}
    \label{tab:valley_planets}
\end{table*}

\subsection{Radius valley depth varies with stellar metallicity} \label{subsect:discussion_feh}
\begin{figure}
    \centering
    \includegraphics[width=\columnwidth]{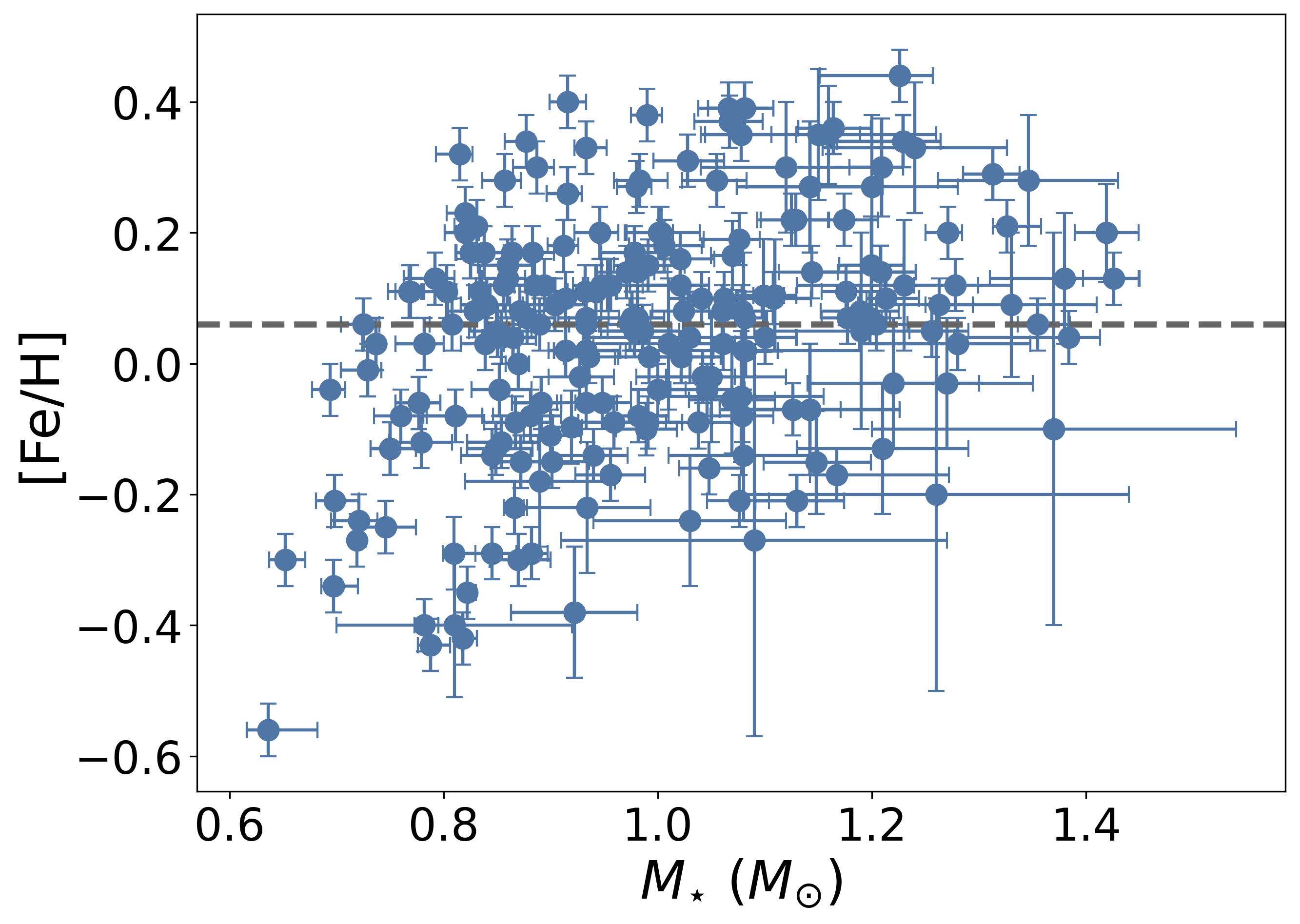}
    \caption{Metallicity and mass of the host stars used in this work. The grey dotted line [Fe/H] $=0.06$ shows the cut-off between stars of low and high metallicity as defined in this work. We note a lack of stars with low metallicity and large stellar mass.}
    \label{fig:fehM_plot}
\end{figure}

In Section~\ref{subsect:results_feh}, we show a higher average $E$ value (i.e. a deeper radius valley) for planets around metal-poor stars. This seems to contradict the suggestion that the radius valley is deeper for planets around metal-rich stars \citep{owen2018metallicity}. However, we note from Figure~\ref{fig:fehM_plot}, that in our sample, the metal-rich host stars span a wider range of stellar masses, due to lack of metal-poor stars with large radii. As from Section~\ref{subsect:ms_results} we observe that the radius valley depends on stellar mass as well, the superposition of the radius valley for different stellar masses potentially smears the gap, making the radius valley appear shallower.

The degeneracy between stellar mass and metallicity is not fully resolved, hence we are unable to determine the sole effect of stellar metallicity on the radius valley in this work. We therefore consider the results related to metallicity to be inconclusive and in need of further future study.

\section{Conclusion} \label{sect:conclusion}
In summary, we performed transit light curve fitting on 431 planets using \textit{Kepler} 1-minute short cadence data, the vast majority of which have not been previously analysed homogeneously using short cadence observations. In this paper, we presented their revised planetary parameters, which in some cases differ substantially from those previously reported. These differences are unrelated to stellar parameters but may be related to the details of the transit fitting approach or the shorter observing cadence, the effects of which should be disentangled in future studies.

By statistically analysing the small close-in planets in our sample, we observed a radius valley which is deeper than that reported in several other studies, although not entirely empty. The valley's depth likely implies a homogeneous initial planetary core composition where the planets are similar in composition at formation, and likely to have similar iron fractions. We provide a table of those planets that appear to be inside the valley, as they may warrant further study. 

The radius valley has a strong dependence on planetary orbital period and the mass of the host star. It also displays a weak dependence on the stellar age. We compared several possible radius valley models using support vector machines. We determined that the radius valley can best be described in four dimensions using the formula
    \begin{equation}
        R_{p, \text{valley}} \propto P^{A}M_{\star}^{B}\left(\text{age}\right)^C
    \end{equation}
    with $A = -0.096_{-0.027}^{+0.023}$, $B = 0.231_{-0.064}^{+0.053}$, and $C = 0.033_{-0.025}^{+0.017}$.

Comparing our radius valley dependencies with theoretical models, we found that in $R_p$--$S$--$M_{\star}$ space, our posterior distributions are most consistent with core-powered mass loss, where they agree within less than 1$\sigma$. The models are also consistent with photoevaporation scenarios at $\approx 2\sigma$. We did not find a significant dependence of the radius valley on stellar metallicity.

With the Transiting Exoplanet Survey Satellite \citep[TESS, e.g.][]{ricker2015transiting} now in its extended mission, and the upcoming launch of the PLAnetary Transits and Oscillations of stars (PLATO) mission \citep[e.g.][]{rauer2014plato}, such future planetary studies could drastically increase the number of planets with radii measurements and hence provide an even more detailed view of the radius valley. This work highlights the impact of careful transit fitting using short, 1-minute cadence observations to obtain precise planetary radii. This will likely be of key importance to derive precise planetary radii using transit observations from ongoing and future missions, which will ultimately allow us to better understand the formation and evolution of small close-in planets.

\section*{Acknowledgements}
C.S.K.H. would like to thank the Science and Technology Facilities Council (STFC) for funding support through a PhD studentship. We would like to thank Erik Petigura, James Rogers, and James Owen for insightful discussions. We also thank the anonymous reviewer for taking their time to review the paper, and for their valuable suggestions which have improved the manuscript.

\section*{Data Availability}
The \textit{Kepler} 1-minute short cadence light curves are available for download on the NASA Mikulski Archive for Space Telescopes (MAST) database\footnote{\url{https://archive.stsci.edu}}. The parameter estimates from HMC posteriors are provided in the Appendix tables.
 



\bibliographystyle{mnras}
\bibliography{paper} 

\begin{thebibliography}{}
\makeatletter
\relax
\def\mn@urlcharsother{\let\do\@makeother \do\$\do\&\do\#\do\^\do\_\do\%\do\~}
\def\mn@doi{\begingroup\mn@urlcharsother \@ifnextchar [ {\mn@doi@}
  {\mn@doi@[]}}
\def\mn@doi@[#1]#2{\def\@tempa{#1}\ifx\@tempa\@empty \href
  {http://dx.doi.org/#2} {doi:#2}\else \href {http://dx.doi.org/#2} {#1}\fi
  \endgroup}
\def\mn@eprint#1#2{\mn@eprint@#1:#2::\@nil}
\def\mn@eprint@arXiv#1{\href {http://arxiv.org/abs/#1} {{\tt arXiv:#1}}}
\def\mn@eprint@dblp#1{\href {http://dblp.uni-trier.de/rec/bibtex/#1.xml}
  {dblp:#1}}
\def\mn@eprint@#1:#2:#3:#4\@nil{\def\@tempa {#1}\def\@tempb {#2}\def\@tempc
  {#3}\ifx \@tempc \@empty \let \@tempc \@tempb \let \@tempb \@tempa \fi \ifx
  \@tempb \@empty \def\@tempb {arXiv}\fi \@ifundefined
  {mn@eprint@\@tempb}{\@tempb:\@tempc}{\expandafter \expandafter \csname
  mn@eprint@\@tempb\endcsname \expandafter{\@tempc}}}

\bibitem[\protect\citeauthoryear{{Astropy Collaboration} et~al.,}{{Astropy
  Collaboration} et~al.}{2013}]{astropy2013}
{Astropy Collaboration} et~al., 2013, \mn@doi [\aap]
  {10.1051/0004-6361/201322068}, \href
  {https://ui.adsabs.harvard.edu/abs/2013A&A...558A..33A} {558, A33}

\bibitem[\protect\citeauthoryear{{Astropy Collaboration} et~al.,}{{Astropy
  Collaboration} et~al.}{2018}]{astropy2018}
{Astropy Collaboration} et~al., 2018, \mn@doi [\aj] {10.3847/1538-3881/aabc4f},
  \href {https://ui.adsabs.harvard.edu/abs/2018AJ....156..123A} {156, 123}

\bibitem[\protect\citeauthoryear{{Berger}, {Huber}, {Gaidos}, {van Saders}  \&
  {Weiss}}{{Berger} et~al.}{2020}]{berger2020gaia2}
{Berger} T.~A.,  {Huber} D.,  {Gaidos} E.,  {van Saders} J.~L.,   {Weiss}
  L.~M.,  2020, \mn@doi [\aj] {10.3847/1538-3881/aba18a}, \href
  {https://ui.adsabs.harvard.edu/abs/2020AJ....160..108B} {160, 108}

\bibitem[\protect\citeauthoryear{Bourque et~al.,}{Bourque
  et~al.}{2021}]{exoctk2021}
Bourque M.,  et~al., 2021, The Exoplanet Characterization Toolkit (ExoCTK),
  \mn@doi{10.5281/zenodo.4556063}, \url
  {https://doi.org/10.5281/zenodo.4556063}

\bibitem[\protect\citeauthoryear{{Claret} \& {Bloemen}}{{Claret} \&
  {Bloemen}}{2011}]{claret2011gravity}
{Claret} A.,  {Bloemen} S.,  2011, \mn@doi [\aap]
  {10.1051/0004-6361/201116451}, \href
  {https://ui.adsabs.harvard.edu/abs/2011A&A...529A..75C} {529, A75}

\bibitem[\protect\citeauthoryear{{Cloutier} \& {Menou}}{{Cloutier} \&
  {Menou}}{2020}]{cloutier2020evolution}
{Cloutier} R.,  {Menou} K.,  2020, \mn@doi [\aj] {10.3847/1538-3881/ab8237},
  \href {https://ui.adsabs.harvard.edu/abs/2020AJ....159..211C} {159, 211}

\bibitem[\protect\citeauthoryear{{David} et~al.,}{{David}
  et~al.}{2021}]{david2021evolution}
{David} T.~J.,  et~al., 2021, \mn@doi [\aj] {10.3847/1538-3881/abf439}, \href
  {https://ui.adsabs.harvard.edu/abs/2021AJ....161..265D} {161, 265}

\bibitem[\protect\citeauthoryear{{Foreman-Mackey} et~al.,}{{Foreman-Mackey}
  et~al.}{2021}]{foremanmackey2021exoplanet}
{Foreman-Mackey} D.,  et~al., 2021, \mn@doi [The Journal of Open Source
  Software] {10.21105/joss.03285}, \href
  {https://ui.adsabs.harvard.edu/abs/2021JOSS....6.3285F} {6, 3285}

\bibitem[\protect\citeauthoryear{{Fulton} \& {Petigura}}{{Fulton} \&
  {Petigura}}{2018}]{fulton2018california}
{Fulton} B.~J.,  {Petigura} E.~A.,  2018, \mn@doi [\aj]
  {10.3847/1538-3881/aae828}, \href
  {https://ui.adsabs.harvard.edu/abs/2018AJ....156..264F} {156, 264}

\bibitem[\protect\citeauthoryear{{Fulton} et~al.,}{{Fulton}
  et~al.}{2017}]{fulton2017california}
{Fulton} B.~J.,  et~al., 2017, \mn@doi [\aj] {10.3847/1538-3881/aa80eb}, \href
  {https://ui.adsabs.harvard.edu/abs/2017AJ....154..109F} {154, 109}

\bibitem[\protect\citeauthoryear{Furlan et~al.,}{Furlan
  et~al.}{2017}]{furlan2017kepler}
Furlan E.,  et~al., 2017, \aj, 153, 71

\bibitem[\protect\citeauthoryear{{Ginsburg} et~al.,}{{Ginsburg}
  et~al.}{2019}]{astroquery2019}
{Ginsburg} A.,  et~al., 2019, \mn@doi [\aj] {10.3847/1538-3881/aafc33}, \href
  {https://ui.adsabs.harvard.edu/abs/2019AJ....157...98G} {157, 98}

\bibitem[\protect\citeauthoryear{{Ginzburg}, {Schlichting}  \&
  {Sari}}{{Ginzburg} et~al.}{2016}]{ginzburg2016super}
{Ginzburg} S.,  {Schlichting} H.~E.,   {Sari} R.,  2016, \mn@doi [\apj]
  {10.3847/0004-637X/825/1/29}, \href
  {https://ui.adsabs.harvard.edu/abs/2016ApJ...825...29G} {825, 29}

\bibitem[\protect\citeauthoryear{{Ginzburg}, {Schlichting}  \&
  {Sari}}{{Ginzburg} et~al.}{2018}]{ginzburg2018core}
{Ginzburg} S.,  {Schlichting} H.~E.,   {Sari} R.,  2018, \mn@doi [\mnras]
  {10.1093/mnras/sty290}, \href
  {https://ui.adsabs.harvard.edu/abs/2018MNRAS.476..759G} {476, 759}

\bibitem[\protect\citeauthoryear{{Gupta} \& {Schlichting}}{{Gupta} \&
  {Schlichting}}{2019}]{gupta2019sculpting}
{Gupta} A.,  {Schlichting} H.~E.,  2019, \mn@doi [\mnras]
  {10.1093/mnras/stz1230}, \href
  {https://ui.adsabs.harvard.edu/abs/2019MNRAS.487...24G} {487, 24}

\bibitem[\protect\citeauthoryear{{Gupta} \& {Schlichting}}{{Gupta} \&
  {Schlichting}}{2020}]{gupta2020signatures}
{Gupta} A.,  {Schlichting} H.~E.,  2020, \mn@doi [\mnras]
  {10.1093/mnras/staa315}, \href
  {https://ui.adsabs.harvard.edu/abs/2020MNRAS.493..792G} {493, 792}

\bibitem[\protect\citeauthoryear{{Holczer} et~al.,}{{Holczer}
  et~al.}{2016}]{holczer2016transit}
{Holczer} T.,  et~al., 2016, \mn@doi [\apjs] {10.3847/0067-0049/225/1/9}, \href
  {https://ui.adsabs.harvard.edu/abs/2016ApJS..225....9H} {225, 9}

\bibitem[\protect\citeauthoryear{{Huber} et~al.,}{{Huber}
  et~al.}{2013}]{huber2013fundamental}
{Huber} D.,  et~al., 2013, \mn@doi [\apj] {10.1088/0004-637X/767/2/127}, \href
  {https://ui.adsabs.harvard.edu/abs/2013ApJ...767..127H} {767, 127}

\bibitem[\protect\citeauthoryear{{Kipping}}{{Kipping}}{2013}]{kipping2013efficient}
{Kipping} D.~M.,  2013, \mn@doi [\mnras] {10.1093/mnras/stt1435}, \href
  {https://ui.adsabs.harvard.edu/abs/2013MNRAS.435.2152K} {435, 2152}

\bibitem[\protect\citeauthoryear{{Lee} \& {Chiang}}{{Lee} \&
  {Chiang}}{2016}]{lee2016breeding}
{Lee} E.~J.,  {Chiang} E.,  2016, \mn@doi [\apj] {10.3847/0004-637X/817/2/90},
  \href {https://ui.adsabs.harvard.edu/abs/2016ApJ...817...90L} {817, 90}

\bibitem[\protect\citeauthoryear{{Lee}, {Chiang}  \& {Ormel}}{{Lee}
  et~al.}{2014}]{lee2014make}
{Lee} E.~J.,  {Chiang} E.,   {Ormel} C.~W.,  2014, \mn@doi [\apj]
  {10.1088/0004-637X/797/2/95}, \href
  {https://ui.adsabs.harvard.edu/abs/2014ApJ...797...95L} {797, 95}

\bibitem[\protect\citeauthoryear{{Lightkurve Collaboration}
  et~al.,}{{Lightkurve Collaboration} et~al.}{2018}]{2018ascl.soft12013L}
{Lightkurve Collaboration} et~al., 2018, {Lightkurve: Kepler and TESS time
  series analysis in Python}, Astrophysics Source Code Library (\mn@eprint
  {ascl} {1812.013})

\bibitem[\protect\citeauthoryear{{Lopez} \& {Fortney}}{{Lopez} \&
  {Fortney}}{2013}]{lopez2013role}
{Lopez} E.~D.,  {Fortney} J.~J.,  2013, \mn@doi [\apj]
  {10.1088/0004-637X/776/1/2}, \href
  {https://ui.adsabs.harvard.edu/abs/2013ApJ...776....2L} {776, 2}

\bibitem[\protect\citeauthoryear{{Lopez} \& {Rice}}{{Lopez} \&
  {Rice}}{2018}]{lopez2018how}
{Lopez} E.~D.,  {Rice} K.,  2018, \mn@doi [\mnras] {10.1093/mnras/sty1707},
  \href {https://ui.adsabs.harvard.edu/abs/2018MNRAS.479.5303L} {479, 5303}

\bibitem[\protect\citeauthoryear{{Loyd}, {Shkolnik}, {Schneider},
  {Richey-Yowell}, {Barman}, {Peacock}  \& {Pagano}}{{Loyd}
  et~al.}{2020}]{loyd2020current}
{Loyd} R.~O.~P.,  {Shkolnik} E.~L.,  {Schneider} A.~C.,  {Richey-Yowell} T.,
  {Barman} T.~S.,  {Peacock} S.,   {Pagano} I.,  2020, \mn@doi [\apj]
  {10.3847/1538-4357/ab6605}, \href
  {https://ui.adsabs.harvard.edu/abs/2020ApJ...890...23L} {890, 23}

\bibitem[\protect\citeauthoryear{{Lundkvist} et~al.,}{{Lundkvist}
  et~al.}{2016}]{lundkvist2016hot}
{Lundkvist} M.~S.,  et~al., 2016, \mn@doi [Nature Communications]
  {10.1038/ncomms11201}, \href
  {https://ui.adsabs.harvard.edu/abs/2016NatCo...711201L} {7, 11201}

\bibitem[\protect\citeauthoryear{{Luque} \& {Pall{\'e}}}{{Luque} \&
  {Pall{\'e}}}{2022}]{luque2022density}
{Luque} R.,  {Pall{\'e}} E.,  2022, \mn@doi [Science]
  {10.1126/science.abl7164}, \href
  {https://ui.adsabs.harvard.edu/abs/2022Sci...377.1211L} {377, 1211}

\bibitem[\protect\citeauthoryear{{MacDonald}}{{MacDonald}}{2019}]{macdonald2019examining}
{MacDonald} M.~G.,  2019, \mn@doi [\mnras] {10.1093/mnras/stz1480}, \href
  {https://ui.adsabs.harvard.edu/abs/2019MNRAS.487.5062M} {487, 5062}

\bibitem[\protect\citeauthoryear{{Martinez}, {Cunha}, {Ghezzi}  \&
  {Smith}}{{Martinez} et~al.}{2019}]{martinez2019spectroscopic}
{Martinez} C.~F.,  {Cunha} K.,  {Ghezzi} L.,   {Smith} V.~V.,  2019, \mn@doi
  [\apj] {10.3847/1538-4357/ab0d93}, \href
  {https://ui.adsabs.harvard.edu/abs/2019ApJ...875...29M} {875, 29}

\bibitem[\protect\citeauthoryear{{Mathur} et~al.,}{{Mathur}
  et~al.}{2017}]{mathur2017revised}
{Mathur} S.,  et~al., 2017, \mn@doi [\apjs] {10.3847/1538-4365/229/2/30}, \href
  {https://ui.adsabs.harvard.edu/abs/2017ApJS..229...30M} {229, 30}

\bibitem[\protect\citeauthoryear{{Mordasini}}{{Mordasini}}{2020}]{mordasini2020planetary}
{Mordasini} C.,  2020, \mn@doi [\aap] {10.1051/0004-6361/201935541}, \href
  {https://ui.adsabs.harvard.edu/abs/2020A&A...638A..52M} {638, A52}

\bibitem[\protect\citeauthoryear{{Mullally} et~al.,}{{Mullally}
  et~al.}{2015}]{mullally2015planetary}
{Mullally} F.,  et~al., 2015, \mn@doi [\apjs] {10.1088/0067-0049/217/2/31},
  \href {https://ui.adsabs.harvard.edu/abs/2015ApJS..217...31M} {217, 31}

\bibitem[\protect\citeauthoryear{{Owen} \& {Adams}}{{Owen} \&
  {Adams}}{2019}]{owen2019effects}
{Owen} J.~E.,  {Adams} F.~C.,  2019, \mn@doi [\mnras] {10.1093/mnras/stz2601},
  \href {https://ui.adsabs.harvard.edu/abs/2019MNRAS.490...15O} {490, 15}

\bibitem[\protect\citeauthoryear{{Owen} \& {Murray-Clay}}{{Owen} \&
  {Murray-Clay}}{2018}]{owen2018metallicity}
{Owen} J.~E.,  {Murray-Clay} R.,  2018, \mn@doi [\mnras]
  {10.1093/mnras/sty1943}, \href
  {https://ui.adsabs.harvard.edu/abs/2018MNRAS.480.2206O} {480, 2206}

\bibitem[\protect\citeauthoryear{{Owen} \& {Wu}}{{Owen} \&
  {Wu}}{2013}]{owen2013kepler}
{Owen} J.~E.,  {Wu} Y.,  2013, \mn@doi [\apj] {10.1088/0004-637X/775/2/105},
  \href {https://ui.adsabs.harvard.edu/abs/2013ApJ...775..105O} {775, 105}

\bibitem[\protect\citeauthoryear{{Owen} \& {Wu}}{{Owen} \&
  {Wu}}{2017}]{owen2017evaporation}
{Owen} J.~E.,  {Wu} Y.,  2017, \mn@doi [\apj] {10.3847/1538-4357/aa890a}, \href
  {https://ui.adsabs.harvard.edu/abs/2017ApJ...847...29O} {847, 29}

\bibitem[\protect\citeauthoryear{{Petigura}}{{Petigura}}{2020}]{petigura2020two}
{Petigura} E.~A.,  2020, \mn@doi [\aj] {10.3847/1538-3881/ab9fff}, \href
  {https://ui.adsabs.harvard.edu/abs/2020AJ....160...89P} {160, 89}

\bibitem[\protect\citeauthoryear{{Petigura} et~al.,}{{Petigura}
  et~al.}{2017}]{petigura2017california}
{Petigura} E.~A.,  et~al., 2017, \mn@doi [\aj] {10.3847/1538-3881/aa80de},
  \href {https://ui.adsabs.harvard.edu/abs/2017AJ....154..107P} {154, 107}

\bibitem[\protect\citeauthoryear{{Petigura} et~al.,}{{Petigura}
  et~al.}{2022}]{petigura2022california}
{Petigura} E.~A.,  et~al., 2022, \mn@doi [\aj] {10.3847/1538-3881/ac51e3},
  \href {https://ui.adsabs.harvard.edu/abs/2022AJ....163..179P} {163, 179}

\bibitem[\protect\citeauthoryear{{Raghavan} et~al.,}{{Raghavan}
  et~al.}{2010}]{raghavan2010survey}
{Raghavan} D.,  et~al., 2010, \mn@doi [\apjs] {10.1088/0067-0049/190/1/1},
  \href {https://ui.adsabs.harvard.edu/abs/2010ApJS..190....1R} {190, 1}

\bibitem[\protect\citeauthoryear{{Rasmussen} \& {Williams}}{{Rasmussen} \&
  {Williams}}{2006}]{rasmussen2006gaussian}
{Rasmussen} C.~E.,  {Williams} C. K.~I.,  2006, {Gaussian Processes for Machine
  Learning}.
The MIT Press

\bibitem[\protect\citeauthoryear{{Rauer} et~al.,}{{Rauer}
  et~al.}{2014}]{rauer2014plato}
{Rauer} H.,  et~al., 2014, \mn@doi [Experimental Astronomy]
  {10.1007/s10686-014-9383-4}, \href
  {https://ui.adsabs.harvard.edu/abs/2014ExA....38..249R} {38, 249}

\bibitem[\protect\citeauthoryear{{Ricker} et~al.,}{{Ricker}
  et~al.}{2015}]{ricker2015transiting}
{Ricker} G.~R.,  et~al., 2015, \mn@doi [Journal of Astronomical Telescopes,
  Instruments, and Systems] {10.1117/1.JATIS.1.1.014003}, \href
  {https://ui.adsabs.harvard.edu/abs/2015JATIS...1a4003R} {1, 014003}

\bibitem[\protect\citeauthoryear{{Rogers} \& {Owen}}{{Rogers} \&
  {Owen}}{2021}]{rogers2021unveiling}
{Rogers} J.~G.,  {Owen} J.~E.,  2021, \mn@doi [\mnras] {10.1093/mnras/stab529},
  \href {https://ui.adsabs.harvard.edu/abs/2021MNRAS.503.1526R} {503, 1526}

\bibitem[\protect\citeauthoryear{{Rogers}, {Gupta}, {Owen}  \&
  {Schlichting}}{{Rogers} et~al.}{2021}]{rogers2021photoevaporation}
{Rogers} J.~G.,  {Gupta} A.,  {Owen} J.~E.,   {Schlichting} H.~E.,  2021,
  \mn@doi [\mnras] {10.1093/mnras/stab2897}, \href
  {https://ui.adsabs.harvard.edu/abs/2021MNRAS.508.5886R} {508, 5886}

\bibitem[\protect\citeauthoryear{{Rowe} et~al.,}{{Rowe}
  et~al.}{2014}]{rowe2014validation}
{Rowe} J.~F.,  et~al., 2014, \mn@doi [\apj] {10.1088/0004-637X/784/1/45}, \href
  {https://ui.adsabs.harvard.edu/abs/2014ApJ...784...45R} {784, 45}

\bibitem[\protect\citeauthoryear{Salvatier, Wiecki  \& Fonnesbeck}{Salvatier
  et~al.}{2016}]{salvatier2016probabilistic}
Salvatier J.,  Wiecki T.~V.,   Fonnesbeck C.,  2016, PeerJ Computer Science, 2,
  e55

\bibitem[\protect\citeauthoryear{{Silva Aguirre} et~al.,}{{Silva Aguirre}
  et~al.}{2015}]{silvaaguirre2015ages}
{Silva Aguirre} V.,  et~al., 2015, \mn@doi [\mnras] {10.1093/mnras/stv1388},
  \href {https://ui.adsabs.harvard.edu/abs/2015MNRAS.452.2127S} {452, 2127}

\bibitem[\protect\citeauthoryear{{Smith} et~al.,}{{Smith}
  et~al.}{2012}]{smith2012kepler}
{Smith} J.~C.,  et~al., 2012, \mn@doi [\pasp] {10.1086/667697}, \href
  {https://ui.adsabs.harvard.edu/abs/2012PASP..124.1000S} {124, 1000}

\bibitem[\protect\citeauthoryear{{Stumpe} et~al.,}{{Stumpe}
  et~al.}{2012}]{stumpe2012kepler}
{Stumpe} M.~C.,  et~al., 2012, \mn@doi [\pasp] {10.1086/667698}, \href
  {https://ui.adsabs.harvard.edu/abs/2012PASP..124..985S} {124, 985}

\bibitem[\protect\citeauthoryear{{Thompson} et~al.,}{{Thompson}
  et~al.}{2018}]{thompson2018planetary}
{Thompson} S.~E.,  et~al., 2018, \mn@doi [\apjs] {10.3847/1538-4365/aab4f9},
  \href {https://ui.adsabs.harvard.edu/abs/2018ApJS..235...38T} {235, 38}

\bibitem[\protect\citeauthoryear{{Van Eylen} \& {Albrecht}}{{Van Eylen} \&
  {Albrecht}}{2015}]{vaneylen2015eccentricity}
{Van Eylen} V.,  {Albrecht} S.,  2015, \mn@doi [\apj]
  {10.1088/0004-637X/808/2/126}, \href
  {https://ui.adsabs.harvard.edu/abs/2015ApJ...808..126V} {808, 126}

\bibitem[\protect\citeauthoryear{{Van Eylen}, {Agentoft}, {Lundkvist},
  {Kjeldsen}, {Owen}, {Fulton}, {Petigura}  \& {Snellen}}{{Van Eylen}
  et~al.}{2018}]{vaneylen2018asteroseismic}
{Van Eylen} V.,  {Agentoft} C.,  {Lundkvist} M.~S.,  {Kjeldsen} H.,  {Owen}
  J.~E.,  {Fulton} B.~J.,  {Petigura} E.,   {Snellen} I.,  2018, \mn@doi
  [\mnras] {10.1093/mnras/sty1783}, \href
  {https://ui.adsabs.harvard.edu/abs/2018MNRAS.479.4786V} {479, 4786}

\bibitem[\protect\citeauthoryear{{Van Eylen} et~al.,}{{Van Eylen}
  et~al.}{2019}]{vaneylen2019orbital}
{Van Eylen} V.,  et~al., 2019, \mn@doi [\aj] {10.3847/1538-3881/aaf22f}, \href
  {https://ui.adsabs.harvard.edu/abs/2019AJ....157...61V} {157, 61}

\bibitem[\protect\citeauthoryear{{Van Eylen} et~al.,}{{Van Eylen}
  et~al.}{2021}]{vaneylen2021masses}
{Van Eylen} V.,  et~al., 2021, \mn@doi [\mnras] {10.1093/mnras/stab2143}, \href
  {https://ui.adsabs.harvard.edu/abs/2021MNRAS.507.2154V} {507, 2154}

\makeatother
\end{thebibliography}




\appendix

\section{Extra Material} \label{appendix}
\begin{landscape}
\begin{table}
    \renewcommand*{\arraystretch}{1.5}
        \centering
        \caption{Host star parameters of planets fitted in this work. $R_{\star}$, $M_{\star}$, $T_{\text{eff}}$, [Fe/H], Ksmag are taken from the following sources: 1: \citet{fulton2018california}, 2: \citet{vaneylen2018asteroseismic}. Stellar ages are taken from \citet{fulton2018california}. Radius correction factors (RCFs) are taken from \citet{furlan2017kepler}; we take $\text{RCF} = 1.0000$ where there are no measurements in \citet{furlan2017kepler}.} $\rho_{\star}$, $u_0$, $u_1$ are resulting parameters from the transit fitting. Only the first 10 host stars are shown here; the full table is available online in a machine-readable format.
        {\scriptsize
        \begin{tabular}{llllllllllllllll}
        \hline 
KOI & $R_{\star}$ ($R_{\sun}$) (1) & $M_{\star}$ ($M_{\sun}$) (1) & $T_{\text{eff}}$ (K) (1) & [Fe/H] (dex) (1) & Ksmag (1) & $R_{\star}$ ($R_{\sun}$) (2) & $M_{\star}$ ($M_{\sun}$) (2) & $T_{\text{eff}}$ (K) (2) & [Fe/H] (dex) (2) & Ksmag (2) & Age (Gyr) & RCF & $\rho_*$ (g cm$^{-3}$) & $u_0$ & $u_1$ \\ 
\hline 
41 & $1.53_{-0.03}^{+0.03}$ & $1.10_{-0.03}^{+0.07}$ & $5854_{-60}^{+60}$ & $0.10 \pm 0.04$ & 9.768 & $1.51_{-0.01}^{+0.01}$ & $1.11_{-0.02}^{+0.02}$ & $5903_{-66}^{+53}$ & $0.10 \pm 0.09$ & 11.197 & $6.92_{-0.80}^{+1.91}$ & 1.0083 & $0.313 \pm 0.015$ & $0.46 \pm 0.08$ & $0.17 \pm 0.11$ \\ 
46 & $1.66_{-0.04}^{+0.05}$ & $1.24_{-0.09}^{+0.03}$ & $5661_{-60}^{+60}$ & $0.39 \pm 0.04$ & 12.011 & N/A & N/A & N/A & N/A & N/A & $5.25_{-1.69}^{+0.73}$ & 1.0000 & $0.279 \pm 0.025$ & $0.52 \pm 0.18$ & $0.27 \pm 0.16$ \\ 
49 & $1.29_{-0.03}^{+0.03}$ & $0.95_{-0.03}^{+0.03}$ & $5779_{-60}^{+60}$ & $-0.06 \pm 0.04$ & 11.917 & N/A & N/A & N/A & N/A & N/A & $10.47_{-1.21}^{+1.21}$ & 1.0000 & $0.441 \pm 0.021$ & $0.34 \pm 0.12$ & $0.25 \pm 0.13$ \\ 
69 & $0.94_{-0.02}^{+0.02}$ & $0.87_{-0.03}^{+0.03}$ & $5594_{-60}^{+60}$ & $-0.09 \pm 0.04$ & 8.37 & $0.91_{-0.02}^{+0.02}$ & $0.89_{-0.07}^{+0.07}$ & $5669_{-75}^{+75}$ & $-0.18 \pm 0.10$ & 9.931 & $9.12_{-2.10}^{+2.94}$ & 1.0000 & $1.044 \pm 0.055$ & $0.44 \pm 0.04$ & $0.16 \pm 0.06$ \\ 
70 & $0.88_{-0.02}^{+0.02}$ & $0.94_{-0.03}^{+0.02}$ & $5508_{-60}^{+60}$ & $0.11 \pm 0.04$ & 10.871 & N/A & N/A & N/A & N/A & N/A & $3.09_{-1.57}^{+1.99}$ & 1.0054 & $1.391 \pm 0.052$ & $0.45 \pm 0.04$ & $0.24 \pm 0.07$ \\ 
72 & $1.08_{-0.03}^{+0.03}$ & $0.87_{-0.01}^{+0.02}$ & $5599_{-60}^{+60}$ & $-0.11 \pm 0.04$ & 9.496 & $1.07_{-0.01}^{+0.01}$ & $0.92_{-0.02}^{+0.01}$ & $5678_{-49}^{+55}$ & $-0.10 \pm 0.06$ & 10.961 & $12.88_{-0.89}^{+1.48}$ & 1.0005 & $0.690 \pm 0.019$ & $0.38 \pm 0.10$ & $0.24 \pm 0.13$ \\ 
82 & $0.72_{-0.02}^{+0.02}$ & $0.80_{-0.02}^{+0.01}$ & $4909_{-60}^{+60}$ & $0.11 \pm 0.04$ & 9.351 & N/A & N/A & N/A & N/A & N/A & $1.07_{-1.58}^{+0.89}$ & 1.0000 & $2.039 \pm 0.055$ & $0.62 \pm 0.06$ & $0.09 \pm 0.07$ \\ 
85 & $1.44_{-0.03}^{+0.03}$ & $1.25_{-0.03}^{+0.01}$ & $6220_{-60}^{+60}$ & $0.13 \pm 0.04$ & 9.806 & $1.40_{-0.01}^{+0.01}$ & $1.20_{-0.03}^{+0.03}$ & $6193_{-52}^{+43}$ & $0.15 \pm 0.08$ & 11.018 & $2.95_{-0.41}^{+0.34}$ & 1.0001 & $0.428 \pm 0.019$ & $0.33 \pm 0.06$ & $0.25 \pm 0.07$ \\ 
92 & $1.06_{-0.02}^{+0.02}$ & $1.08_{-0.05}^{+0.03}$ & $5923_{-60}^{+60}$ & $0.09 \pm 0.04$ & 10.3 & $1.05_{-0.03}^{+0.03}$ & $1.08_{-0.11}^{+0.11}$ & $5952_{-119}^{+119}$ & $0.02 \pm 0.15$ & 11.667 & $2.34_{-1.40}^{+1.35}$ & 1.0000 & $0.894 \pm 0.049$ & $0.41 \pm 0.12$ & $0.26 \pm 0.13$ \\ 
94 & $1.37_{-0.03}^{+0.03}$ & $1.18_{-0.02}^{+0.03}$ & $6181_{-60}^{+60}$ & $0.07 \pm 0.04$ & 10.926 & N/A & N/A & N/A & N/A & N/A & $3.47_{-0.56}^{+0.64}$ & 1.0000 & $0.462 \pm 0.021$ & $0.33 \pm 0.11$ & $0.37 \pm 0.16$ \\ 
... & ... & ... & ... & ... & ... & ... & ... & ... & ... & ... & ... & ... & ... & ... \\ 
\hline
        \end{tabular}
        }
        \label{tab:appendix_stellar1}
\end{table}

\begin{table}
    \renewcommand*{\arraystretch}{1.5}
    \centering
    \caption{Planetary parameters from transit fits performed in this work. $P$, $t_0$, $R_p/R_{\star}$, $b$, $e$, $\omega$ are obtained directly from fitting, $a/R_{\star}$ and $S$ are indirectly calculated. The $R_p/R_{\star}$ values from 1: \citet{fulton2018california} and 2: \citet{vaneylen2018asteroseismic} are also included for comparison. Only the first 10 planets are shown here; the full table is available online in a machine-readable format.}
    {\scriptsize
    \begin{tabular}{llllllllllll}
    \hline 
KOI & Kepler name & $P$ (days) & $t_0$ (BJD-2454833) & $R_p/R_{\star}$ & $R_p/R_{\star}$ (1) & $R_p/R_{\star}$ (2) & $b$ & $e$ & $\omega$ ($^{\circ}$) & $a/R_*$ & $S$ ($S_{\earth}$) \\ 
\hline 
K00041.01 & Kepler-100 c & $12.815893 \pm 0.000008$ & $122.9476 \pm 0.0005$ & $0.0138 \pm 0.0002$ & $0.0140_{-0.0006}^{+0.0001}$ & $0.0134_{-0.0001}^{+0.0001}$ & $0.39 \pm 0.13$ & $0.06 \pm 0.04$ & $30 \pm 96$ & $14.36 \pm 1.20$ & $245.37 \pm 20.59$ \\ 
K00041.02 & Kepler-100 b & $6.887062 \pm 0.000007$ & $133.1781 \pm 0.0008$ & $0.0082 \pm 0.0001$ & $0.0081_{-0.0002}^{+0.0013}$ & $0.0079_{-0.0002}^{+0.0004}$ & $0.63 \pm 0.05$ & $0.05 \pm 0.04$ & $3 \pm 107$ & $9.26 \pm 0.88$ & $590.40 \pm 56.33$ \\ 
K00041.03 & Kepler-100 d & $35.333093 \pm 0.000019$ & $153.9835 \pm 0.0009$ & $0.0104 \pm 0.0002$ & $0.0092_{-0.0003}^{+0.0023}$ & $0.0092_{-0.0002}^{+0.0004}$ & $0.81 \pm 0.02$ & $0.05 \pm 0.04$ & $1 \pm 104$ & $27.50 \pm 2.51$ & $66.87 \pm 6.13$ \\ 
K00046.02 & Kepler-101 c & $6.029792 \pm 0.000020$ & $132.4816 \pm 0.0010$ & $0.0073 \pm 0.0007$ & $0.0069_{-0.0005}^{+0.0003}$ & N/A & $0.41 \pm 0.21$ & $0.05 \pm 0.04$ & $-0 \pm 103$ & $8.13 \pm 0.77$ & $646.80 \pm 61.97$ \\ 
K00049.01 & Kepler-461 b & $8.313784 \pm 0.000015$ & $175.9915 \pm 0.0008$ & $0.0287 \pm 0.0010$ & $0.0259_{-0.0003}^{+0.0003}$ & N/A & $0.75 \pm 0.11$ & $0.23 \pm 0.13$ & $77 \pm 77$ & $14.75 \pm 2.22$ & $213.71 \pm 32.30$ \\ 
K00069.01 & Kepler-93 b & $4.726739 \pm 0.000001$ & $134.9265 \pm 0.0001$ & $0.0151 \pm 0.0001$ & $0.0159_{-0.0008}^{+0.0002}$ & $0.0149_{-0.0001}^{+0.0001}$ & $0.23 \pm 0.13$ & $0.20 \pm 0.09$ & $105 \pm 58$ & $13.06 \pm 1.39$ & $252.14 \pm 27.04$ \\ 
K00070.01 & Kepler-20 A c & $10.854089 \pm 0.000003$ & $138.6082 \pm 0.0002$ & $0.0290 \pm 0.0002$ & $0.0300_{-0.0006}^{+0.0018}$ & N/A & $0.18 \pm 0.11$ & $0.09 \pm 0.03$ & $86 \pm 39$ & $22.49 \pm 0.80$ & $75.81 \pm 2.80$ \\ 
K00070.02 & Kepler-20 A b & $3.696115 \pm 0.000001$ & $134.5021 \pm 0.0002$ & $0.0182 \pm 0.0002$ & $0.0209_{-0.0023}^{+0.0014}$ & N/A & $0.49 \pm 0.08$ & $0.05 \pm 0.04$ & $26 \pm 101$ & $10.24 \pm 0.83$ & $365.46 \pm 29.71$ \\ 
K00070.03 & Kepler-20 A d & $77.611598 \pm 0.000019$ & $164.7274 \pm 0.0005$ & $0.0263 \pm 0.0003$ & $0.0259_{-0.0003}^{+0.0005}$ & N/A & $0.36 \pm 0.13$ & $0.06 \pm 0.04$ & $33 \pm 96$ & $78.67 \pm 6.41$ & $6.20 \pm 0.51$ \\ 
K00070.05 & Kepler-20 A f & $19.577627 \pm 0.000020$ & $135.2063 \pm 0.0009$ & $0.0091 \pm 0.0004$ & $0.0098_{-0.0003}^{+0.0004}$ & N/A & $0.65 \pm 0.06$ & $0.05 \pm 0.04$ & $7 \pm 104$ & $30.66 \pm 2.84$ & $40.78 \pm 3.81$ \\ 
... & ... & ... & ... & ... & ... & ... & ... & ... & ... \\ 
\hline
    \end{tabular}}
    \label{tab:appendix_planet}
\end{table}
\end{landscape}

\begin{table}
    \centering
    \caption{Transit jitter and GP parameters from transit fitting of planetary systems in this work. Only the first 10 systems are shown here; the full table is available online in a machine-readable format.}
    \begin{tabular}{llll}
    \hline
        KOI & $\log{\sigma_{\text{lc}}}$ & $\log{\sigma_{\text{gp}}}$ & $\log{\rho_{\text{gp}}}$ \\ 
    \hline
        41 & $-8.4379 \pm 0.0016$ & $-9.5822 \pm 0.0105$ & $-4.5121 \pm 0.0335$ \\ 
46 & $-7.1119 \pm 0.0037$ & $-9.1794 \pm 0.1025$ & $-4.3449 \pm 0.2492$ \\ 
49 & $-7.3268 \pm 0.0044$ & $-8.8595 \pm 0.0680$ & $-1.2568 \pm 0.1547$ \\ 
69 & $-8.9861 \pm 0.0019$ & $-10.1248 \pm 0.0133$ & $-4.8968 \pm 0.0382$ \\ 
70 & $-7.7689 \pm 0.0012$ & $-9.4665 \pm 0.0150$ & $-3.7110 \pm 0.0415$ \\ 
72 & $-8.4175 \pm 0.0053$ & $-9.6473 \pm 0.0392$ & $-4.0440 \pm 0.1139$ \\ 
82 & $-8.1671 \pm 0.0018$ & $-8.6193 \pm 0.0116$ & $-3.2278 \pm 0.0240$ \\ 
85 & $0.0000 \pm 0.0000$ & $0.0000 \pm 0.0000$ & $0.0000 \pm 0.0000$ \\ 
92 & $-8.1505 \pm 0.0062$ & $-9.8036 \pm 0.0880$ & $-3.9919 \pm 0.2448$ \\ 
94 & $-7.8819 \pm 0.0017$ & $-9.8028 \pm 0.0308$ & $-2.9898 \pm 0.1183$ \\ 
... & ... & ... & ... \\ 
\hline
    \end{tabular}
    \label{tab:appendix_gp}
\end{table}



\bsp	
\label{lastpage}
\end{document}